\documentclass[modern]{aastex631}

\shorttitle{Quantum mechanical simulations of the radical-radical chemistry on icy surfaces}
\shortauthors{Enrique-Romero et al.}

\usepackage{graphicx}
\usepackage{txfonts}
\hypersetup{final}
\usepackage{multirow}
\usepackage{makecell}
\usepackage{colortbl}
\usepackage[english]{babel}
\usepackage{xcolor}
\usepackage{rotating}
\usepackage{fancyvrb}
\usepackage{tikz}
\usepackage{titletoc}
\usepackage{hyperref}

\begin{document}

   \title{Quantum mechanical simulations of the radical-radical chemistry on icy surfaces}

\author[0000-0002-2147-7735]{Joan Enrique-Romero}
\affiliation{Univ. Grenoble Alpes, CNRS, Institut de Plan\'{e}tologie et d'Astrophysique de Grenoble (IPAG), 38000 Grenoble, France}
\affiliation{Departament de Qu\'{i}mica, Universitat Aut\`{o}noma de Barcelona, Bellaterra, 08193, Catalonia, Spain}

\author[0000-0002-9637-4554]{Albert Rimola}
\affiliation{Departament de Qu\'{i}mica, Universitat Aut\`{o}noma de Barcelona, Bellaterra, 08193, Catalonia, Spain}

\author[0000-0001-9664-6292]{Cecilia Ceccarelli}
\affiliation{Univ. Grenoble Alpes, CNRS, Institut de Plan\'{e}tologie et d'Astrophysique de Grenoble (IPAG), 38000 Grenoble, France}

\author[0000-0001-8886-9832]{Piero Ugliengo}
\affiliation{Dipartimento di Chimica and Nanostructured Interfaces and Surfaces (NIS) Centre, Universit\`{a} degli Studi di Torino, via P. Giuria 7, 10125, Torino, Italy}

\author[0000-0001-5121-5683]{Nadia Balucani}
\affiliation{Dipartimento di Chimica, Biologia e Biotecnologie, Universit\`{a} di Perugia, Via Elce di Sotto 8, 06123 Perugia, Italy}
\affiliation{Osservatorio Astrosico di Arcetri, Largo E. Fermi 5, 50125 Firenze, Italy}
\affiliation{Univ. Grenoble Alpes, CNRS, Institut de Plan\'{e}tologie et d'Astrophysique de Grenoble (IPAG), 38000 Grenoble, France}

\author{Dimitrios Skouteris}
\affiliation{Master-Tech, I-06123 Perugia, Italy}

\date{Received --; accepted March --}

\begin{abstract}

The formation of the interstellar complex organic molecules (iCOMs) is a hot topic in astrochemistry. One of the main paradigms trying to reproduce the observations postulates that iCOMs are formed on the ice mantles covering the interstellar dust grains as a result of radical--radical coupling reactions.
 
We investigate iCOMs formation on the icy surfaces by means of computational quantum mechanical methods. In particular, we study the coupling and direct hydrogen abstraction reactions involving the CH$_3$ + X systems (X = NH$_2$, CH$_3$, HCO, CH$_3$O, CH$_2$OH) and HCO + Y (Y = HCO, CH$_3$O, CH$_2$OH), plus the CH$_2$OH + CH$_2$OH and CH$_3$O + CH$_3$O systems.

We computed the activation energy barriers of these reactions as well as the binding energies of all the studied radicals, by means of density functional theory (DFT) calculations on two ice water models, made of 33 and 18 water molecules. Then, we estimated the efficiency of each reaction using the reaction activation, desorption and diffusion energies and derived kinetics with the Eyring equations. 

We find that radical--radical chemistry on surfaces is not as straightforward as usually assumed. In some cases, direct H abstraction reactions can compete with radical--radical couplings, while in others they may contain large activation energies. Specifically, we found that (i) ethane, methylamine and ethylene glycol are the only possible products of the relevant radical--radical reactions; (ii) glyoxal, methyl formate, glycolaldehyde, formamide, dimethyl ether and ethanol formation is likely in competition with the respective H-abstraction products, and (iii) acetaldehyde and dimethyl peroxide do not seem a likely grain surface products.

\end{abstract}

\keywords{Interstellar molecules --- Interstellar dust processes --- Dense interstellar clouds --- Surface ices}

\newpage

\section{Introduction} \label{sec:intro}

After the unexpected detection of diatomic molecules in the late 1930s and early 1940s, it was believed that those were the most complex molecules that could be present in the interstellar medium (ISM).
The belief was so strong that the searches for more complex interstellar species with radio telescopes were systematically rejected by the telescope allocation committees because they were considered too speculative \citep{Snyder2006}.
It was only after the 1968/1969, two years that were revolutionary in so many aspects, that polyatomic molecules were detected \citep{Cheung1968,Cheung1969} and even organic molecules \citep{Snyder1969}.

Among the more than 200 molecular species detected hitherto in the ISM, about one third contains at least six atoms, among which one or more are carbon atoms.
This class of molecules is called interstellar Complex Organic Molecules in the literature \citep[COMs, or iCOMs:][]{Herbst2009,ceccarelli2017} and are prevalently, but not exclusively, detected in star forming regions.
They attract a lot of attention for two major reasons. 
First, they represent the dawn of organic chemistry and could be involved in the emergence of life \citep[e.g.][]{deDuve2005Book,deDuve2011,ceccarelli2017}.
Second, their formation in the harsh ISM environment represents a challenge to astrochemists \citep[e.g.][]{Vasyunin2013,balucani2015,ceccarelli2017,JinGarrod2020}.
Since their discovery in the 70s \citep{Rubin1971}, two competing theories have been proposed to explain the presence and abundances of iCOMs.
Either iCOMs are synthesised on the interstellar grain surfaces or in gas-phase by reactions involving simpler grain surface-chemistry products.
The debate is still vivid, with the weight moving from one to the other with time.

In both theories, the first step is the formation of icy mantles by hydrogenation reactions of simple species, atoms or molecules, frozen onto the grain surfaces, such as O and CO \citep{Tielens_Hagen_1982}.
As a result, the grain mantles dominant species is water, followed by other less abundant species like CO$_2$, ammonia, methane and methanol.
The subsequent evolution of the icy mantles diverges in the two theories.
In the first one, iCOMs are formed in the gas-phase during the hot ($\geq100$ K) protostellar stage, by reactions involving the components of the sublimated icy mantles \citep[e.g.][]{Charnley1997,Taquet2016,Skouteris2018,Skouteris2019,Vazart2020MNRAS}.
In the second theory, it is postulated that, during the cold pre-protostellar stage, the ice components are partially photo-dissociated by UV photons, generated by the interaction of cosmic-rays (CR) or X-rays with the hydrogen atoms in the gas phase, creating radicals that remain trapped into the ices.
Once the protostar gradually warms up its surroundings \citep{viti2004}, these radicals can diffuse over the ice, meet and react forming iCOMs \cite{Garrod2006,Garrod2008b,Herbst2009,kalvans2018}.
Additional processes have been considered to boost the iCOMs formation on the icy grain surfaces, such as the the reactivity of gas-phase C atoms \citep{Ruaud_ER_2015} or CN \citep{Rimola2018} landing on the icy surfaces, the formation of glyoxal by the coupling of two HCO radicals formed one next to the other on CO ices, followed by its hydrogenation that leads to glycoaldehyde and ethylene glycol \cite{Simons2020}, or the reactions induced by landing cations on negatively charged icy grains \citep{Rimola2021_ions}. 

In this work, we focus on the reactivity on the grain icy surfaces between radicals, arguably the most crucial step of this theory.
In astrochemical models, it is usually assumed that, when two radicals meet on the grain surfaces, the reaction coupling them into an iCOM is barrierless.
Numerically, this is obtained by assuming that the reaction efficiency $\varepsilon$ is equal to unity.
Here, we present new quantum chemical simulations on nine systems postulated to synthesize iCOMs, and observed in the ISM, such as dimethyl ether, methyl formate and ethanol, in which $\epsilon$ is not always equal to one.
Our aim is to compute the activation energy barriers of the reactions forming iCOM and the respective competitive channels employing two ice models representing two different surface environments and the same methodology for all systems.
In addition, we compute the approximate efficiency $\varepsilon$ of the studied reactions based on the binding and activation barrier energies, and provide hints on the possible expected output of other radical-radical systems, not studied here and that are of relevance in the formation of iCOMs on the icy surfaces.

The article is organised as follows.
In Sect. \ref{sec:systems}, we present the systems studied in this work and the previous studies on which the present one is based on;
in Sect. \ref{sec:methods} and \ref{sec:results} we describe the adopted methodology and the results of the new computations, respectively; in Sect. \ref{sec:disc} we discuss the results, and Sect. \ref{sec:concl} concludes the article.

\section{Previous works and present studied systems} \label{sec:systems}

Several theoretical studies on the interstellar icy surfaces chemistry have appeared in the literature during the last decade.
A recent general review can be found in \cite{Zamirri2019Review}.
Here we focus on the studies involving radical--radical reactions on the icy surfaces.
Other studies have considered pure CO ices \citep{Lamberts2019}, where the bonds between the species and the surface are substantially different and much less strong with respect to water ices.

The new study presented in this work is based on previous ones from our groups, described here.
We presented a first pioneer study of the two systems HCO + CH$_3$ and HCO + NH$_2$ in previous works \citep{Rimola2018,ER2019}, as well as an in-depth study on the accuracy of the adopted methodology \citep{ER2020}.
In the present work, we expand the number of systems studied in \citet{ER2019} adding nine more cases.
Specifically, we selected a subset of the radical species considered by \cite{Garrod2008b} and that are the photolysis products of the closed-shell species formaldehyde, methane, methanol and ammonia: CH$_3$, HCO, CH$_3$O, CH$_2$OH and NH$_2$.
Among those radicals, here we focus on the CH$_3$ + X and HCO + Y systems, where X = NH$_2$, CH$_3$, HCO, CH$_3$O, CH$_2$OH and Y = HCO, CH$_3$O, CH$_2$OH.
The list of studied systems and the possible products (from radicals combination and direct H-abstraction, respectively) are summarised in Table \ref{tab:studied_systems}. 

Our first goal here is to provide the potential energy surface (PES) of the reactions of the above systems, namely the energetics of the radical-radical coupling (hereinafter Rc) reaction, leading to the formation of an iCOM, as well as the possible competitive channels.
On this respect, the previous studies mentioned above have shown that the H-abstraction reactions can potentially be more energetically favorable than the simple combination of the two radicals.
In general, for H-abstraction reactions to take place, an H-donor and an H-acceptor radicals are needed.
In some cases, such as CH$_3$ + CH$_3$, it does not happen.
In others cases, such as HCO + CH$_3$O, both radicals can act as either H-acceptors/donors so that two direct H-abstraction channels might exist with different products.

In \citet{ER2019}, we carried out the calculations considering two models for the amorphous water surfaces (AWS): with 18 (W18) and 33 (W33) waters, respectively.
The W33 model is large enough to possess a geometrical cavity where radicals can lie, whereas the W18 model is too small for that purpose and only a "flat" surface is possible.
These two models are obviously rough analogues of the ices that cover the interstellar grains.
Nonetheless, they allow to have estimates of the energetics of reactions of radicals sitting on flat surfaces and inside a cavity, respectively.
The latter is particularly interesting to describe the likely situation of most frozen radicals, as they are believed to be formed by the UV irradiation of the ice bulk.
In other words, reactions among radicals are much more likely to occur in situations where they are surrounded by water molecules than on a flat surface exposed to the gas-phase.
Actually, it is even possible that our W33 cavity description provides an optimistic view, as radicals may even be trapped in frozen water cages.
For this reasons, in this work, we will pay special attention to the reactions occurring in the W33 cavity.

Providing the energetic of the process is a first mandatory step but it is not the end of the story.
Following the study by \citet{ER2021}, further kinetics calculations will then provide estimates of the efficiency of the formation of the iCOMs on the icy grain surfaces \textit{via} the radical-radical coupling as well as the H-abstraction competitive products, respectively.
This study on the kinetics is postponed to a forthcoming article.

It is worth mentioning that the present work holds some limitations. Radical--radical chemical reactions on top of these two cluster surface models were explored by considering a single reaction site on each model for reaction. This is motivated by the large number of investigated reactions (see Table~\ref{tab:studied_systems}) and the relatively high computational cost of the simulations we have carried out, especially when dealing with reactivity. However, by means of this approach, we represent two different surface morphological situations, namely, reactions taking place on a rather flat surface and on a small cavity (W18 and W33, respectively).
Similarly, for the computations on the binding energies (see also \S~\ref{sec:be_methods}) of each radical interacting with the surfaces, a single binding site for each radical on top of each ice model was investigated. This is indeed a simplistic assumption given that different surface binding sites are available and, accordingly, a distribution of binding energies exist \citep[e.g.][]{Ferrero_2020_BE, Bovolenta2020, Duflot2021}.
Finally, diffusion has not explicitly been studied in this work given the rather small size of our ice models. In order to properly study surface diffusion, one would need a much larger ice model with well characterized diffusion barriers \citep[see e.g.][]{Senevirathne2017}.
In those cases where diffusion energies where needed, they were just approximated as a fraction of the calculated binding energies (see~\ref{sec:react_efficiencies}), as usually done in astrochemical modelling.

\begin{table*}[!htb]
\centering
\caption{Summary of the systems and reactions studied in this work.}
\label{tab:studied_systems}
\resizebox{0.99\textwidth}{!}{%
\begin{tabular}{c|ccc}
\hline
System &
  \makecell[c]{Radical\\coupling (Rc)} &
  \makecell[c]{Direct H-abstraction\\product 1} &
  \makecell[c]{Direct H-abstraction\\product 2} \\ \hline
CH$_3$ + CH$_3$     & \makecell[c]{C$_2$H$_6$\\(Ethane)}                  &                      &                      \\ \hline
CH$_3$ + NH$_2$     & \makecell[c]{CH$_3$NH$_2$\\(Methylamine)}           &                      &                      \\ \hline 
CH$_3$ + CH$_3$O    & \makecell[c]{CH$_3$OCH$_3$\\(Dimethyl ether)}       & CH$_4$   + H$_2$CO   &                      \\ \hline 
CH$_3$ + CH$_2$OH   & \makecell[c]{CH$_3$CH$_2$OH\\(Ethanol)}             & CH$_4$   + H$_2$CO   &                      \\ \hline 
HCO + HCO           & \makecell[c]{HCOCHO\\(Glyoxal)}                   & CO   + H$_2$CO       &                      \\ \hline
HCO + CH$_3$O       & \makecell[c]{HC(O)OCH$_3$\\(Methyl formate)}        & CO   + CH$_3$OH      & H$_2$CO  + H$_2$CO   \\ \hline 
HCO + CH$_2$OH      & \makecell[c]{HC(O)CH$_2$OH\\(Glycolaldehyde)}       & CO   + CH$_3$OH      & H$_2$CO  + H$_2$CO   \\ \hline
CH$_3$O + CH$_3$O   & \makecell[c]{CH$_3$OOCH$_3$\\(Dimethyl peroxide)}     & H$_2$CO   + CH$_3$OH &                      \\ \hline
CH$_2$OH + CH$_2$OH & \makecell[c]{CH$_2$(OH)CH$_2$OH\\(Ethylene glycol)} & H$_2$CO + CH$_3$OH   &                      \\ \hline
\end{tabular}%
}
\end{table*}

\section{Methods} \label{sec:methods}

In this section, we present the adopted ASW ice models, the methods employed for the electronic structure calculations, and finally, how binding energies were calculated.

\subsection{Water ice models}

\begin{figure}[!htb]
    \centering
    \includegraphics[width=0.39\textwidth]{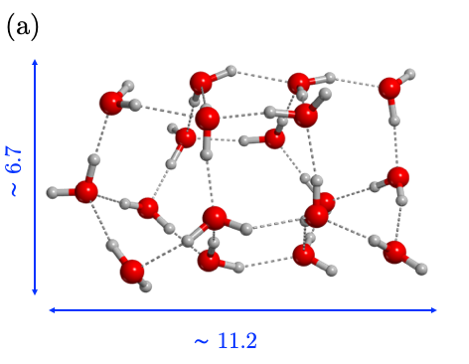}
    \includegraphics[width=0.49\textwidth]{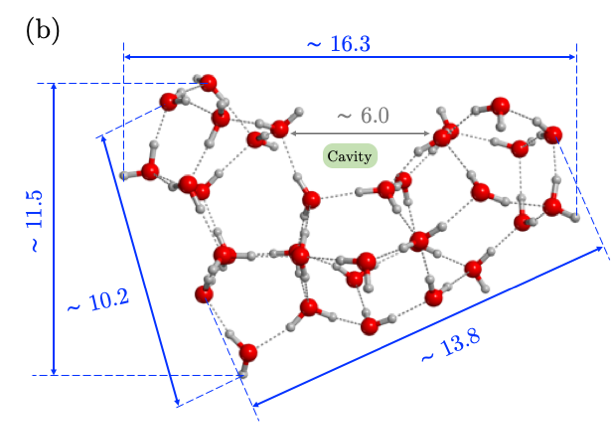}
    \caption{ASW models used in this work. The geometries were optimised at BHLYP-D3(BJ)/6-31+G(d,p) level. Distances in \r{A}.}
    \label{fig:ASWmodels}
\end{figure}

Following \cite{ER2019}, two cluster models have been used to simulate the surfaces of interstellar ASW (shown in Figure \ref{fig:ASWmodels}). 
They consist of 18 and 33 water molecules, which will be hereafter referred to as W18 and W33, respectively. 
The dimensions of the ice models slightly changed with respect to those reported in \cite{ER2019} due to using an improved dispersion correction term in the geometry optimisations (namely, the D3(BJ) dispersion instead of the bare D3 one). 
Interestingly, as highlighted in Figure \ref{fig:ASWmodels}, while W18 presents a flat surface morphology (of $\sim$11.2$\times$6.7 \r{A}), this is not the case for W33, which presents two different regions: a 6 \r{A} wide cavity, and its elongated side (13.8 \r{A} long). 
However, in this work, at variance with \cite{ER2019}, for W33 calculations, only the cavity structure has been considered. 
The reason of this choice relies on the fact that, as shown in \cite{ER2019}, results provided by W18 and by W33 elongated side are very similar. 
Additionally, we have also found that, for some radical-radical reactions, dramatic structural changes occurred on the W33 cluster model, in which the cavity collapsed when reactions between highly bound species were simulated. 
Both ice surfaces have a thickness of about 6--7 \r{A}.\\

\subsection{Electronic structure calculations} \label{sec:elec_struct}

All DFT calculations were run with the \textsc{Gaussian16} software package \citep{g16}. Following our previous work, the BHLYP functional \citep{becke1993,LYP88} was used, in which the Grimme's 3-body dispersion correction alongside the Becke-Johnson damping function (D3(BJ)) \citep{D3-grimme2010,d3bj_grimme} was introduced in \textit{a posteriori} manner.

The radical-radical reactions were studied both on amorphous solid water surface models (ASW) and in the absence of water molecules\footnote{I.e. the reaction between the two radicals alone, in order to assess whether radicals are able to directly react or not.}.
For the former, demanding calculations like geometry optimisations or frequency calculations were run by using the double-$\zeta$ Pople's basis set 6-31+G(d,p) \citep{hehre_selfconsistent_1972,hariharan_influence_1973}, which were later refined by single point calculations with the 6-311++G(2df,2pd) \citep{krishnan_selfconsistent_1980} basis set. For the latter, due to their less demanding computational effort, all geometry optimizations were performed with the 6-311++G(2df,2pd) basis set.

All stationary points were characterized by the analytical calculation of the harmonic frequencies as minima (reactants, products, and intermediates) and saddle points (transition
states). 
Intrinsic reaction coordinate (IRC) calculations at the same level of theory were carried out (when needed) to ensure that the transition states connect with the corresponding minima.
Thermochemical corrections to the potential energy values were carried out using the standard rigid rotor/harmonic oscillator formulas to compute the zero point energy (ZPE) corrections \cite{mcquarrie2000}.
In order to properly simulate singlet electronic state biradical systems, we used the unrestricted formalism alongside the broken (spin) symmetry (BS) approach \citep[e.g.][]{Neese2004}, which has been proven to be a cost-effective methodology to properly describe the electronic structure of this kind of systems reaching good agreement with highly correlated methods \citep{ER2020, ER2021}. In the BS approach, singlet and triple states are mixed. This allows each unpaired electron to be localized on top of each radical \citep[see Table \ref{tab:initial_spin_dens} and the supporting Figures available on Zenodo][]{zenodo_data}, at the expense of having a non-uniform spin density with positive spin density on one of the radicals and negative on the other. The BS singlet state is not a solution of the $\hat{S}^{2}$ operator, and therefore $\langle S^{2} \rangle$ is not equal to 0 but it is 1 for asinglet biradical system, i.e. a mixture between singlet and triplet states.
Spin contamination issues may appear; however, we do not observe it in any of our systems as indicated by the spin annihilation step automatically run by \textsc{Gaussian16}, with errors lower than a 10\% with respect to the pure singlet, as can be seen in Tables \ref{tab:S2_w18_w33} and \ref{tab:S2_TSs} in the appendix.

Finally, we also calculated the tunneling crossover temperatures ($T_c$) following \cite{FermannAuerbach2000_Tc} (see \S \ref{sec:appendix:crossover_temp} in the annex for more details) for those reaction steps where a hydrogen atom is transferred.

\subsection{Binding energies}\label{sec:be_methods}

The calculations of the radical--surface binding energies adopted the same electronic structure methodology as for reactivity. 
That is, for each radical--surface complex and isolated components (i.e. radicals and surfaces), geometry optimizations and frequency calculations (and hence ZPE corrections) were computed at the BHLYP-D3(BJ)/6-31+G(d,p) level, which were followed by single point energy calculations at the improved BHLYP-D3(BJ)/6-311++G(2df,2pd) level to refine the potential energy values. 
With the obtained values we calculated the dispersion and deformation corrected interaction energies ($\Delta E_{ads}$). 

Subsequently, basis set superposition error (BSSE) corrections were obtained by running single point energy calculations at the BHLYP-D3(BJ)/6-311++G(2df,2pd) theory level. This means that BSSE was corrected in a posteriori fashion to the optimization of the complexes, i.e., it was not accounted for during the geometry relaxation.
The final, corrected, interaction energy ($\Delta E_{ads}^{CP}$) was calculated using the following equation:
%
%
\begin{equation}
        \Delta E_{ads}^{CP} (AB) =  \Delta E_{ads} + {BSSE}(A) + {BSSE}(B) + \Delta ZPE
     \label{eq:binding_bsse}
\end{equation}

Note that we used the same sign convention as in \cite{ER2019}, namely $\Delta E_{ads}^{CP} = - \Delta E_{bind}^{CP}$, and that $\Delta E_{ads}$ already contain the contributions of the deformation due to the formation of the surface--radical complex.

\subsection{Reaction efficiencies}\label{sec:react_efficiencies}

Astrochemical models compute the abundance of species by solving time-dependent equations that equate formation and destruction rates for each species.
For grain-surface reactions, the formation rate is determined by the rate of encounters of the two reactants on the reaction site multiplied by the efficiency of the reaction, $\varepsilon$, which is the probability that when the two reactants meet they also react \citep{Hasegawa1993,Garrod2006}.

In order to provide a rough estimate of $\varepsilon$, we used the activation energy barriers and binding energies, following the schemes commonly used in astrochemical models \citep[see][for a detailed discussion]{ER2021}:

\begin{equation}
    \varepsilon = \frac{k_{aeb}}{k_{aeb} + k_{diff,1} + k_{des,1} + k_{diff,2} + k_{des,2}}
    \label{eqn:efficiency}
\end{equation}

\noindent where $k_{aeb}$ are the rate constants related to the activation energy barrier and $k_{diff,i}$, $k_{des,i}$ are the diffusion and desorption rate constants of the species $i$.

All of these rate constants were derived using the Eyring equation:
\begin{equation}
    k=(k_BT/h)\exp(-E_a/k_BT)
\end{equation}
\noindent where $k_B$ and $h$ are the Boltzmann and Planck constants, $T$ is the (surface) temperature and $E_a$ is the activation energy of the process, i.e. the activation energy barrier for reactions or the diffusion $E_{diff}$ and desorption $E_{des}$ energies. It is worth noting that in Eq. \ref{eqn:efficiency}, entropic effects are neglected, which is consistent with the very low temperature at which the processes (chemical reaction, diffusion, desorption) take place.
The desorption energy, responsible of the k$_{des,1/2}$ terms, is just the opposite of the binding energy of each species, while the diffusion energy is taken to be a fraction of the desorption one.

In the literature, the $E_{diff}/E_{des}$ ratio is usually assumed to be in the range of 0.3 and 0.4 for molecules \cite[e.g.][]{HHL1992,Karssemeijer_2014, Penteado2017, ruaud2016_nautilus, Aikawa2020, JinGarrod2020}. 
Recently, \cite{He_Vidali_2018} were able to measure the diffusion barrier of a number of molecules on ASW ices. 
They found $E_{diff}/E_{des}$ ratios ranging between $\sim$0.3 and 0.6, depending on the coverage of admolecules, so that little coverage (sub monoloayer, ML, of admolecules) corresponds to the lower end of the $E_{diff}/E_{des}$ ratio range, while higher coverages ($>$1ML) correspond to the higher end of the ratio range.
Larger values of $E_{diff}/E_{des}$ are normally assumed for atomic species (e.g. \cite{Minissale2016} found experimentally a $E_{diff}/E_{des}$ value of 0.55 for N and O), and in the literature there is a fairly large amount of work in which a value of 0.5 is assumed (e.g. \cite{Garrod2006,Garrod2008b,GP2011,Ruaud_ER_2015,Vasyunin2017,Jensen2021}. 
This can, however, cause surface reactions to be much more efficient than using the recommended 0.3--0.4 range, as shown and discussed in detail by \cite{ER2021}.
For this reason, we used an intermediate value for $E_{diff}/E_{des}$ of 0.35 in order to calculate the efficiency of the radical-radical reactions presented in this work.

\section{Results} \label{sec:results}

\subsection{Binding energies} \label{sec:BE}

\begin{figure*}[!htbp]
    \centering
    \includegraphics[width=0.99\textwidth]{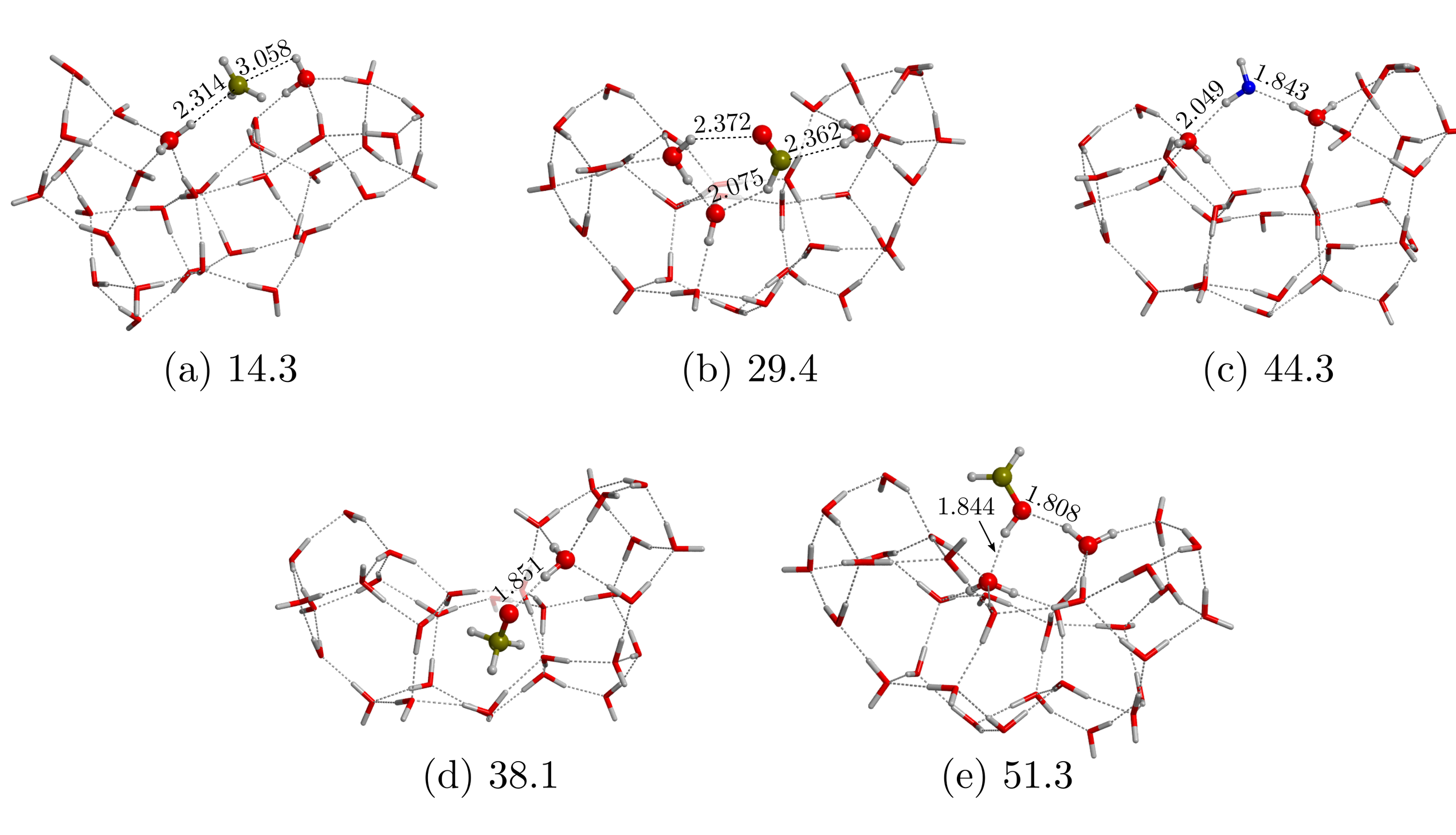}
    \caption{Geometries of the five studied radicals, CH$_3$ (a), HCO (b), NH$_2$ (c), CH$_3$O (d) and CH$_2$H (e); adsorbed on W33 fully optimised at the UBHLYP-D3(BJ)/6-31+G(d,p) theory level. Energy values in kJ mol$^{-1}$ are those refined at BHLYP-D3(BJ)/6-311++G(2df,2pd) level with the ZPE- and BSSE- corrections. Distances in \r{A}.}
    \label{fig:ANNEX_W33_ads}
\end{figure*}

The computed binding energies (BEs) of the studied radicals with W33 and W18 are reported in Table \ref{tab:BEs}. 
Optimized geometries for W33 are reported in Figure \ref{fig:ANNEX_W33_ads} while those for W18 are available in Figure \ref{fig:ANNEX_W18_ads} in the annex.
For CH$_3$, HCO and NH$_2$, complexes in \cite{ER2019} were re-optimized at the current theory level. 
For CH$_3$O and CH$_2$OH, the initial structures were constructed by maximising the inter-molecular interactions between the radicals and the cluster models.

Computed BEs follow the order of CH$_3$ $<$ HCO $<$ NH$_2$ $<$ CH$_3$O $<$ CH$_2$OH.

Differences with respect to BE values in \cite{ER2019} arise from the different dispersion terms used in the two works, namely, D3(BJ) here versus D3 in \cite{ER2019}, in which the former is understood to be more accurate as the components defining the D3(BJ) term enter in an optimisation processes in agreement with the particular system to simulate. 
Nevertheless, the same BE trends are obtained for the CH$_3$, HCO and NH$_2$ cases. 
\begin{table}[!htb]
    \centering
    \begin{tabular}{|l|ccccc|}
    \hline
    $\Delta E_{bind}^{CP}$ & CH$_3$ & HCO  & NH$_2$ & CH$_3$O & CH$_2$OH \\\hline
    W33               & 14.3   & 29.4 & 44.3  & 38.1    & 51.3     \\\hline
    W18               & 8.1    & 20.5  & 31.8 & 26.1    & 45.9 \\ \hline
    \end{tabular}
    \caption{Computed corrected binding energies ($\Delta E_{bind}^{CP}$) for the radicals interacting with the W18 and W33 cluster models. Units are in kJ mol$^{-1}$.}
    \label{tab:BEs}
\end{table}
Interestingly, BEs on W33 are about a 12--76 \% higher than on W18, showing the importance of the larger number of inter-molecular interactions formed in the former cluster, as well as the larger dispersion interactions originated when the radicals adsorb in the cavity.

The reliability of our methodology in computing these BE values is evidenced by comparing the results at BHLYP-D3(BJ)/6-311++G(2df,2pd) with those at CCSD(T)/aug-cc-pVTZ level (single point energy calculations on the BHLYP-D3(BJ) optimized geometries), in which a very good correlation between values is obtained (see Figure \ref{fig:interaction_energies_correlation} in the annex). 

\subsection{Radical-radical reactivity}

In this section, the reactivity of the different sets of radical pairs on the W33 and W18 surface cluster models is presented. 
For the sake of clarity and brevity, along this section only structures involving W33 are shown.
However, all computed structures, i.e., all the stationary points for both W33 and W18, are available in the the online version of the as Figure Sets 1 and 2 (available in the online version of the published paper).

\figsetstart
\figsetnum{1}
\figsettitle{Stationary points of the radical--radical reactions on the W18 ice model.}

\figsetgrpstart
\figsetgrpnum{1.1}
\figsetgrptitle{ZPE-corrected critical points of the Rc and dHa1/2 PESs for CH$_2$OH + CH$_2$OH on W18 fully optimised at the UBHLYP-D3(BJ)/6-31+G(d,p) theory level with the DFT energy refined to UBHLYP-D3(BJ)/6-311++G(2df,2pd). Energy units are in kJ/mol and distances in \r{A}.}
\figsetplot{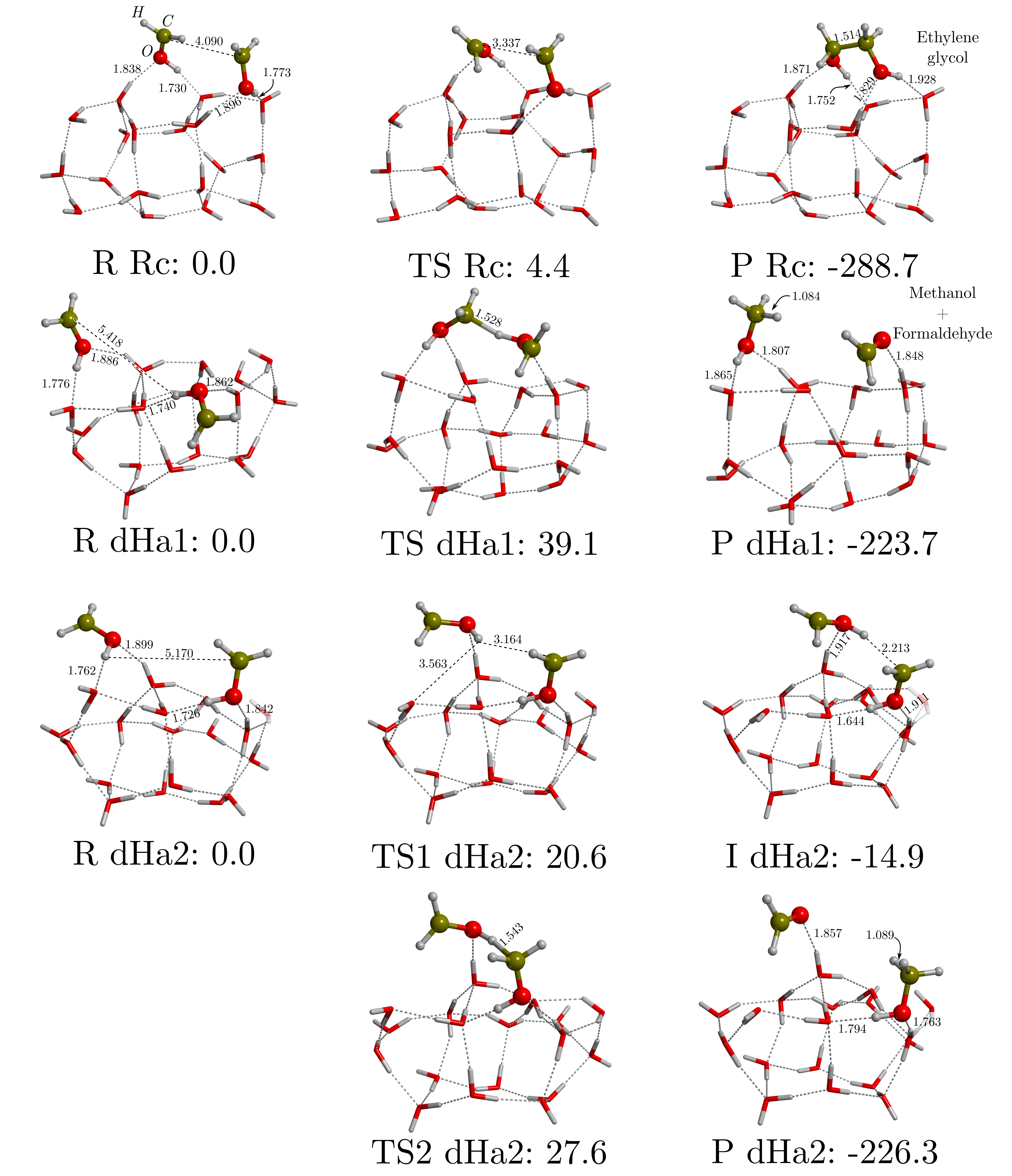}
\figsetgrpnote{Stationary points of radical--radical reactions on W18}
\figsetgrpend

\figsetgrpstart
\figsetgrpnum{1.2}
\figsetgrptitle{ZPE-corrected critical points of the Rc and dHa PESs for CH$_3$ + CH$_2$OH on W18 fully optimised at the UBHLYP-D3(BJ)/6-31+G(d,p) theory level with the DFT energy refined to UBHLYP-D3(BJ)/6-311++G(2df,2pd). Energy units are in kJ/mol and distances in \r{A}.}
\figsetplot{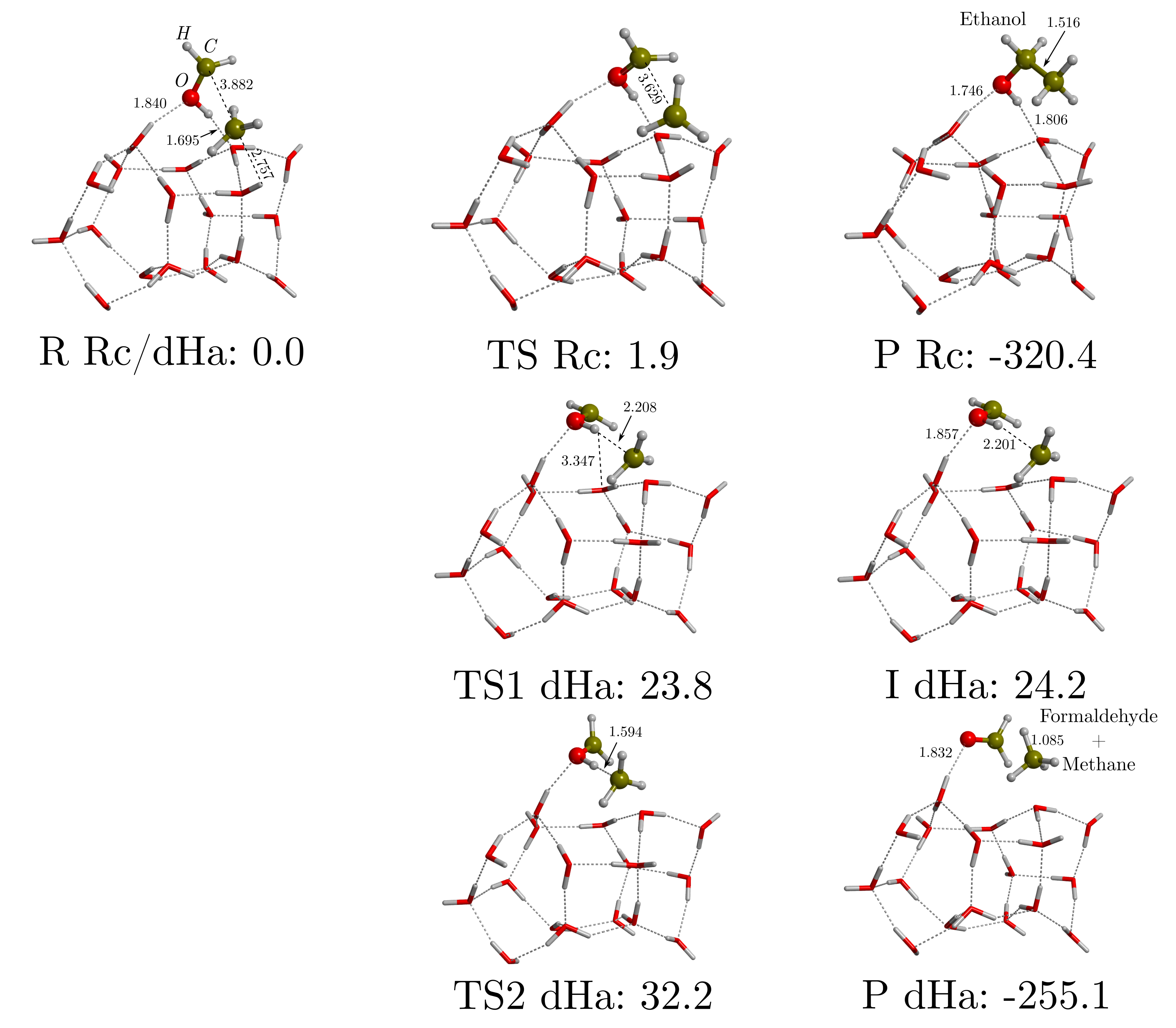}
\figsetgrpnote{Stationary points of radical--radical reactions on W18}
\figsetgrpend

\figsetgrpstart
\figsetgrpnum{1.3}
\figsetgrptitle{ZPE-corrected critical points of the Rc PES for CH$_3$ + CH$_3$ on W18 fully optimised at the UBHLYP-D3(BJ)/6-31+G(d,p) theory level with the DFT energy refined to UBHLYP-D3(BJ)/6-311++G(2df,2pd). Energy units are in kJ/mol and distances in \r{A}.}
\figsetplot{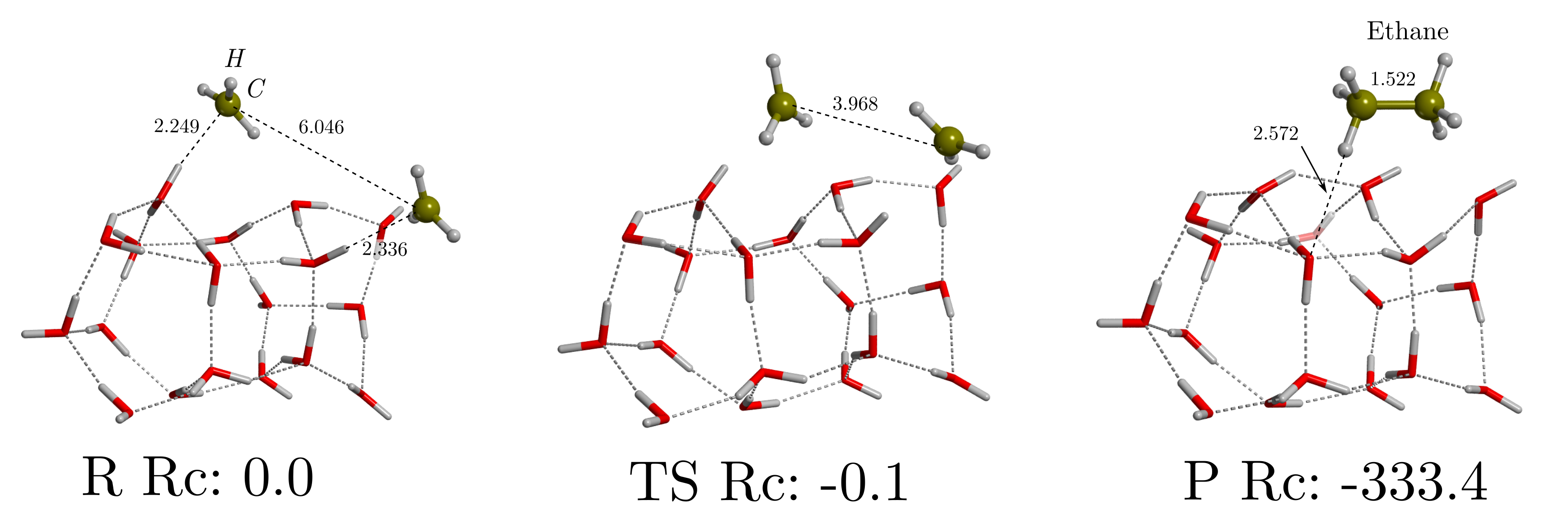}
\figsetgrpnote{Stationary points of radical--radical reactions on W18}
\figsetgrpend

\figsetgrpstart
\figsetgrpnum{1.4}
\figsetgrptitle{ZPE-corrected critical points of the Rc and dHa PESs for CH$_3$ + CH$_3$O on W18 fully optimised at the UBHLYP-D3(BJ)/6-31+G(d,p) theory level with the DFT energy refined to UBHLYP-D3(BJ)/6-311++G(2df,2pd). Energy units are in kJ/mol and distances in \r{A}.}
\figsetplot{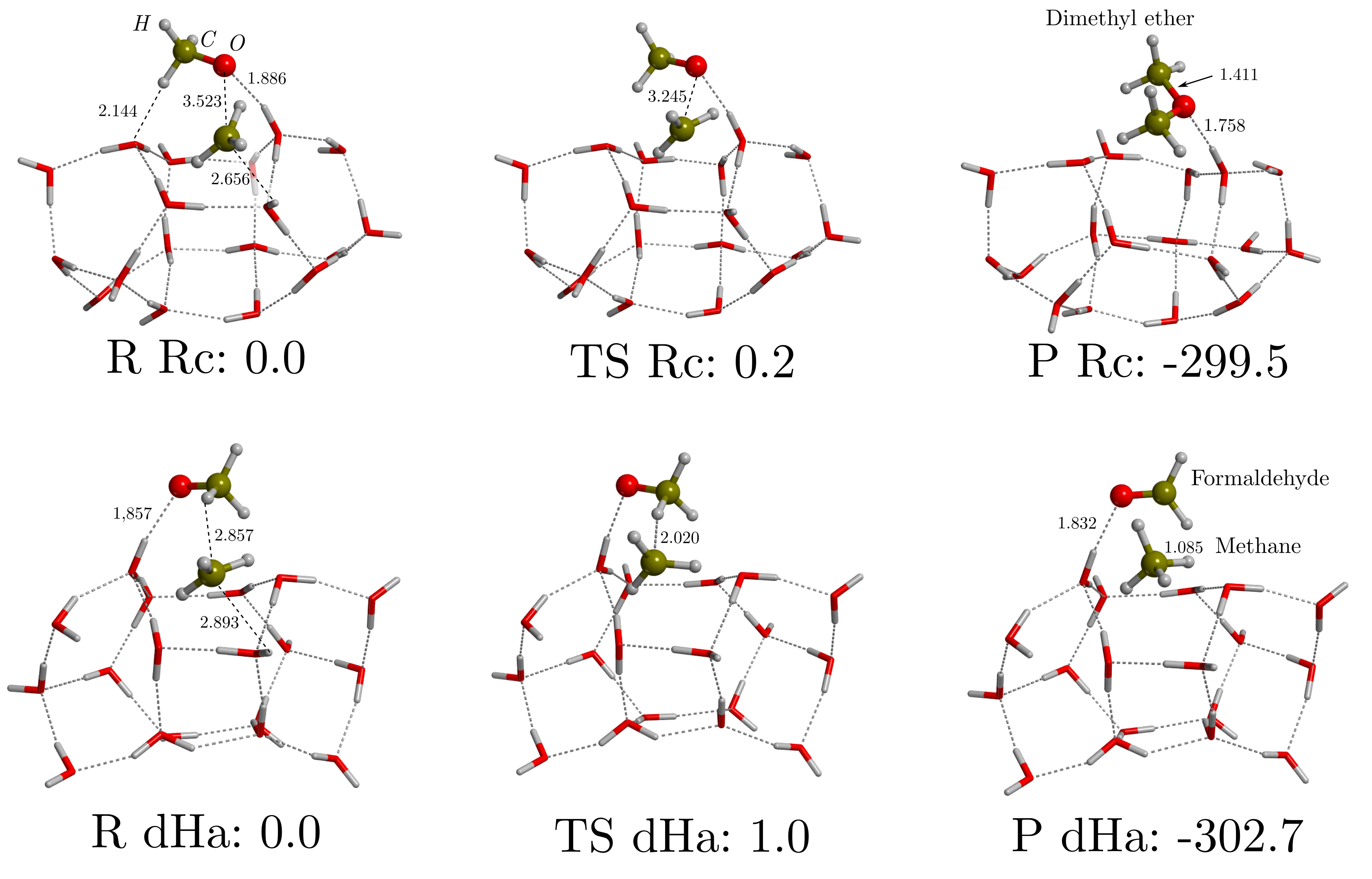}
\figsetgrpnote{Stationary points of radical--radical reactions on W18}
\figsetgrpend

\figsetgrpstart
\figsetgrpnum{1.5}
\figsetgrptitle{ZPE-corrected critical points of the Rc PES for CH$_3$ + NH$_2$ on W18 fully optimised at the UBHLYP-D3(BJ)/6-31+G(d,p) theory level with the DFT energy refined to UBHLYP-D3(BJ)/6-311++G(2df,2pd). Energy units are in kJ/mol and distances in \r{A}.}
\figsetplot{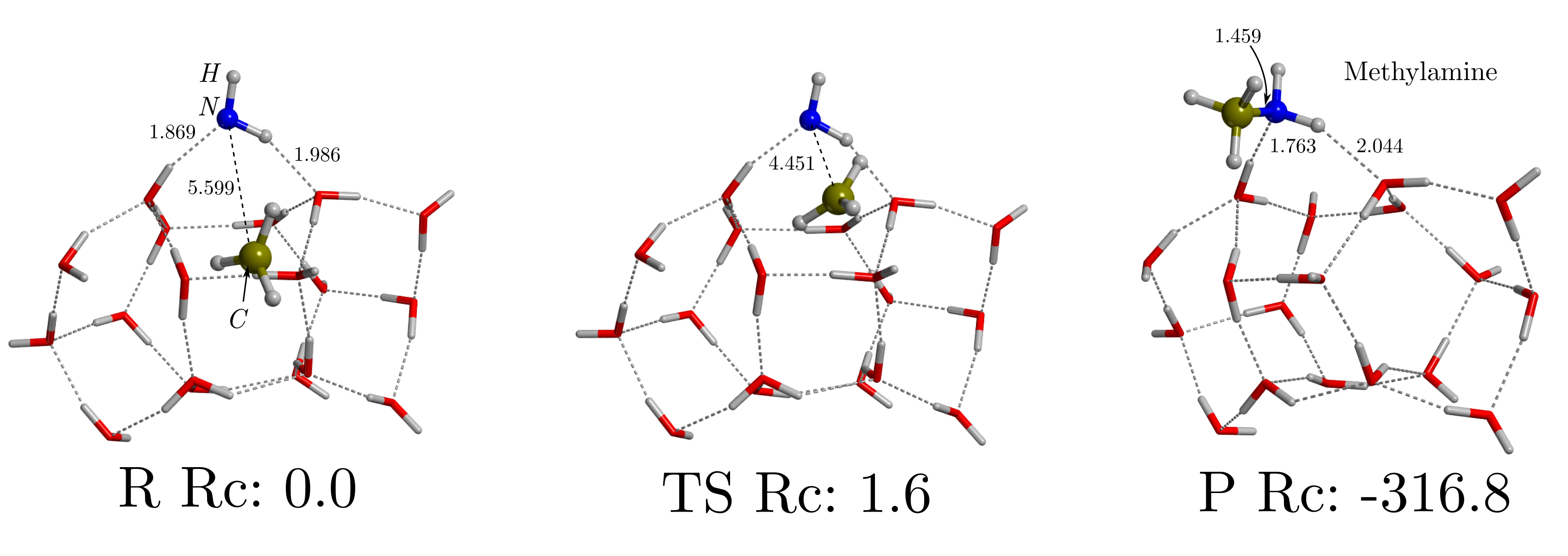}
\figsetgrpnote{Stationary points of radical--radical reactions on W18}
\figsetgrpend

\figsetgrpstart
\figsetgrpnum{1.6}
\figsetgrptitle{ZPE-corrected critical points of the Rc and dHa1/2 PESs for CH$_3$O + CH$_3$O on W18 fully optimised at the UBHLYP-D3(BJ)/6-31+G(d,p) theory level with the DFT energy refined to UBHLYP-D3(BJ)/6-311++G(2df,2pd). Energy units are in kJ/mol and distances in \r{A}.}
\figsetplot{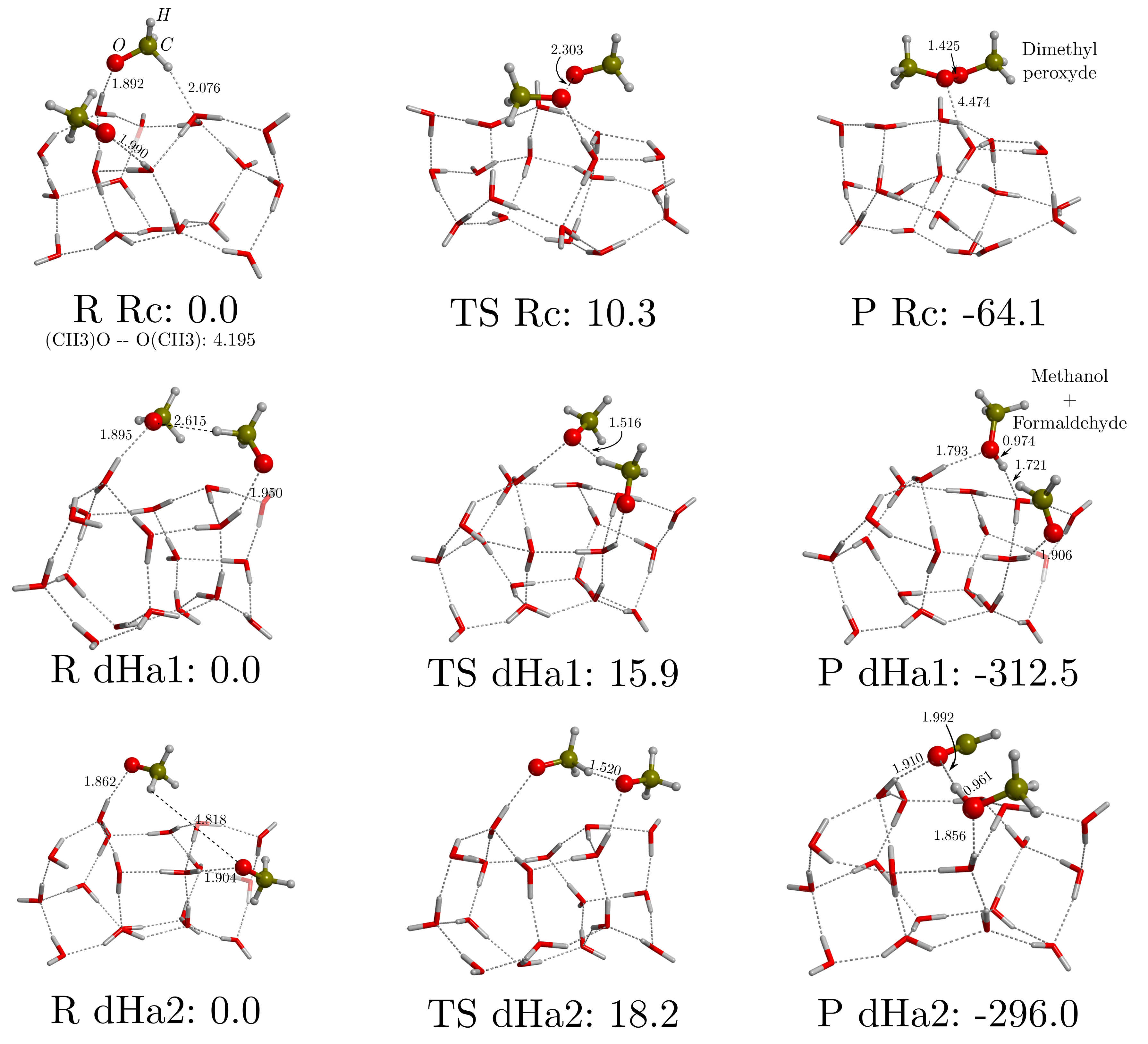}
\figsetgrpnote{Stationary points of radical--radical reactions on W18}
\figsetgrpend

\figsetgrpstart
\figsetgrpnum{1.7}
\figsetgrptitle{ZPE-corrected critical points of the Rc and dHa1/2 PESs for HCO + CH$_2$OH on W18 fully optimised at the UBHLYP-D3(BJ)/6-31+G(d,p) theory level with the DFT energy refined to UBHLYP-D3(BJ)/6-311++G(2df,2pd). Energy units are in kJ/mol and distances in \r{A}.}
\figsetplot{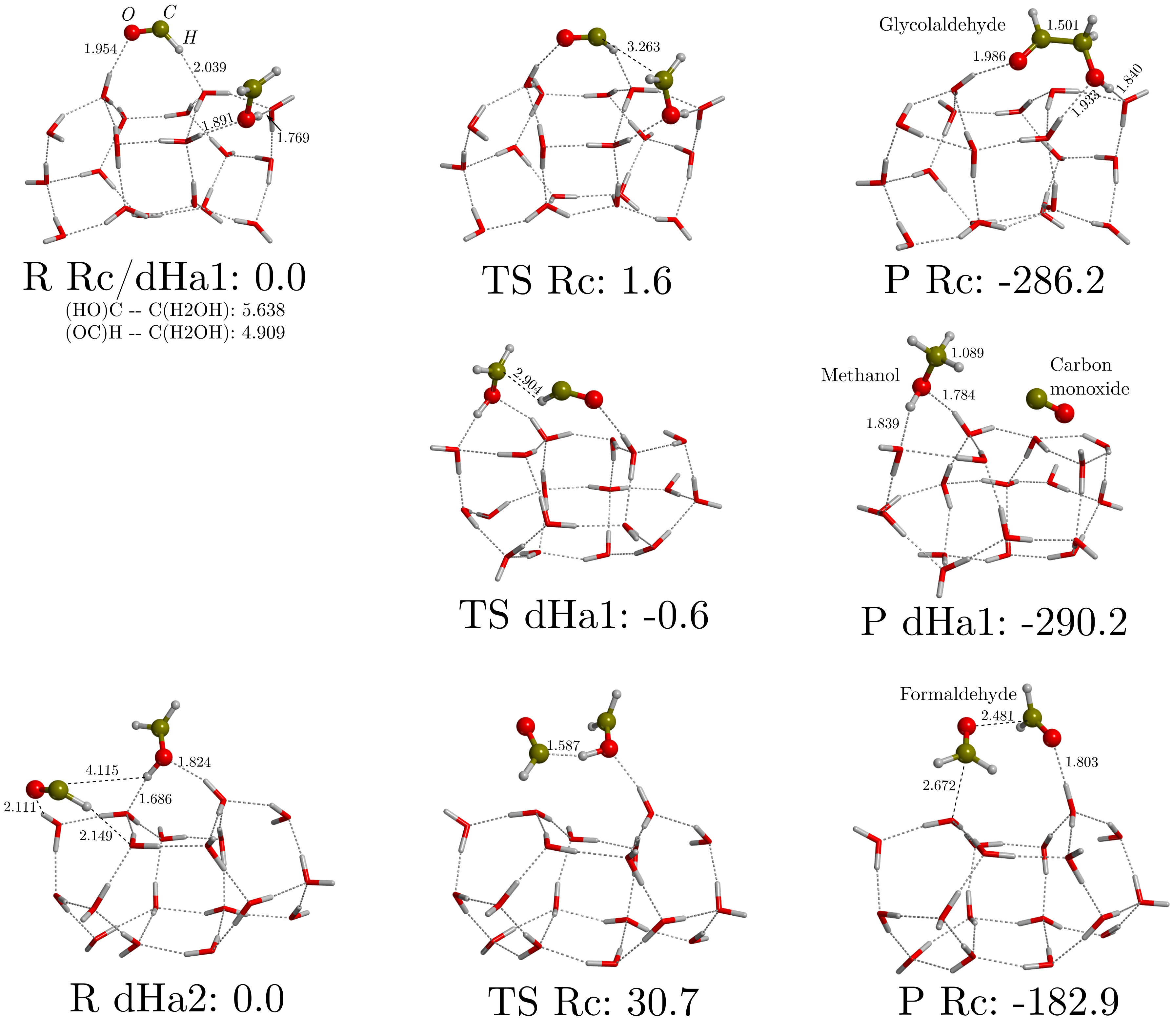}
\figsetgrpnote{Stationary points of radical--radical reactions on W18}
\figsetgrpend

\figsetgrpstart
\figsetgrpnum{1.8}
\figsetgrptitle{ZPE-corrected critical points of the Rc and dHa1/2 PESs for HCO + CH$_3$O on W18 fully optimised at the UBHLYP-D3(BJ)/6-31+G(d,p) theory level with the DFT energy refined to UBHLYP-D3(BJ)/6-311++G(2df,2pd). Energy units are in kJ/mol and distances in \r{A}.}
\figsetplot{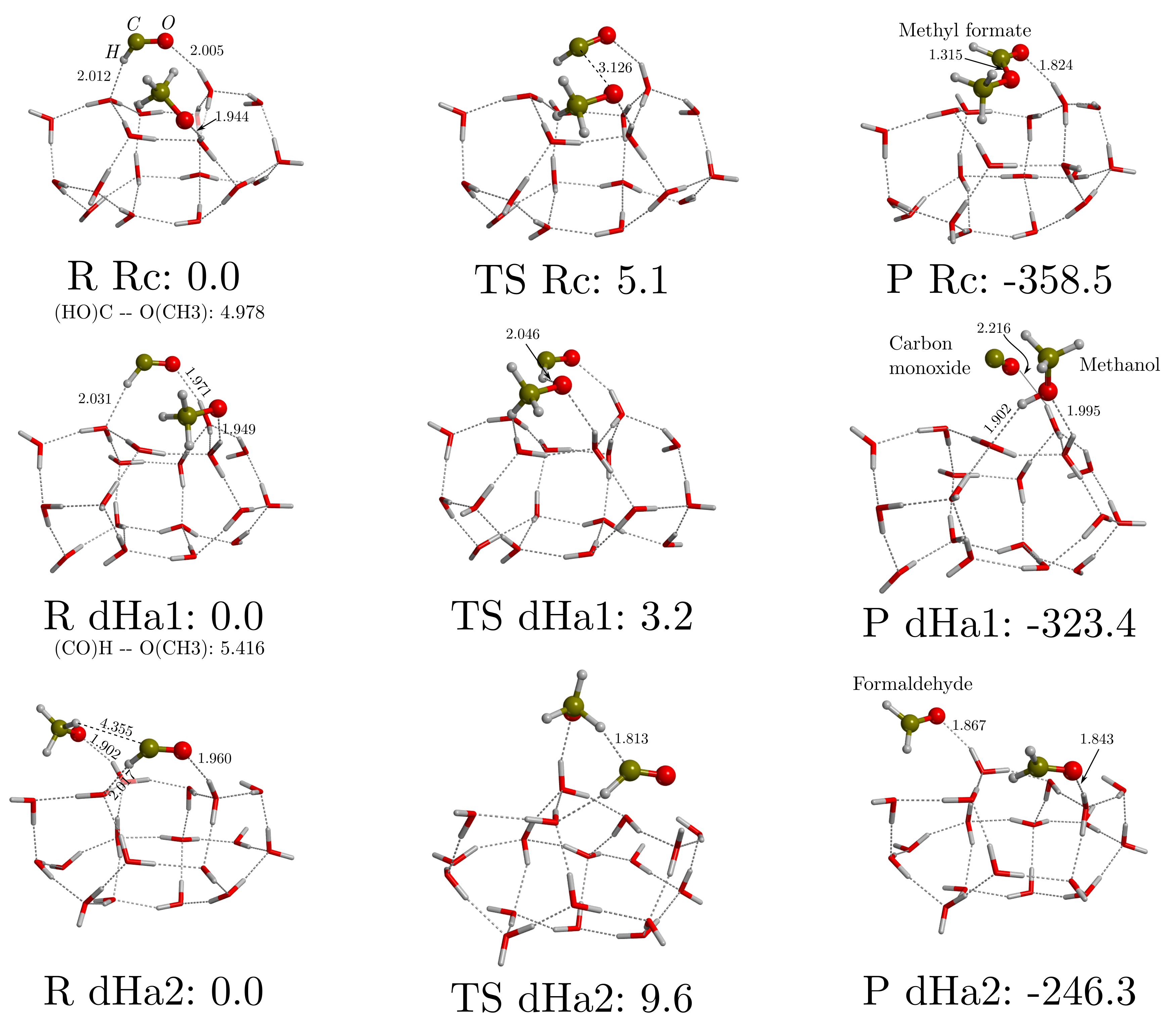}
\figsetgrpnote{Stationary points of radical--radical reactions on W18}
\figsetgrpend

\figsetgrpstart
\figsetgrpnum{1.9}
\figsetgrptitle{ZPE-corrected critical points of the Rc and dHa1/2 PESs for HCO + HCO on W18 fully optimised at the UBHLYP-D3(BJ)/6-31+G(d,p) theory level with the DFT energy refined to UBHLYP-D3(BJ)/6-311++G(2df,2pd). Energy units are in kJ/mol and distances in \r{A}.}
\figsetplot{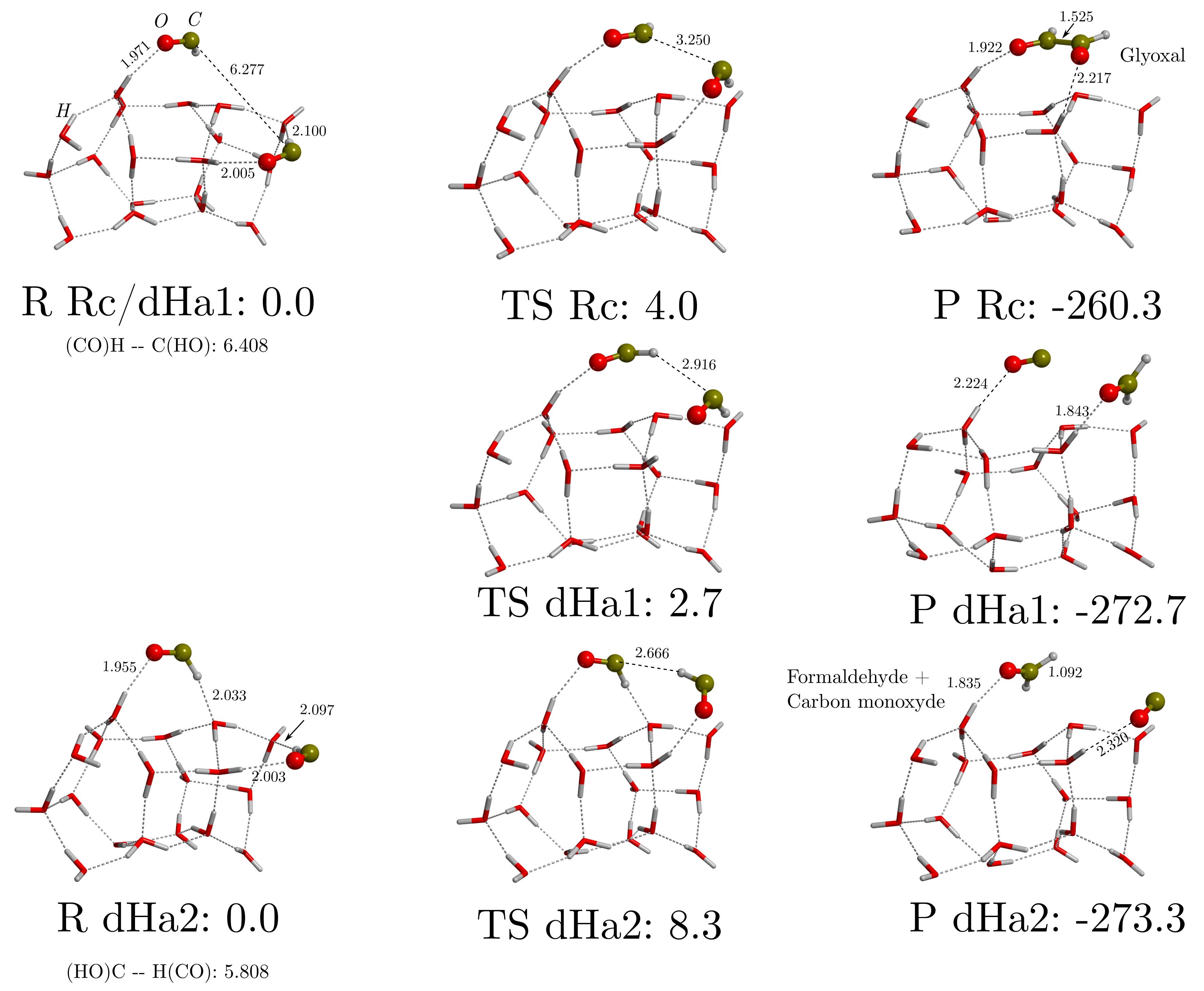}
\figsetgrpnote{Stationary points of radical--radical reactions on W18}
\figsetgrpend

\figsetend

\figsetstart
\figsetnum{2}
\figsettitle{Stationary points of the radical--radical reactions on the W33 ice model from the main text.}

\figsetgrpstart
\figsetgrpnum{2.1}
\figsetgrptitle{ZPE-corrected dHa2 PES critical points for HCO + CH$_3$O on W33-cav fully optimised at the BHLYP-D3(BJ) theory level. For dHa we report the two possibilities, the transfer from HCO and the one from CH$_3$O. Energy units are in kJ/mol and distances in \r{A}.}
\figsetplot{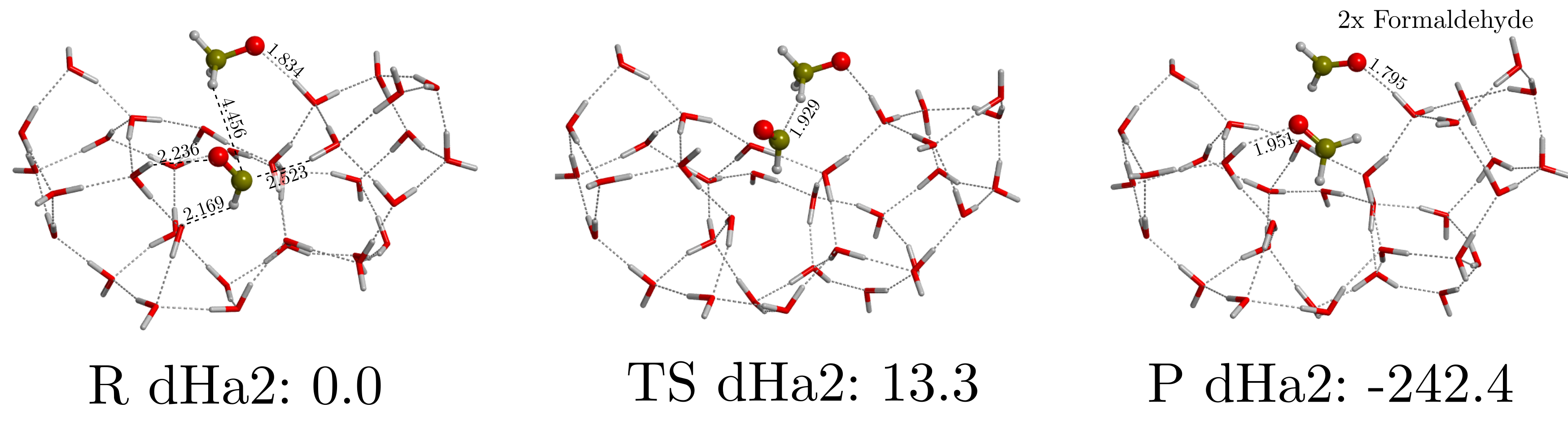}
\figsetgrpnote{More stationary points of radical--radical reactions on W33}
\figsetgrpend

\figsetgrpstart
\figsetgrpnum{2.2}
\figsetgrptitle{ZPE-corrected dHa PES critical points from radical 1 to radical 2 for CH$_2$OH + CH$_2$OH on W33-cav fully optimised at the BHLYP-D3(BJ) theory level. Energy units are in kJ/mol and distances in \r{A}.}
\figsetplot{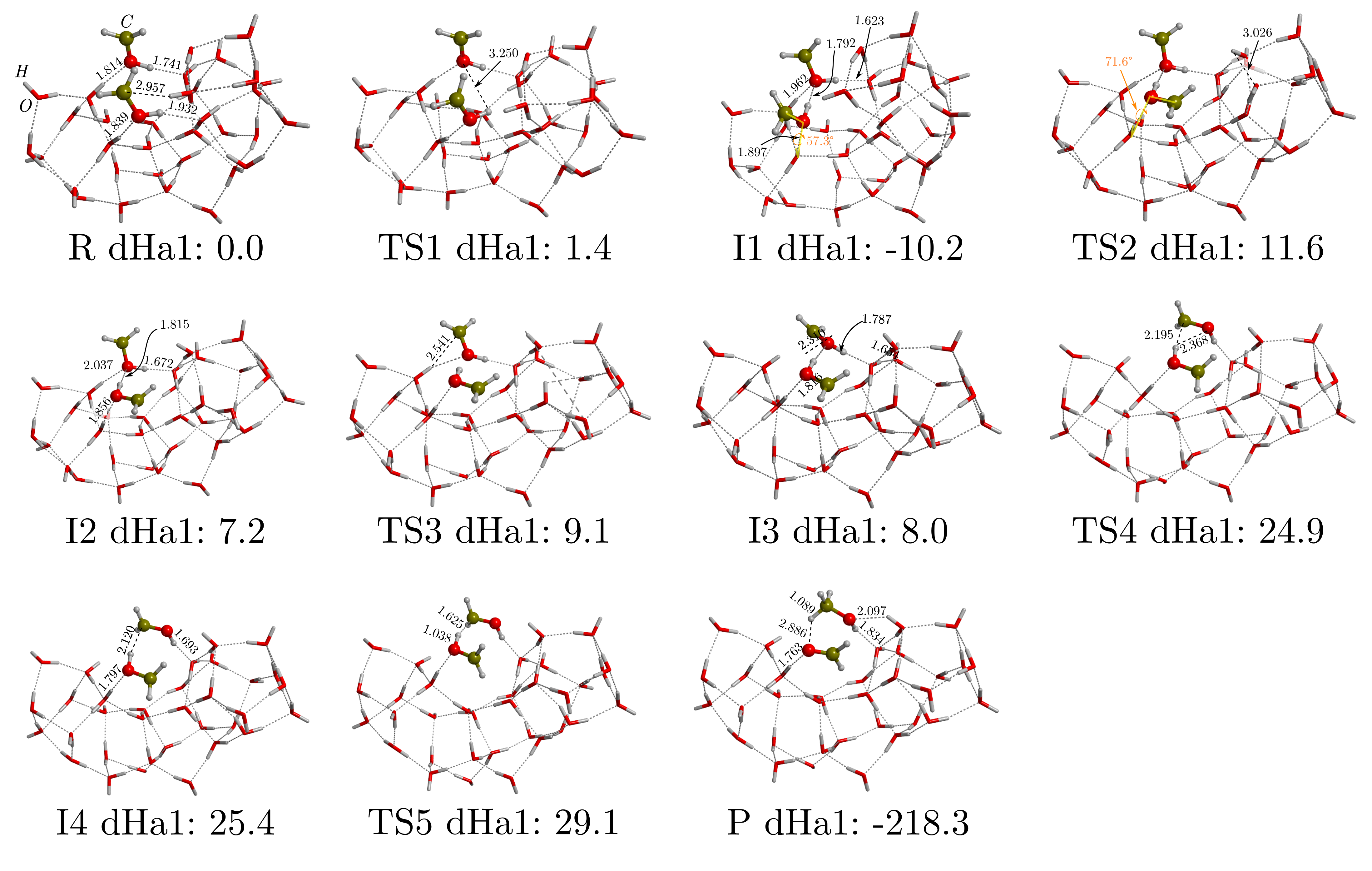}
\figsetgrpnote{More stationary points of radical--radical reactions on W33}
\figsetgrpend

\figsetgrpstart
\figsetgrpnum{2.3}
\figsetgrptitle{ZPE-corrected HCO + HCO on W33-cav PES critical points for the dHa2 transition state, which in this case connects a Rc and a dHa-like paths. The grey-shaded structure corresponds to an intermediate position in the IRC path leading from the TS to the Rc minimum (M1). a plot with the IRC (not ZPE-corrected) can be found in the lower right side, in which the grey dot corresponds to the intermediate structure and the black one to the transition state. The minima and maxima of the TS were fully optimised at the BHLYP-D3(BJ) theory level. Energy units are in kJ mol$^{-1}$ and distances in \r{A}.}
\figsetplot{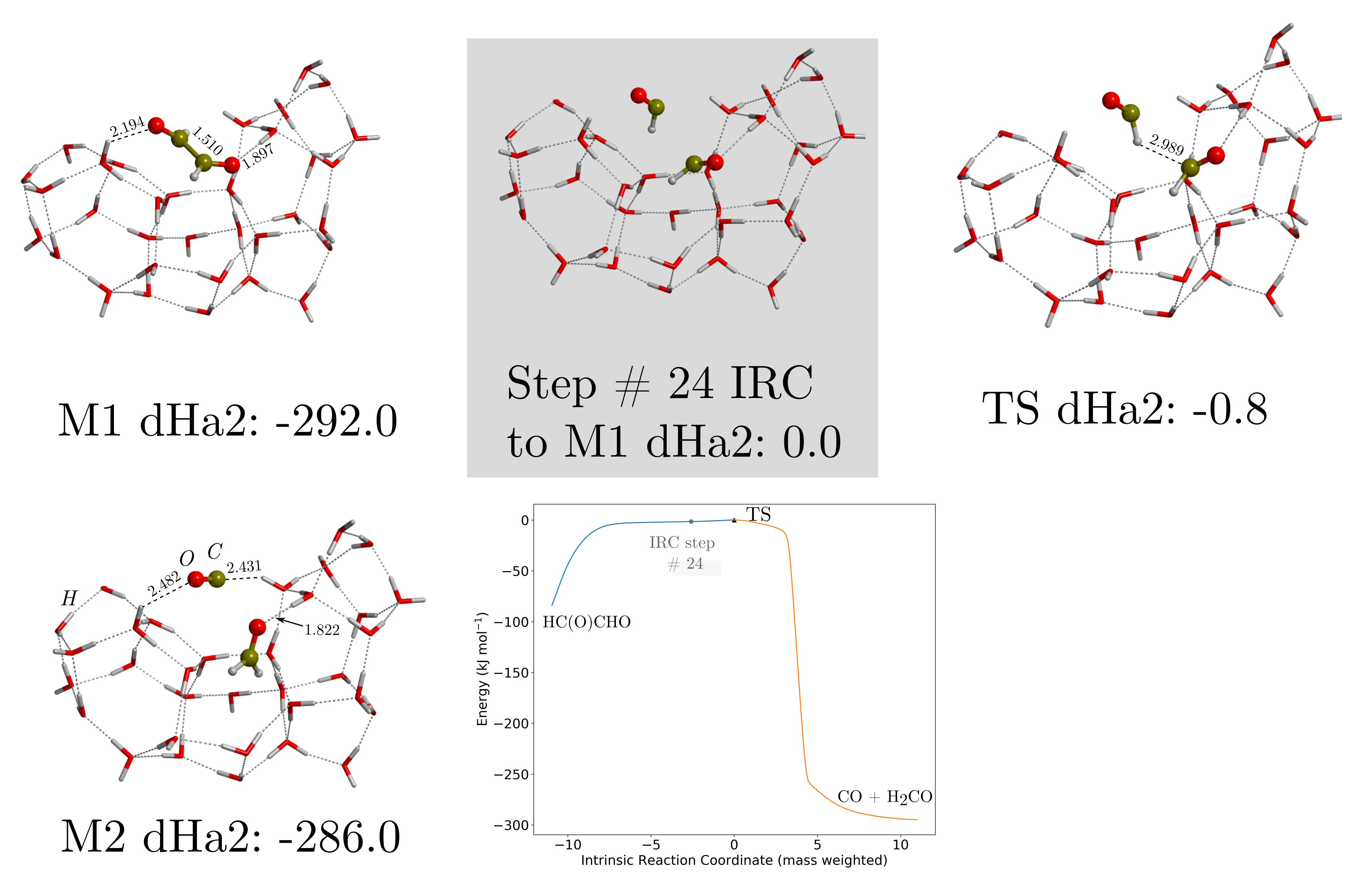}
\figsetgrpnote{More stationary points of radical--radical reactions on W33}
\figsetgrpend

\figsetgrpstart
\figsetgrpnum{2.4}
\figsetgrptitle{ZPE-corrected dHa2 PESs critical points for HCO + CH$_2$OH on W33-cav fully optimised at the BHLYP-D3(BJ) theory level. Energy units are in kJ/mol and distances in \r{A}.}
\figsetplot{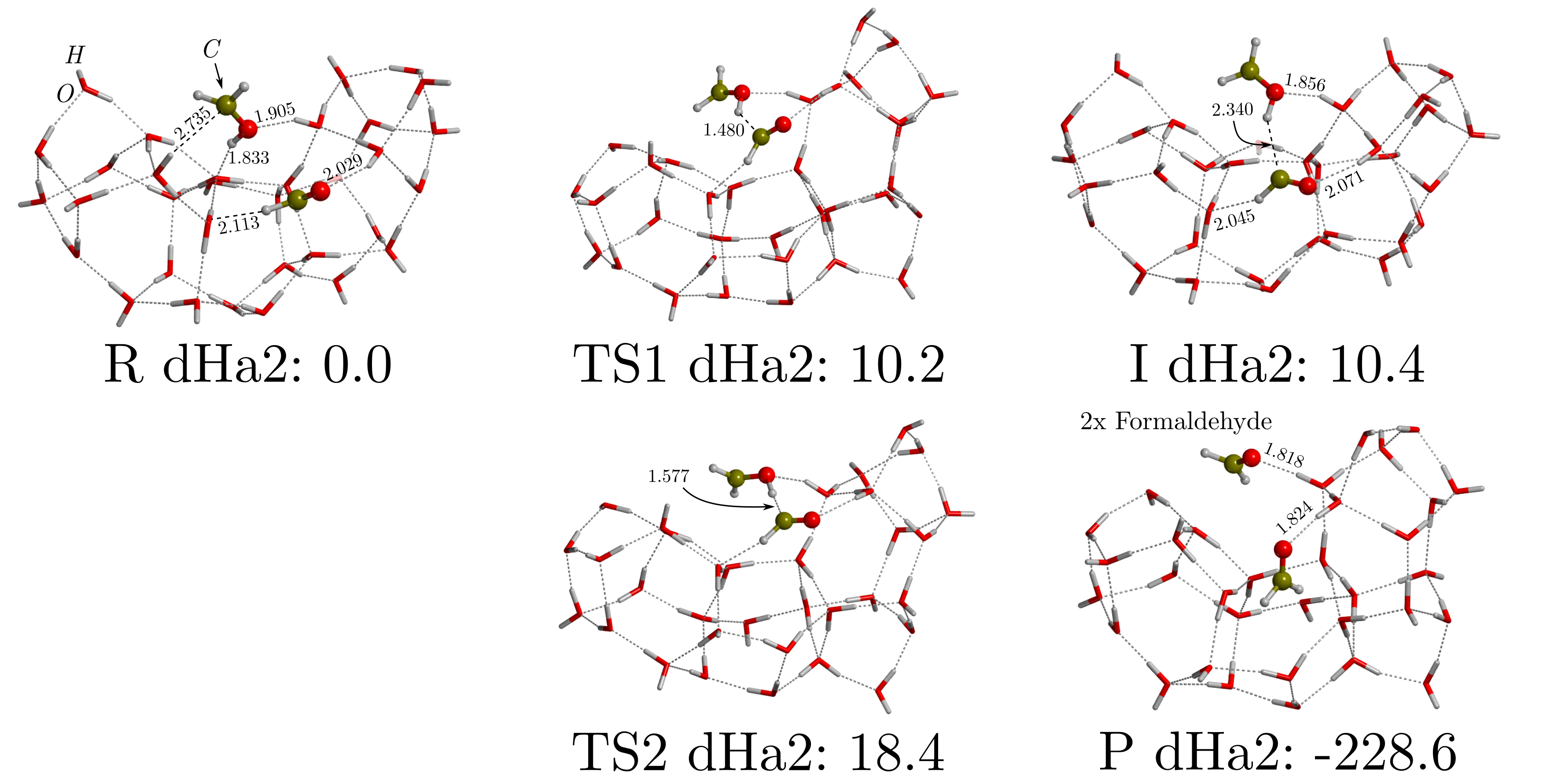}
\figsetgrpnote{More stationary points of radical--radical reactions on W33}
\figsetgrpend

\figsetgrpstart
\figsetgrpnum{2.5}
\figsetgrptitle{ZPE-corrected dHa from radical 2 to radical 1 and \textit{vice-versa} PESs critical points for CH$_3$O + CH$_3$O on W33-cav fully optimized at the BHLYP-D3(BJ) theory level. Energy units are in kJ/mol and distances in \r{A}.}
\figsetplot{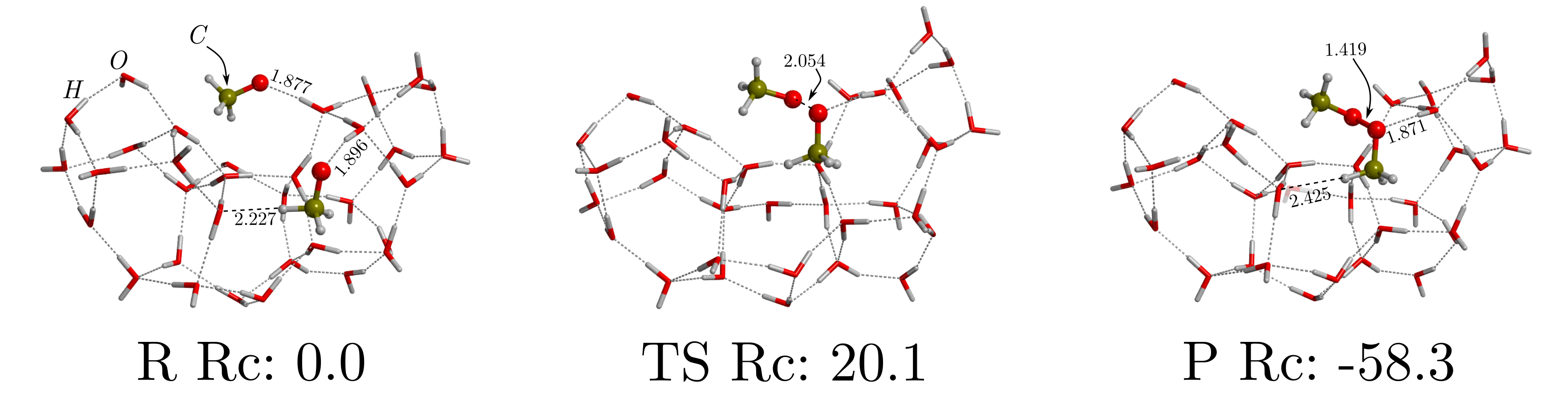}
\figsetgrpnote{More stationary points of radical--radical reactions on W33}
\figsetgrpend

\figsetend 

Remarkably, we showed in previous publications \citep{ER2019,ER2020} that water assisted hydrogen transfer reactions present multiple steps and exceedingly high activation energies to be surmountable at interstellar conditions, so that these paths have been excluded in this work. 
Therefore, here we focus on the radical-radical coupling (Rc) leading to the formation of iCOMs ( e.g. CH$_3$ + CH$_3$O $\to$ CH$_3$OCH$_3$), and the direct hydrogen abstraction (hereinafter dHa) leading to simpler products (e.g. CH$_3$ + CH$_3$O $\to$ CH$_4$ + H$_2$CO). 

It is worth reminding that dHa reactions are not possible in the cases of CH$_3$ + CH$_3$ and CH$_3$ + NH$_2$, since none of these radicals can behave as H-donor (i.e. these reactions would be endothermic).
In contrast, radical pairs in which both reactants exhibit properties of H-acceptors and H-donors, such as HCO, CH$_3$O and CH$_2$OH, two possible dHa processes are investigated (from each species, respectively). 
However, here we only show the most energetically favorable Rc and dHa channels.
The energetic data of all the computed reactions on both ice models are available in Tables \ref{tab:EnergiesW18}, \ref{tab:W18_energies} and \ref{tab:All_TS} in the annex. 

In the following, we will discuss the results separating the reactions into three groups, for better clarity: 
(i) reactions of CH$_3$ + X, 
(ii) reactions of HCO + X, and 
(iii) reactions of CH$_3$O + CH$_3$O and CH$_2$OH + CH$_2$OH.

\subsubsection{\texorpdfstring{CH$_3$ + X reactions}{CH3 + X}}

In general, this kind of reactions have very low energy barriers unless CH$_3$ is trapped by surrounding water molecules in the ice structure incrementing the energy barriers due to steric effects, as it is the case of CH$_3$ + CH$_3$O (both Rc and dHa channels) and CH$_3$ + CH$_3$. Another source of high energy barriers is when the radical partner experiences a strong attachment with the surface, which has to be broken for the reaction to take place. This is the case of the CH$_3$ + CH$_2$OH dHa reaction.

\begin{figure*}[!htbp]
    \centering
    \includegraphics[width=\textwidth]{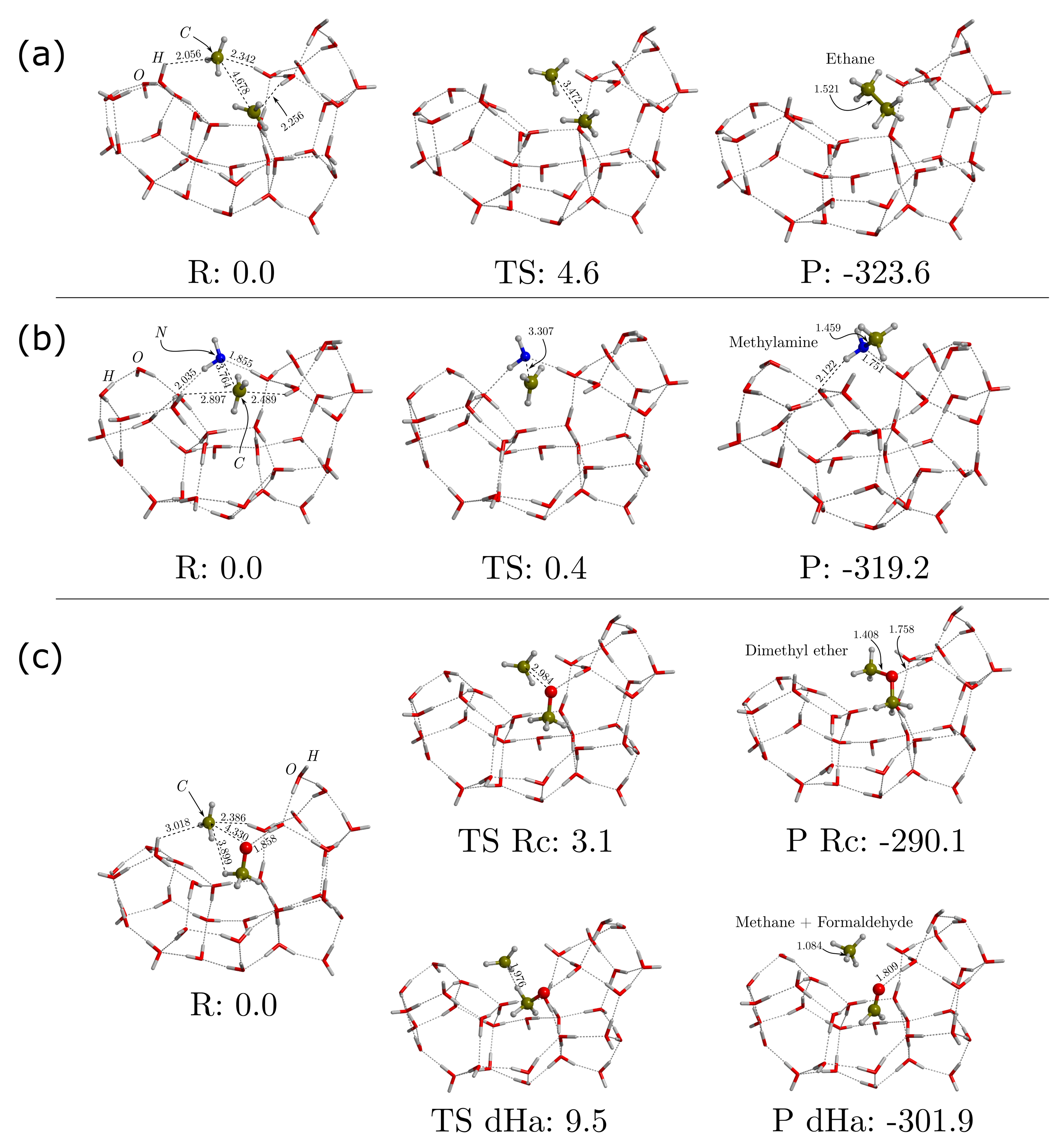}
    \caption{Relative ZPE-corrected potential energies of the stationary points for (a) CH$_3$/CH$_3$, (b) CH$_3$/NH$_2$, (c) CH$_3$/CH$_3$O and (d) CH$_3$/CH$-2$OH on W33-cav fully optimized at the BHLYP-D3(BJ)/6-31+G(d,p) theory level. DFT energies where further refined at BHLYP-D3(BJ)/6-311++G(2df,2pd) theory level. Energy units are in kJ mol$^{-1}$ and distances in \r{A}. [This Figure has a continuation in another page]}
    \label{fig:ch3_X}
\end{figure*}
\begin{figure*}[!htbp]
    \centering
    \includegraphics[width=\textwidth]{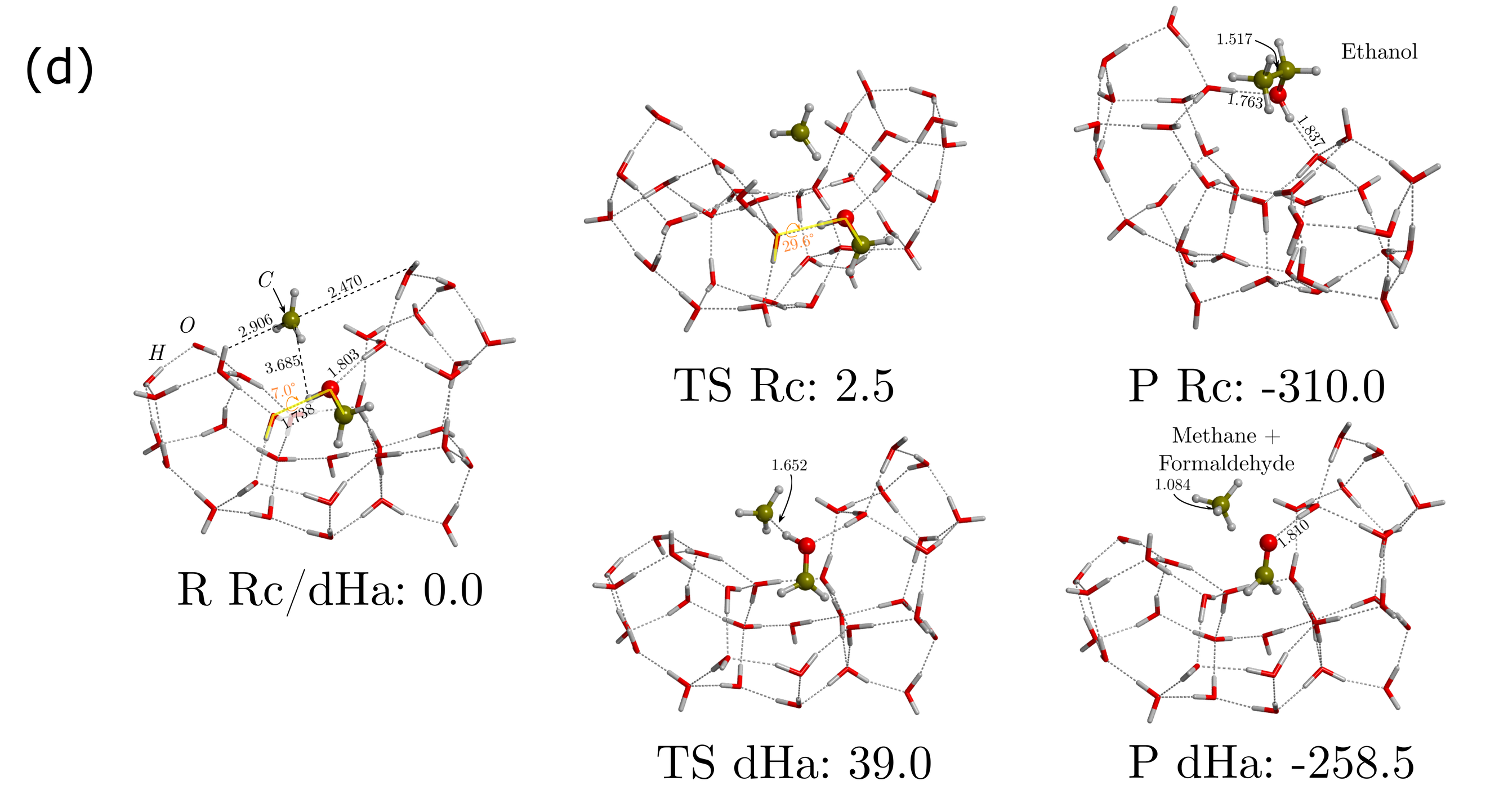}
    Figure \ref{fig:ch3_X}. Continued
\end{figure*}

\paragraph{\texorpdfstring{CH$_3$ + CH$_3$}{CH3 + CH3}}
This reaction, which can only lead to ethane formation through the Rc channel, is barrierless on W18. 
In contrast, on W33, despite that the CH$_3$/surface interactions are essentially via dispersive forces \citep{ER2019}, the reaction has a net energy barrier of 4.6 kJ mol$^{-1}$. 
The origin of this energy barrier arises from the interaction of one CH$_3$ with the water molecules of the surfaces. 
Indeed, in the reactant structure, one CH$_3$ is trapped by two dangling water surface H atoms, this way establishing weak H-bond interactions and reinforcing the dispersion interaction contribution. This ``blocked'' CH$_3$ radical needs to move against the surrounding water molecules, which requires a certain amount of energy. This can be seen by the change in the bare interaction energies (i.e. non-corrected for ZPE and BSSE) of the blocked CH$_3$ radical with the surface in reactants (24.6 kJ mol$^{-1}$) and in the TS (15.6 kJ mol$^{-1}$), where the TS has a much less stable CH$_3$ adsorption situation (by 9.1 kJ mol$^{-1}$).
 
Interestingly, these two weakly C--H$\cdots$O interactions are only possible on W33 due to the surface morphology of the cluster. 
Conversely, on W18, this interaction is not present (the two C--H$\cdots$O dangling bonds are missing) and, accordingly, the Rc reaction proceeds in a barrierless fashion.

\paragraph{\texorpdfstring{CH$_3$ + NH$_2$}{CH3 + NH2}}
As in the previous case, reaction between CH$_3$ and NH$_2$ only leads to the formation of the iCOM (Rc channel), in this case methylamine (CH$_3$NH$_2$). 
For this radical pair, product formation presents very low energy barriers (0.4/1.6 kJ mol$^{-1}$ on W33/W18). 
This is because the transition states mainly involve a translation/rotation of CH$_3$ towards NH$_2$.

\paragraph{\texorpdfstring{CH$_3$ + CH$_3$O}{CH3 + CH3O}} The formation of dimethyl ether (CH$_3$OCH$_3$) through the Rc channel on W18 has a barrier 0.2 kJ mol$^{-1}$ while on W33 of 3.1 kJ mol$^{-1}$. 
This reaction is barrierless when we do not consider any ASW model (as in practice on W18), therefore, the origin of the barrier on W33 is caused by the morphology of the cluster model, similarly to what happens to the CH$_3$ + CH$_3$ case.
Indeed, also in this case, the CH$_3$ establishes weak H-bond interactions, which have to be broken to couple with the O of CH$_3$O. The difference in the interaction energies of the CH$_3$ radical with the surface between reactants (interaction energy of 24.4 kJ mol$^{-1}$) and the transition state (interaction energy of 17.1 kJ mol$^{-1}$) is 7.3 kJ mol$^{-1}$ (not corrected for ZPE and BSSE), while for the CH$_3$O radical is rather small, just 0.4 kJ mol$^{-1}$.
The dHa channel leading to CH$_4$ + H$_2$CO presents activation barriers of 9.5 and 1.0 kJ mol$^{-1}$ on W33 and W18, respectively. 
In the absence of water molecules this reaction has a barrier of 1.7 kJ mol$^{-1}$ associated with the H--CH$_2$O bond breaking. 
This indicates that the energy barrier on W18 arises from this weak H-bond breaking, while on W33 it also has contributions from cavity effects. In this case, it is the combination of both the blocking of the CH$_3$ radical and the steric effects arising from the dispersion forces between the --CH$_3$ moiety of CH$_3$O and the surface in order to reach a proper orientation allowing the H transfer. Overall, the CH$_3$ radical experiences a change in interaction energies with the surface of 13.5 kJ mol$^{-1}$, and CH$_3$O of 6.0 kJ mol$^{-1}$.

\paragraph{\texorpdfstring{CH$_3$ + CH$_2$OH}{CH3 + CH2OH}} This radical pair presents very low energy barriers in the Rc channel to form ethanol (CH$_3$CH$_2$OH), of 2.5 and 1.9 kJ mol$^{-1}$ on W33 and W18, respectively. The opposite occurs for the dHa channel to form CH$_4$ + H$_2$CO (39.0 and 32.2 kJ mol$^{-1}$ on W33 and W18, respectively).
Although the interaction of CH$_2$OH with the surface is strong due to stable H-bonds, these H-bonds do not affect the Rc channel, since the unpaired electron is on the C atom, which is freely accessible for the coupling.
Accordingly, the Rc channel only requires overcoming the CH$_3$/surface dispersion interactions to form the C--C bond. 
In contrast, for the dHa channel, the strong CH$_2$OH/surface H-bond interactions inhibit the reaction since the H to be transferred is participating in the H-bonds, requiring the breaking of these interactions. 
The cost of this action is reflected by the fact that, in the absence of water molecules, the dHa channel has a lower energy barrier, of 9.8 kJ mol$^{-1}$.

\subsubsection{HCO + X}

At difference from the previous set of reactions, HCO + X have slightly higher energy barriers due to the higher binding energy of HCO. Nevertheless, HCO is a relatively good H-donor, and therefore, dHa reactions in which HCO donates its H atom have similar energy barriers as those of Rc.

\paragraph{HCO + HCO} The energy barriers for Rc forming glyoxal (HCOCHO) and for dHa forming CO + H$_2$CO are very similar, i.e. 4.1 and 4.0 kJ mol$^{-1}$ for Rc, and 4.0 and 2.7 kJ mol$^{-1}$ for dHa, on the W33 and W18 surfaces, respectively. 
Here, the energy barriers are very similar because in both paths the structural reorganization of the reactants leading to products is also similar. 
Indeed, the reactions mainly involve the rotation of one of the two HCO radicals to arrive at the proper orientation to form either HCOCHO or CO + H$_2$CO, with a similar energy cost. 

\paragraph{\texorpdfstring{HCO + CH$_3$O}{HCO + CH3O}} For this system, the Rc and dHa channels (forming methyl formate and CO + CH$_3$OH, respectively) on W33 and W18 present similar energy barriers, of 3.5 and 5.1 and 2.0 and 3.2 kJ mol$^{-1}$, respectively. 
This is because the reactions proceed either through the translation (Rc) or the rotation (dHa) of the HCO radical, which present a similar energy cost. It is worth mentioning that this biradical system can also present another dHa channel, in which the CH$_3$O transfers its H atom to HCO to form H$_2$CO + H$_2$CO. However, this channel has a higher energy barrier (13.3/9.6 kJ mol$^{-1}$ on W33/W18) because the orientation of CH$_3$O to transfer its H requires the breaking of the CH$_3$O/surface interactions. This was also observed in other radical pairs in which CH$_3$O is the H-donor in dHa processes (e.g., CH$_3$ + CH$_3$O).


\begin{figure*}[!htbp]
    \centering
    \includegraphics[width=\textwidth]{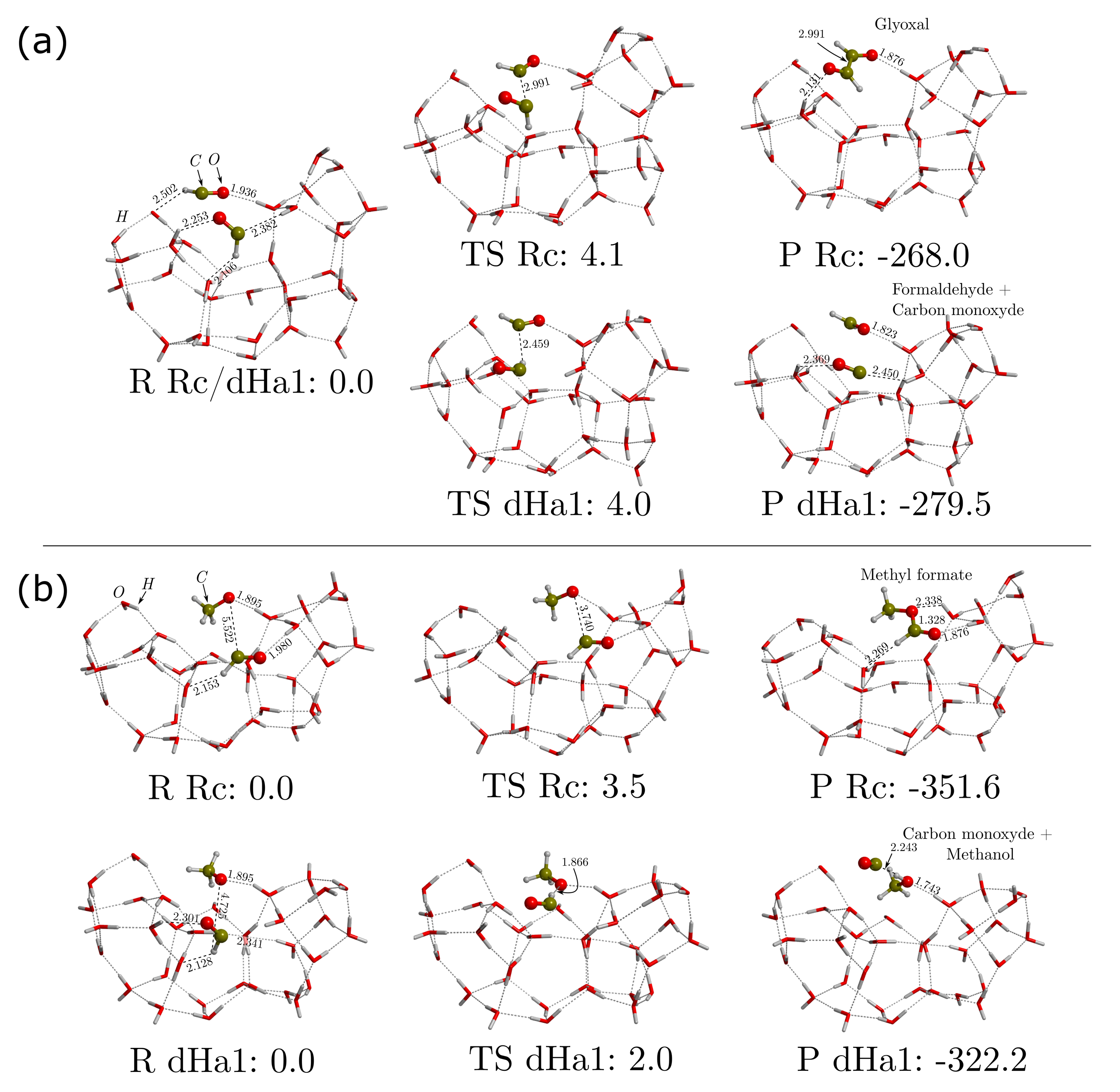}
    \caption{Relative ZPE-corrected potential energies of the stationary points for (a) HCO/HCO, (b) HCO/CH$_3$O and (c) HCO/CH$_2$OH on W33-cav fully optimized at the BHLYP-D3(BJ)/6-31+G(d,p) theory level. DFT energies where further refined at BHLYP-D3(BJ)/6-311++G(2df,2pd) theory level. Energy units are in kJ mol$^{-1}$ and distances in \r{A}. [This Figure has a continuation in another page]}
    \label{fig:hco_X}
\end{figure*}
\begin{figure*}[!htbp]
    \centering
    \includegraphics[width=\textwidth]{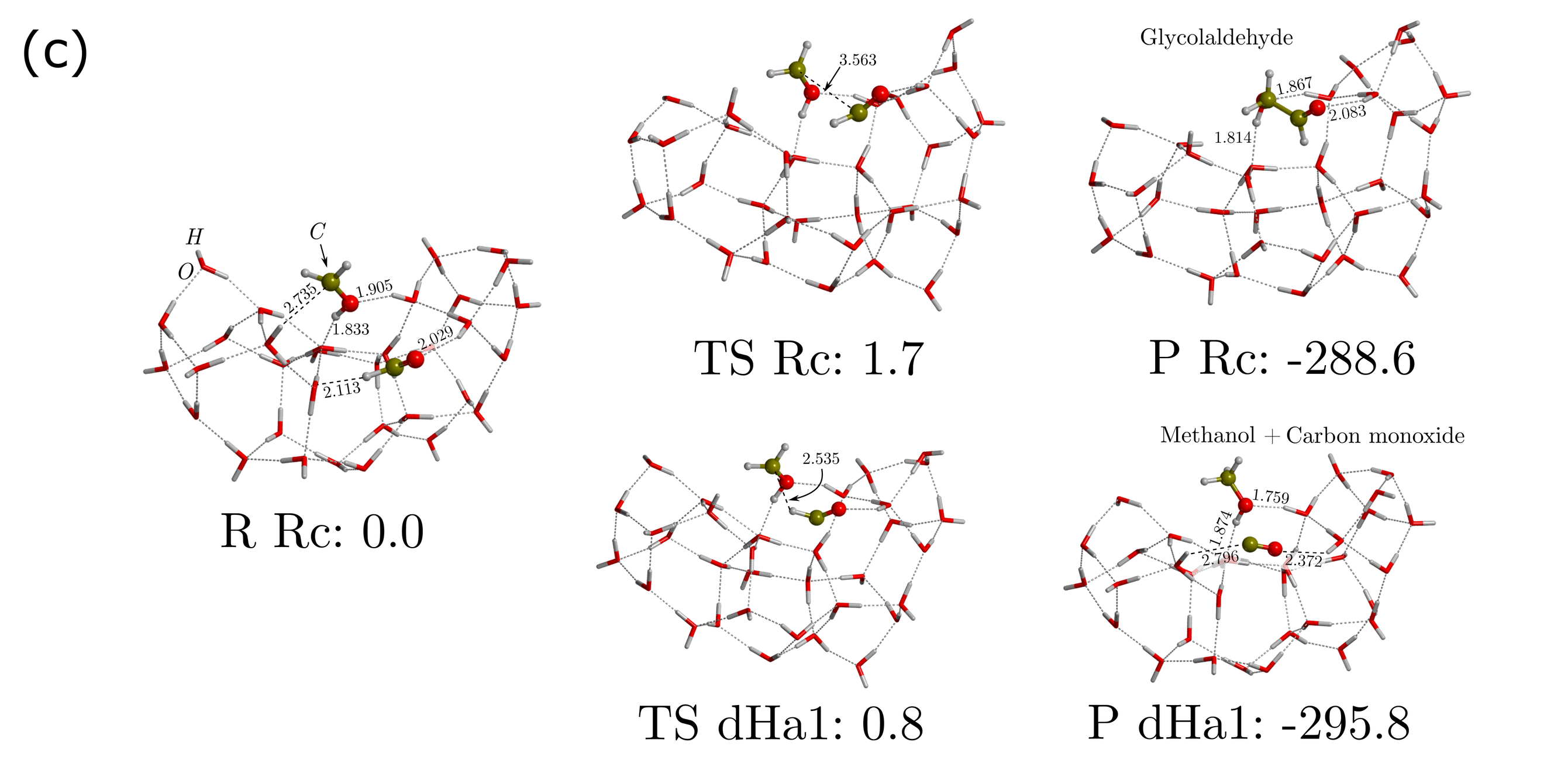}
Figure \ref{fig:hco_X}. Continued
\end{figure*}



\paragraph{\texorpdfstring{HCO + CH$_2$OH}{HCO + CH2OH}} The reactivity of this biradical system is similar to the previous one. That is, both the Rc channel (forming glycolaldehyde) and the dHa channel (in this case forming CO + CH$_3$OH due the H transfer from HCO to CH$_2$OH) present energy barriers below $\sim$2 kJ mol$^{-1}$, irrespective of the surface model where they are calculated. The explanation is the same: since CH$_2$OH is firmly attached by H-bonds on the surface, the reactions are driven by the motion of HCO, which in practice does not present any energy cost. Similarly to the previous biradical system, another dHa channel has been identified: that in which the H transfer takes place from CH$_2$OH to HCO forming H$_2$CO + H$_2$CO. Also in this case, the energy barriers are as high as $\sim$18 kJ mol$^{-1}$ (on W33, see Figure Set 1 in the online version) due to the energy cost to break the CH$_2$OH/surface interactions, which is mandatory to transfer the H atom. The same reaction in the absence of water presents a barrier of 8.8 kJ mol$^{-1}$, indicating that the this structural reorganization is hindered by the CH$_2$OH/surface H-bonds.


\subsubsection{\texorpdfstring{CH$_3$O + CH$_3$O and CH$_2$OH + CH$_2$OH}{CH3O + CH3O and CH2OH + CH2OH}}

As it was seen in above, the reactivity of CH$_2$OH + X and CH$_3$O + X (where X = CH$_3$ and HCO) share some similar aspects, namely Rc and dHa (where neither CH$_3$O nor CH$_2$OH act as an H donor) reactions tend to have low energy barriers. However, high energy barriers appear when either CH$_3$O or CH$_2$OH act as H-donors in dHa reactions.

In the CH$_3$O + CH$_3$O and CH$_2$OH + CH$_2$OH cases, we observed a clear different reactivity for their coupling reactions. In the CH$_3$O + CH$_3$O system, the Rc channel presents high energy barriers, given that the unpaired electrons on the O atoms are less reactive as a consequence of the H-bonding interaction with the surface, while for the CH$_2$OH + CH$_2$OH case, very low activation energies are obtained for the Rc channel on either surface models as a consequence of CH$_2$OH binding mode.
Regarding their dHa reactions, we observe high energy barriers for both systems, as observed for CH$_2$OH + CH$_3$/HCO and CH$_3$O + CH$_3$/HCO where CH$_2$OH/CH$_3$O act as the H-donor.

\begin{figure*}[!htb]
    \centering
    \includegraphics[width=\textwidth]{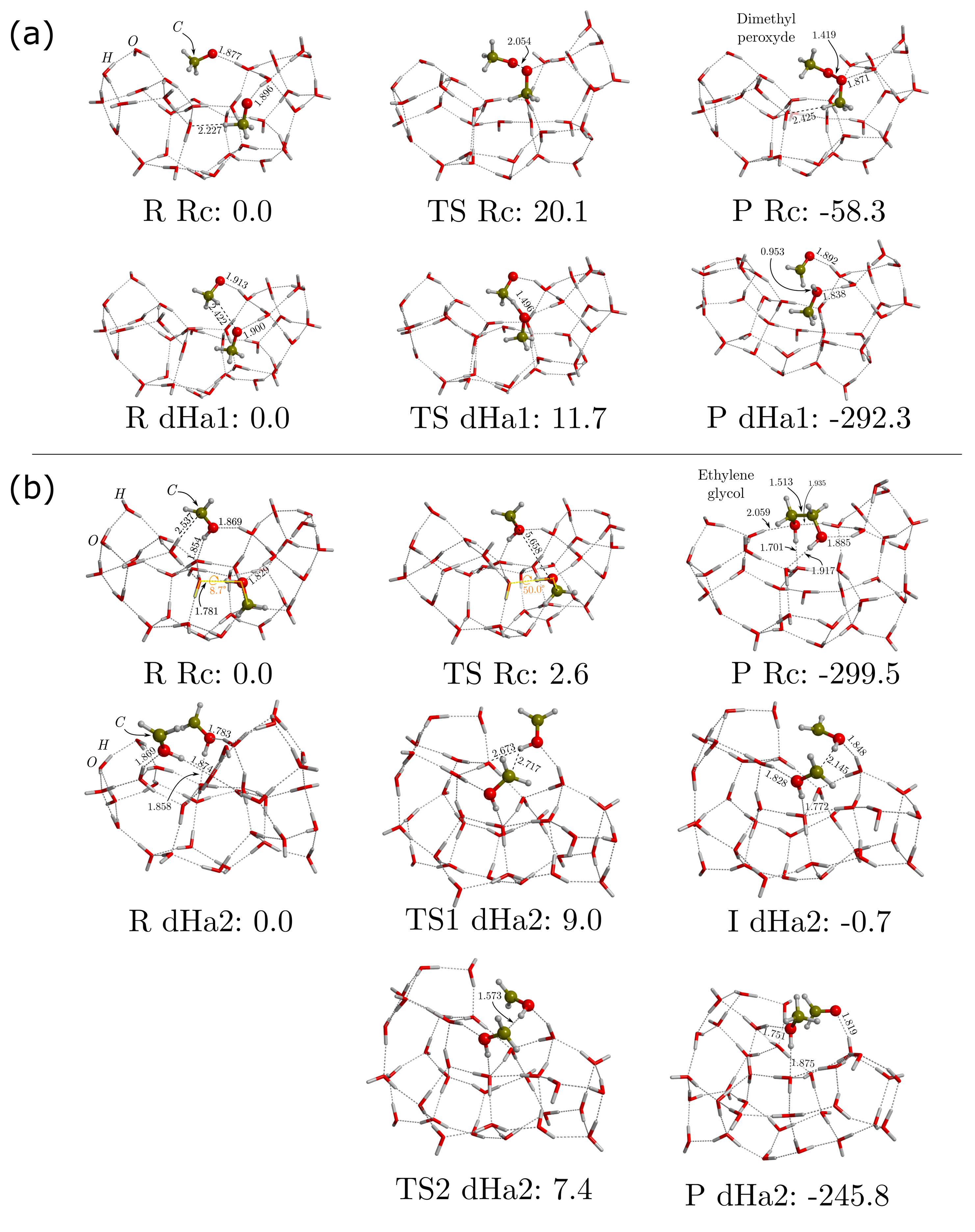}
    \caption{Relative ZPE-corrected potential energies of the stationary points for (a) CH$_3$O/CH$_3$O and (b) CH$_2$OH/CH$_2$OH on W33-cav fully optimized at the BHLYP-D3(BJ)/6-31+G(d,p) theory level. DFT energies where further refined at BHLYP-D3(BJ)/6-311++G(2df,2pd) theory level. Energy units are in kJ mol$^{-1}$ and distances in \r{A}.}
    \label{fig:ch3o_ch3o__ch2oh_ch2oh}
\end{figure*}

\paragraph{\texorpdfstring{CH$_3$O + CH$_3$O}{CH3O + CH3O}} The Rc channel between two CH$_3$O radicals on W33 has a higher energy barrier than the dHa one (20.1 and 11.7 kJ mol$^{-1}$, respectively). This is because, in the reactant structure, both CH$_3$O radicals establish H-bond interactions with the surface through their O atoms. Since the Rc channel involves the coupling of the unpaired electrons of the two O atoms, the reaction requires the breaking of these H-bonds in both species. The contribution of these H-bond interactions in this energy barrier is demonstrated by the calculated value of the barrier in absence of water, of 1.8 kJ mol$^{-1}$. In the dHa channel, in contrast, a H atom is transferred from one CH$_3$O to the other without the need to break these H-bonds. In this case, the reorientation of the radicals is enough to facilitate the H transfer.

However, on W18 we observe the opposite trend: the Rc channel presents a lower energy barrier than the dHa one (10.3 and 15.9 kJ mol$^{-1}$, respectively). This is because in the reactant structure there are less intermolecular interactions and the two radicals are well oriented for the coupling, something that cannot take place in the cavity model due to its size and the lack of well oriented binding sites.

\paragraph{\texorpdfstring{CH$_2$OH + CH$_2$OH}{CH2OH + CH2OH}} This system is a paradigmatic case in which Rc channel has a low energy barrier while the dHa ones do not. Indeed, the lowest energy path is the Rc one, with 2.6/4.4 kJ mol$^{-1}$ on W33/W18. On both clusters, the reaction involves a simple rotation around the intermolecular C--C dihedral angle (e.g. see Rc path in Figure \ref{fig:ch3o_ch3o__ch2oh_ch2oh}(b)) in such a way that once the C atoms of each radical are faced one to each other the system easily evolves to form CH$_2$(OH)CH$_2$OH (ethylene glycol) as a product.

In contrast, dHa reactions present higher energy barriers, about 9.1/20.6 kJ mol$^{-1}$ on W33/W18, and often have multiple reorientation steps before the actual H-abstraction takes place, see for example the dHa1 and dHa2 channels on W33 shown in Figure Set 1 (find it in the online version of the journal) and and Figure \ref{fig:ch3o_ch3o__ch2oh_ch2oh}(b), respectively.
These are the consequences of the intrinsic stability of the CH$_2$OH radical (in the absence of water molecules, two CH$_2$OH radicals are able to form stable dimers), and the high capacity of this radical 

\subsubsection{Summary of radical--radical reactivity}

The results on the activation barriers for each studied system and reaction are listed in Table \ref{tab:sum-best-Eact}. 
In this table, we report the highest energy barriers (including ZPE corrections), if any, of the least energetic path for Rc and dHa reactions, respectively, for both W33 and W18 ice models and in the absence of any ASW model.
For those cases in which the PES has more than one reaction step (for example the  dHa-CH$_2$OH/CH$_2$OH$\cdots$W33: Fig. \ref{fig:ch3o_ch3o__ch2oh_ch2oh}), only the highest barrier of the sequence is reported. 
Likewise, for those cases where two dHa channels exist, only the highest barrier of the most favourable channel (according to the activation energies) is reported.

Table \ref{tab:sum-best-Eact} also reports the reaction energy for all the studied reactions.

\begin{table*}[!htbp]
\centering
    \caption{\footnotesize Summary of the theoretical results for radical-radical reactivity.
    First column reports the radical-radical system and column (2) the ice model to which the computations apply, i.e. W33 or W18 ice models or absence of water molecules (noW).
    Columns from (3) to (5) report the Radical coupling (Rc) product (col. 3) with the (ZPE-corrected) activation energy ($\Delta H^{\ddagger}$: col. 4) and the reaction energy ($\Delta H^{reac}$: col. 5).
    Columns from (6) to (9) report the Direct H-abstraction (dHa) product (col. 6) with the (ZPE-corrected) activation energy ($\Delta H^{\ddagger}$: col. 7), the reaction energy ($\Delta H^{reac}$: col. 8) and the crossover temperature ($T_c$, see \S \ref{sec:appendix:crossover_temp} in the annex: col. 9).
    The last column reports the category to which the reaction belongs (see text), based on the the efficiencies computed in Eq. \ref{eqn:efficiency} (assuming a diffusion-to-desorption barrier ratio of 0.35) and the crossover temperatures: (1) Rc plausible \& dHa not plausible/possible; (2) Rc--dHa competition; (3) Rc--dHa competition at low temperatures thanks to tunneling; (4) Rc not plausible, dHa only plausible at low temperatures thanks to tunneling. Energy units are kJ mol$^{-1}$, temperatures in K.}
\label{tab:sum-best-Eact}
\resizebox{1.0\textwidth}{!}{%
\hskip-2.5cm \begin{tabular}{l|c|ccc|cccc|c}
\hline
System & Ice & \multicolumn{3}{c|}{Radical-radical coupling} & \multicolumn{4}{c|}{Direct H-abstraction} &  Reaction  \\  
       & Model & Product & $\Delta H^{\ddagger}$ & $\Delta H^{reac}$ & Product & $\Delta H^{\ddagger}$ & $\Delta H^{reac}$ & $T_c$ [K] & category\\ \hline \hline
\multirow{3}{*}{CH$_3$ + CH$_3$} 
      & W33  & CH$_3$CH$_3$ & 4.6 & -323.6 &          &     &     &         &  1  \\
      & W18  &              & NB  & -333.4 & None             &     &     &         &  1 \\
      & noW  &              & NB  & -338.1 &              &     &     &         &    \\ \hline 
\multirow{3}{*}{CH$_3$ + NH$_2$}
      & W33  & CH$_3$NH$_2$ & NB  & -319.2 &          &     &     &         &  1 \\
      & W18  &              & NB  & -316.8 &  None            &     &     &         &  1 \\
      & noW  &              & NB  & -309.1 &              &     &     &         &     \\ \hline 
\multirow{3}{*}{CH$_3$ + CH$_3$O}  
      & W33  & CH$_3$OCH$_3$& 3.1 & -290.1 & CH$_4$ + H$_2$CO & 9.5 & -301.9 & 47.1 &  1, 3 \\
      & W18  &              & NB  & -299.5 &                  & 1.0 & -302.7 & 36.1 & 2  \\
      & noW  &              & NB  & -298.3 &                  & 1.7 & -297.5        &    \\ \hline
\multirow{3}{*}{CH$_3$ + CH$_2$OH}  
      & W33  & CH$_3$CH$_2$OH& 2.5 & -310.0             & CH$_4$ + H$_2$CO & 39.0 & -258.5 & 242.0 & 1, 3(?) \\
      & W18  &               & 1.9 & -320.4             &                  & 23.9 & -255.1 & 295.7 & 1, 3(?) \\
      & noW  &               & NB*  & -327.5$^{\nabla}$ &                  & 9.8  & -273.1        &      \\ \hline
\multirow{3}{*}{HCO + HCO} 
     & W33 & CHOCHO & 4.1 & -268.0            & CO + H$_2$CO & 4.0 & -279.5           & 28.4 & 2 \\
     & W18 &        & 4.0 & -260.3            &              & 2.7 & -272.7           & 10.8 & 2 \\
     & noW &        & NB  & -275.2$^{\nabla}$ &              & NB  & -278.5$^{\nabla}$&      & \\ \hline
\multirow{3}{*}{HCO + CH$_3$O} 
     & W33 & CH$_3$OCHO & 3.5 & -351.6                 & CO + CH$_3$OH & 2.0        & -322.2 & 34.6 & 2 \\
     & W18 &            & 5.1 & -358.5                 &               & 3.2        & -323.4 & 57.8 & 2 \\
     & noW &            & NB$^{*}$ & -358.6$^{\nabla}$ &               & NB$^{*;\#}$ & -326.3 &      &   \\ \hline
\multirow{3}{*}{HCO + CH$_2$OH} 
     & W33 & CHOCH$_2$OH & 1.7 & -288.6                 & CO + CH$_3$OH & NB$^{*}$       & -295.8 & - & 2 \\
     & W18 &             & 1.6 & -286.2                 &               & NB$^{*}$       & -290.2 & - & 2 \\
     & noW &             & NB$^{*}$ & -303.8$^{\nabla}$ &               & NB$^{*;\#}$ & -297.8    &   &   \\\hline \hline
\multirow{3}{*}{HCO + CH$_3$}
     & W33 & CH$_3$CHO & 5.5 & -324.5            & CH$_4$ + CO & 7.2 & -328.9            & 40.0 &  4    \\  
     & W18 &           & 1.8 & -329.5            &             & 5.0 & -321.5            & 8.0  &  1 \\ 
     & noW &           & NB  & -326.1$^{\nabla}$ &             & NB  & -340.4$^{\nabla}$ &      &     \\ \hline
\multirow{3}{*}{HCO + NH$_2$}
     & W33 & NH$_2$CHO & 2.1 & -385.7            &  NH$_3$ + CO & 1.4 & -335.0            & 28.8 & 2 \\   
     & W18 &           & 3.8 & -364.9            &              & 4.9 & -338.4            & 7.3  & 2 \\  
     & noW &           & NB  & -388.4$^{\nabla}$ &              & NB & -344.0$^{\nabla}$ &      &   \\ \hline
\hline
\multirow{3}{*}{CH$_3$O + CH$_3$O}
     & W33 & (CH$_3$O)$_2$ & 20.1 & -58.3 & CH$_3$OH + H$_2$CO  & 11.7 & -292.3 & 225.7 & 4 \\
     & W18 &               & 10.3 & -64.1 &                     & 15.9 & -312.5 & 212.9 & 4, 3(?) \\
     & noW &               & 1.8  & -75.0 &                     & 8.1  & -284.1 &       &   \\ \hline
\multirow{3}{*}{CH$_2$OH + CH$_2$OH}
     & W33 & (CH$_2$OH)$_2$ & 2.6  & -299.5 & CH$_3$OH + H$_2$CO & 9.0  & -245.8 & 296.2 & 1  \\
     & W18 &                & 4.4  & -288.7 &                    & 27.6 & -226.3 & 394.3 & 1  \\
     & noW &                & 10.7 & -295.1 &                    & 6.0  & -241.0 &       &    \\  \hline
\end{tabular}%
}
\centering
\begin{minipage}{14cm}
    \scriptsize 
    Note: CH$_3$/HCO and NH$_2$/HCO values on W18 and W33 were recalculated from those of \cite{ER2019}. The main difference is on the dispersion correction used in this work \citep[see also][]{ER2021}.\\
    NB$^*$ indicates that the reaction has no effective barrier ($<$ 1 kJ mol$^{-1}$), although a transition state was found, which after correcting for ZPE goes below the energy of the reactants.\\
    $^{\#}$ Regarding the dHa reactions of HCO + CH$_3$O/CH$_2$OH in the absence of water molecules we report those for the dHa1 channel (i.e. HCO transfers its H atom to the partner radical). The dHa2 channels where CH$_3$O or CH$_2$OH transfer its H atoms to HCO have higher barriers, of 5.3 and 8.8 kJ mol$^{-1}$, respectively, similar to CH$_3$ + CH$_3$O/CH$_2$OH.\\
    $^{\nabla}$ Calculated with respect to the asymptote (i.e. the sum of the energy of both radicals alone).
\end{minipage}
\end{table*}

\subsection{Reaction efficiencies}
As described in Sect. \ref{sec:methods}, we computed a rough estimate of the efficiency of the reactions, $\varepsilon$, following the Eq. \ref{eqn:efficiency}, using the computed binding energies of the radicals (Tab. \ref{tab:BEs}, assuming a $E_{diff}/E_{des}$ ratio of 0.35) and the activation energy barriers of the reactions (Tab. \ref{tab:sum-best-Eact}).
Quantum tunneling effects are included in a qualitative manner on dHa reactions \textit{via} their crossover temperatures ($T_c$, see \S \ref{sec:appendix:crossover_temp} in the annex).

With these calculations we aim to provide a simple means to discriminate which radical-radical processes are likely efficient from those that are not.
In order to do this, we provide the efficiency values at the highest temperature possible for each reaction. In the absence of a full astrochemical model, we calculate this upper limit temperature as the temperature at which one would expect radicals to disappear from the surface due to thermal desorption. This is achieved by matching the desorption time-scale (proportional to $1/k_{des}$) to a value of 1 Myr (corresponding to the typical age expected for a protostar), which provides us with a temperature value. 
We label these temperatures by $T_{des}$, and they are listed in Tabs. \ref{tab:efficiencies} and \ref{tab:my-table} for the W33 and W18 ice models, respectively, together with the efficiencies for the radical coupling and direct H-abstraction reactions at these temperatures, and at 10 K. 

Please note that Tabs. \ref{tab:efficiencies} and \ref{tab:my-table} also report the efficiencies and $T_{des}$ of the HCO + CH$_3$/NH$_2$ systems.
The first one, leading to acetaldehyde, has been fully studied in \cite{ER2020,ER2021}, while the energetics of second system, leading to formamide, was presented in \citep{Rimola2018}.

Finally, it must be noted that the reported efficiencies are not the same as branching ratios. The latter take into account the rate at which radicals meet on the surface together with the efficiencies themselves, and provide a perspective of the relative importance of the different reaction channels that two reactants can follow (e.g. \cite{ER2021}). On the other hand, the efficiencies tell us what is the probability that two reactants will react in a given reaction site on the surface before one of the two reactants diffuses or ultimately desorbs.

\paragraph{W33 ice ASW model} Table \ref{tab:efficiencies} shows that out of the eleven reported systems, nine iCOMs forming reactions have efficiencies close to 1 by the end of the shortest radical residence on the ice surfaces, namely at temperatures $T_{des}$, while two (forming dimethyl peroxide and acetaldehyde) have efficiencies less than 0.1.
Similarly, out of the nine systems where H-abstraction reactions are possible, five have efficiencies close to 1 at $T_{des}$ (basically those with the form HCO + X $\to$ CO + HX, thanks to the low energy barriers involved). 
On the contrary, the efficiency of four H-abstraction reactions (CH$_3$ + CH$_3$O, CH$_3$ + CH$_2$OH, HCO + CH$_3$, CH$_3$O + CH$_3$O) fall below 1. Notice that if instead of using $E_{diff}/E_{des}=0.35$, a value of 0.3 is used, all efficiencies are lowered, on the contrary, an increment to 0.4 makes them to increase. In fact, increasing this ratio one allows the radicals to stay longer in a given binding site, increasing the reaction efficiency. For example, the efficiency of the CH$_3$ + CH$_3$ $\to$ C$_2$H$_6$ reaction would go down from 0.7 (using a ratio of 0.35) to $\sim$0.1 for a ratio of 0.3 and up to $\sim$1 for a ratio of 0.4. On the same vein, setting the $E_{diff}/E_{des}$ ratio equal to 0.5 makes almost all reactions to have an efficiency of 1, due to the longer time that radicals would remain together as a result of the lower diffusion. Changing the temperature at which $\varepsilon$ is computed has, in most cases, no effect, indicating that the reaction and diffusion processes are not competitive even at such low temperatures. The exceptions are those reactions that have a small efficiency at $T_{des}$ in Table \ref{tab:efficiencies}.


\begin{table*}[!htbp]
\centering
\caption{Efficiencies, $\epsilon$, of radical-radical reactions on the W33 ASW ice model. They are calculated using Equation \ref{eqn:efficiency}, setting the temperature to $T_{des}$ and 10 K, respectively, and considering that the diffusion barriers are equal to 0.35 times those of desorption. Note that the quoted values do not take into account quantum tunnelling, which could make efficiency larger at very low temperatures. }
\label{tab:efficiencies}
\resizebox{\textwidth}{!}{%
\hskip-2.5cm \begin{tabular}{|l|l|c|c|c|c|c|c|c|}
\hline
                    &            & \multicolumn{1}{l|}{}                                    &\multicolumn{3}{c|}{$T$=$T_{des}$} & \multicolumn{3}{c|}{$T$=10 K} \\ \hline
                    & Fastest    &$T_{des}$& Rc         & dHa case 1            & dHa case 2          & Rc         & dHa case 1   & dHa case 2  \\
System              & hopper     & [K]     & efficiency & efficiency            & efficiency          & efficiency & efficiency   & efficiency  \\ \hline
CH$_3$ + CH$_3$     & CH$_3$     & 29      & 0.7        & --                    & --                  & 1.0       &   --      & --              \\ \hline
CH$_3$ + NH$_2$     & CH$_3$     & 29      & 1.0*       & --                    & --                  & 1.0*      &   --      & --              \\ \hline
CH$_3$ + CH$_3$O    & CH$_3$     & 29      & 1.0        & 0.0                   & --                  & 1.0       &  0.0      & --              \\ \hline
CH$_3$ + CH$_2$OH   & CH$_3$     & 29      & 1.0        & 0.0                   & --                  & 1.0       &  0.0      & --              \\ \hline
HCO + HCO           & HCO        & 60      & 1.0        & 1.0                   & --                  & 1.0       &  1.0      & --              \\ \hline
HCO + CH$_3$O       & HCO        & 60      & 1.0        & 1.0                   & 2.9$\times 10^{-3}$ & 1.0       &  1.0      & 0.0             \\ \hline
HCO + CH$_2$OH      & HCO        & 60      & 1.0        & 1.0*                  & **                  & 1.0       &  1.0*     & **              \\ \hline
CH$_3$O + CH$_3$O   & CH$_3$O    & 77      & 1.3$\times 10^{-5}$    & 0.9       & --                  & 0.0       &  1.0      & --              \\ \hline
CH$_2$OH + CH$_2$OH & CH$_2$OH   & 103     & 1.0        & **                    & **                  & 1.0       & **        & **              \\ \hline
HCO + CH$_3$        & CH$_3$     & 29      & 0.1       & 1.1$\times 10^{-4}$   & --                  & 2.6$\times 10^{-3}$& 0.0 & --           \\ \hline
HCO + NH$_2$        & HCO        & 60      & 1.0        & 1.0                   & --                  & 1.0       & 1.0             & --        \\ \hline
\end{tabular}%
}
\hspace{1cm}
\resizebox{0.6\textwidth}{!}{%
\begin{tabular}{l}
* Barrierless, therefore efficiency is 1 and no crossover temperature can be calculated.\\
** Multiple steps and high barriers, therefore very little efficiency. 
No crossover\\
temperature is listed, as only the very last step will actually benefit from tunneling.
\end{tabular}%
}
\end{table*}

\paragraph{W18 ASW ice model} The reaction efficiency calculations were also carried out for the systems on the W18 ice ASW model, where radical mobilities are higher due to the overall lower binding energies, and the simpler reaction mechanisms (usually single step reactions). The results are reported in Tab. \ref{tab:my-table}. Both at $T_{des}$ and 10 K, all iCOMs forming reactions have efficiencies close to 1, except the one from the CH$_3$O + CH$_3$O system, which due to its high barrier has an efficiency of the order of 10$^{-2}$.
On the contrary, five H-abstraction reactions have efficiencies close to 1, while four have efficiencies close to zero, both at $T_{des}$ and 10 K.

\begin{table*}[!htbp]
\centering
\caption{Efficiencies, $\epsilon$, of radical-radical reactions on the W18 ASW ice model. They are calculated using Equation \ref{eqn:efficiency}, setting the temperature to $T_{des}$ and 10 K, respectively, and considering that the diffusion barriers are equal to 0.35 times those of desorption. Note that the quoted values do not take into account quantum tunnelling, which could make efficiency larger at very low temperatures.}
\label{tab:my-table}
\resizebox{\textwidth}{!}{%
\hskip-2.5cm \begin{tabular}{|l|l|c|c|c|c|c|c|c|}
\hline
              &         &                    & \multicolumn{3}{c|}{$T=T_{des}$}     & \multicolumn{3}{c|}{$T=10$ K}            \\ \hline
              & Fastest & $T_{des}$          & Rc         & dHa case 1 & dHa case 2 & Rc         & dHa case 1 & dHa case 2     \\
System        & hopper  & [K]                & efficiency & efficiency & efficiency & efficiency & efficiency & efficiency     \\ \hline
CH$_3$ + CH$_3$     & CH$_3$     & 17        & 1.0*       & --         & --         & 1.0*       & --         & --    \\ \hline
CH$_3$ + NH$_2$     & CH$_3$     & 17        & 1.0*       & --         & --         & 1.0*       & --         & --    \\ \hline
CH$_3$ + CH$_3$O    & CH$_3$     & 17        & 1.0*       & 1.0        & --         & 1.0*       & 1.0        & --    \\ \hline
CH$_3$ + CH$_2$OH   & CH$_3$     & 17        & 1.0        & 0.0        & --         & 1.0        & 0.0        & --    \\ \hline
HCO + HCO           & HCO        & 42        & 1.0        & 1.0        & --         & 1.0        & 1.0        & --    \\ \hline
HCO + CH$_3$O       & HCO        & 42        & 1.0        & 1.0        & 9.6$\times 10^{-4}$  & 1.0     & 1.0 & 0.0   \\ \hline
HCO + CH$_2$OH      & HCO        & 42        & 1.0        & 1.0*       & 0.0        & 1.0        & 1.0*       & 0.0   \\ \hline
CH$_3$O + CH$_3$O   & CH$_3$O    & 53        & 0.03       & 0.0        & --         & 0.0        & 0.0        & --    \\ \hline
CH$_2$OH + CH$_2$OH & CH$_2$OH   & 92        & 1.0        & 0.0        & **         & 1.0        & 0.0        & **    \\ \hline
HCO + CH$_3$        & CH$_3$     & 17        & 1.0        & 0.0        & --         & 1.0        & 0.0        & --    \\ \hline
HCO + NH$_2$        & HCO        & 42        & 1.0        & 1.0        & --         & 1.0        & 1.0        & --    \\ \hline
\end{tabular}%
}
\hspace{1cm}
\resizebox{0.6\textwidth}{!}{%
\begin{tabular}{l}
* Barrierless, therefore efficiency is 1 and no crossover temperature can be calculated\\
** Multiple steps and high barriers, therefore very little efficiency. 
No crossover\\
temperature is given, as only the very last step will actually benefit from tunneling.
\end{tabular}%
}
\end{table*}


\section{Discussion} \label{sec:disc}

\subsection{iCOMs formation versus H-abstraction}\label{sub:discussion-competition}

In this work, two radical-radical surface reactions-types have been investigated: radical coupling (Rc) and direct hydrogen abstraction (dHa). 
The former leads to the formation of iCOMs, while the latter does not lead to a increase in chemical complexity as the products are as simple as the reactants. 
Using the binding energies (Tab. \ref{tab:BEs}), activation energy barriers (Tab. \ref{tab:sum-best-Eact}) and the reaction efficiencies $\varepsilon$ (Tabs. \ref{tab:efficiencies}), here we discuss which radical-radical reaction will likely take place and if there will be a competition between the Rc and dHa channels. 

Depending on the values of the efficiencies and the crossover temperature for the Rc and dHa reactions (Tabs. \ref{tab:efficiencies} and \ref{tab:my-table}), we can define four categories:\\
(1) Rc plausible and dHa not plausible/possible: the reaction will lead to the iCOM with no competition channel.\\
(2) Rc--dHa competition: both reactions are possible and are in competition.\\
(3) Rc--dHa competition at low temperatures because of the tunneling taking over in the dHa reactions.\\
(4) Rc not plausible and dHa only plausible at low temperatures thanks to tunneling: the reaction will not form the iCOM and, except at low temperatures, not even the competing channel will occur.\\

Based on these reaction categories, here we briefly discuss, in a qualitative manner, which iCOMs are likely to be formed, and in which cases the Rc processes may be in direct competition with the dHa ones.

In three cases (CH$_3$ + CH$_3$, CH$_3$ + NH$_2$ and CH$_2$OH + CH$_2$OH), the only possible product is the iCOM (namely ethane, methylamine, and ethylene glycol, respectively). 
We have categorized these reactions as category 1 on both W33 and W18 ASW ice model surfaces given their high efficiencies even at low temperatures ($\varepsilon >$0.7). 
In four cases, (HCO + HCO, HCO + CH$_3$O, HCO + CH$_2$OH and HCO + NH$_2$), the formation of the iCOM (glyoxal, methyl formate, glycolaldehyde and formamide) and the dHa products are likely competing processes on both the W33 and W18 ice models.
In practice, all reactions involving HCO \citep[except HCO + CH$_3$; see][]{ER2021} have Rc and dHa (of the HCO + X $\to$ CO + HX type) as competitive channels. 
Therefore, while iCOMs can be formed, a significant part of the reactants could be lost \textit{via} these dHa reactions.
In other two cases (CH$_3$ + CH$_3$O and CH$_3$ +CH$_2$OH), iCOM formation (dimethyl ether and ethanol) are the dominant process except at low temperature, where H-abstraction can take over and become competitive thanks to quantum tunneling.

Finally, in two systems (HCO + CH$_3$ and CH$_2$OH + CH$_2$OH), iCOM formation (acetaldehyde and dimethyl peroxide) are unlikely to be formed. This is in agreement with the recent experiments by \citet{2021_Gutierrez_Quintanilla_iCOM}, who did not observe the formation neither acetaldehyde nor dimethyl ether despite of the presence of the radical reactants.

In summary, assuming that radicals have already diffused and have encountered in a specific place similar to those represented by our ASW model ices, most iCOMs in the systems studied in this work are likely to be formed on the icy surfaces.
However, while ethane, methylamine and ethylene glycol are the only possible products, glyoxal, methyl formate, glycolaldehyde, formamide, dimethyl ether and ethanol are likely in competition with the respective H-abstraction products.
On the other end, acetaldehyde and dimethyl peroxide do not seem a likely grain surface products.

Finally, we caution that this just represents one part of the Langmuir-Hinshelwood reactivity.
As discussed also in \cite{ER2021}, the binding and diffusion energies are crucial parameters.
In this study, we assumed (computed) a single value for the binding energy but it is now clear that it depends on the site where the species lands
\citep[e.g.][]{Ferrero_2020_BE, Bovolenta2020}.
Also, as already mentioned, the diffusion energy is poorly known.
Since both parameters enter in an exponential way in the computation of the efficiencies, more theoretical studies are necessary to firmly draw a conclusion, which depends on the fraction of sites with low or high binding and diffusion energies.
Nonetheless, this study shows that these computations are absolutely necessary in order to have quantitative and reliable astrochemical models.

\subsection{Where do barriers come from?}

There are two factors affecting the energy barriers of the reactions. The first one is related to the adsorption of the radicals on the surface, i.e., the way how they adsorb and the strength of this adsorption. All the studied radicals in this work interact with the water molecules exposed on the ice surfaces \textit{via} H-bond and dispersion interactions, in which CH$_3$ and CH$_2$OH present the weakest and the strongest binding, respectively. Because of these interactions, radical-radical reactions on water ice surfaces exhibit energy barriers, as the reactions require the breaking of these radical/surface interactions. Remarkably, since the radical/surface interactions dictate the geometries of the adsorbed radicals, these interactions have also repercussions on the structural reorganization of the reactants necessary for the occurrence of the reactions. Take for example the dHa channels in which CH$_2$OH transfers its H atom. CH$_2$OH interacts with the surface mainly through two strong H-bonds involving the --OH group. Accordingly, dHa reactions require a large re-orientation, including the breaking of the CH$_2$-OH/surface H-bonds, to proceed with the reaction, which is accompanied by a high energy barrier. The second factor is related to the intrinsic feasibility of the reactions, that is, how stable against reaction are the biradical systems.
To assess this point, we have investigated the reaction in absence of the water cluster, in this way to know the intrinsic energy cost (i.e., without the presence of external agents like the water clusters) of the reactions (see Sec. Methods). 
Results are shown in Table \ref{tab:sum-best-Eact} (``noW'' rows), while the structures of the optimized geometries are available in \cite{zenodo_data}.
We have detected that dHa channels involving either CH$_3$O or CH$_2$OH in which they transfer the H atom, irrespective of the other radical, all present energy barriers. 
This means that the H transfer from these two radicals is intrinsically associated with an energy cost. 
In contrast, this is not the case for dHa channels in which HCO transfers its H atom, since all these processes are barrierless in absence of the water clusters. 
Thus, HCO is a better H atom donor than CH$_3$O and CH$_2$OH and, accordingly, dHa channels involving HCO are more favourable than those involving CH$_3$O and CH$_2$OH. 
This is indeed reflected in the energy barriers of the dHa processes on W33/W18, which are lower for cases with HCO than for those with CH$_3$O or CH$_2$OH. 

For Rc channels in the absence of water molecules, those in which either HCO or CH$_3$ participate, irrespective of the other radical, are barrierless, indicating that couplings involving these two radicals are largely favorable. 
This is reflected in the energy barriers on W33/W18, which are in most of the cases very low. 
In contrast, the Rc channels of CH$_3$O + CH$_3$O and CH$_2$OH + CH$_2$OH in the absence of water molecules do have energy barriers, showing that these two couplings are intrinsically less favourable than those involving HCO and CH$_3$. 
The reason why the CH$_3$O + CH$_3$O and CH$_2$OH + CH$_2$OH Rc reactions are not barrierless in the absence of an ice surface is the high stability of their biradical van der Waals complexes, i.e., CH$_3$O$\cdots$CH$_3$O and CH$_2$OH$\cdots$CH$_2$OH.
On the W33/W18 ASW ice models, these van der Waals cannot be formed as a consequence of the interaction with the surface. 
Indeed, the CH$_2$OH + CH$_2$OH Rc reactions on W33/W18 have similar energy barriers to those of HCO + X and CH$_3$ + X, because the coupling does not require neither a strong structural reorganization nor the breaking of the CH$_2$OH/surface interactions, this way rendering the C--C bond formation energetically easy. 
And for the case of CH$_3$O + CH$_3$O Rc channel on W33/W18, energy barriers are much higher than in the absence of water molecules, due to the energetic cost of breaking the CH$_3$O/surface interactions and re-orientate the radicals to reach the coupling. 

\subsection{\texorpdfstring{CH$_3$O and CH$_2$OH: who stays and who goes?}{CH3O and CH2OH: who stays and who goes?}}
CH$_3$O and CH$_2$OH radicals are chemical isomers but exhibit different adsorption features and different radical--radical reactivity on ASW surfaces. 
Both radicals present high binding energies (CH$_2$OH larger than CH$_3$O) due to their capability to establish strong H-bonds with the surfaces.
Nevertheless, the way how they are established (i.e., atoms involved and number of H-bonds formed) is rather different, and this yields differences in their reactivity.
CH$_2$OH interacts with the surface through two strong H-bonds involving only the --OH group, this way leaving its C atom (namely, the radical center) unprotected and available to react. 
This has important consequences on the reactivity of the CH$_2$OH radical. Indeed, most of the CH$_2$OH + X reactions (X = CH$_2$OH, CH$_3$ and HCO) are Rc plausible as they present smaller energy barriers and often have less reaction steps than dHa.
The unique exception is the CH$_2$OH + HCO reaction, which belongs to the Rc--dHa competition category due to the intrinsic ease of HCO to transfer its H atom (see above).

In contrast, the unpaired electron in CH$_3$O is on the O atom, which in turn is the atom through which the radical establishes H-bonds with the surface. 
Because of that, the O atom is blocked toward chemical reactivity and this is shown by the trends in the CH$_3$O + X reactions (X = CH$_3$O, CH$_3$ and HCO).
CH$_3$O + CH$_3$O presents very high energy barriers, irrespective of the reaction channel and the ASW model. 
This is because CH$_3$O radicals have to reorganize structurally (namely, to break the interactions with the surface) in order to be ready to react. 
Reactions with CH$_3$ and HCO show in general Rc-dHa competition, except for the CH$_3$ + CH$_3$O case on W33 (Rc plausible) due to the high dHa barrier (9.5 kJ mol$^{-1}$) caused by the structural reorientation of CH$_3$O. Nevertheless, at very low temperatures ($<$ 40 K) they could be in competition due to the increased tunneling probability.

Interestingly, these trends gain relevance if we extrapolate them in the plausible scenario of hydrogenation of CH$_2$OH and CH$_3$O, both cases leading to the formation of methanol (CH$_3$OH). 
According to our results, an incoming H atom will react easier with CH$_2$OH than with CH$_3$O, since the C atom of the former is available while the O atom of the later is blocked. 
Remarkably, the propensity of the radicals to react is dictated by the geometrical constraints imposed by their interaction with the surface. 
Thus, in the presence of H atoms, CH$_2$OH would consume better than CH$_3$O, which could partly explain why CH$_3$O is detected and CH$_2$OH is not \citep[e.g.][]{cernicharo2012}.

\subsection{Influence of the water ice surface model}

The W33 ASW ice surface model presents a $\sim$6 \r{A} wide cavity where both radicals can be adsorbed. 
In contrast, the W18 model does not exhibit a cavity, resembling instead a rather flat surface.
As shown in \S ~\ref{sec:BE}, the binding energies on W33 (i.e., adsorption on the cavity) are larger (by a 12--76\%) than on W18, due to the larger number of radical/surface interactions formed on the former cluster model.

These effects are particularly important in those reactions in which the CH$_3$ radical participates. 
Indeed, in different reactions on W33, CH$_3$ is engaged by two weakly H-bonds, this way either hampering its motion towards the other radical (hence disfavouring Rc channels) or inhibiting its capability to receive an H atom from the other radical (hence disfavouring dHa channels: see Fig. \ref{fig:ch3_X}).
Indeed, the average energy barriers for Rc reactions involving CH$_3$ + X $\to$ X--CH$_3$ on W33 is 3.1 kJ mol$^{-1}$.
In contrast, on W18, CH$_3$ adsorbs essentially through dispersion forces and accordingly it is relatively free to translate/rotate to favour the Rc and dHa channels, so that CH$_3$ + X Rc reactions have a barrier of 0.7 kJ mol$^{-1}$ on average. 

For the other radicals, we did not find so clear effects on the energy barriers due to by the water ice surface morphology. 
These are indeed more complex cases than the CH$_3$ ones, since the reacting radicals can adsorb in different ways, establish different radical/surface interactions with different efficiencies, and require different structural reorganizations to react. 
Yet, we find that HCO + X $\to$ X--CHO Rc reactions have average barriers of 3.4/3.3 kJ mol$^{-1}$ on W33/W18 and average dHa HCO + X $\to$ CO + HX reaction energy barriers of 2.9/3.2 kJ mol$^{-1}$ on W33/W18.

Finally, some words related to our cluster models deserve to be mentioned. The first aspect is that they are rather small, although they are capable to host two small radicals on the surface. However, the limited sizes infer that, in the initial states, the reacting radicals are in close proximity. Thus, the predicted energy barriers concern only the chemical reactions between the radicals and not other surface phenomena like diffusion. In real systems, the two radicals will be likely separated by longer distances and, thus, diffusion is necessary. The second aspect is that the clusters composition is purely water, while actual ice mantles will contain other species. Thus, the interaction of the radicals with the surface can be different, affecting the diffusion, the reaction energies and the survival of the radicals against hydrogenation reactions.
Therefore, the energy barriers reported in this work are constrained within these two aspects, assuming that ice composition plus extensive radical diffusion are actually needed for a more realistic modelling.

\subsection{Predictions for other radical-radical systems}

There are several other radical--radical systems proposed in the literature \cite[e.g.][]{Garrod2008b} that we did not study in the present work, for example: OH/NH + X (X = CH$_3$, HCO, OH, NH, NH$_2$, CH$_3$O, CH$_2$OH), NH$_2$ +  Y (Y = NH$_2$, CH$_3$O, CH$_2$OH) and CH$_3$O + CH$_2$OH. However, studying such systems is a highly time consuming task.
Therefore, here we propose to use the trends for the studied reactions and the classification in categories 1 to 4 discussed in \S ~\ref{sub:discussion-competition} as a predictive tool to estimate a likely category of these other radical-radical reactions.

\begin{figure*}[!htb]
    \centering
    \includegraphics[width=1.0\columnwidth]{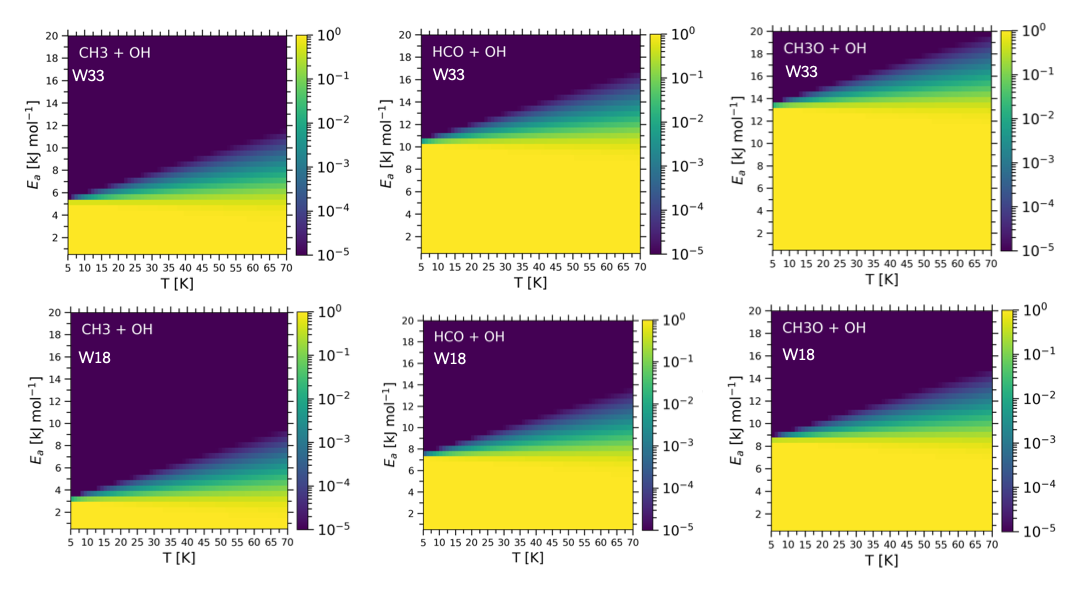}
    \caption{Reaction efficiencies on W33 (upper panels) and W18 (lower panels) as a function of the activation energy and temperature of a subset of the radical--radical systems in \cite{Garrod2008b} not explicitly studied in this work and for which we guess their efficiency (see text). These calculations do not include tunneling effects.}
    \label{fig:guesses_subset}
\end{figure*}

As for the reactions studied in this work, the category classification is based on the reaction efficiency (\S ~\ref{sec:react_efficiencies}), which depends on the reaction energy barrier, radical binding and diffusion energies.
In order to guess the reaction category, we apply the following set of considerations: 
\begin{itemize}
    \item If the reaction mechanism only involves translations/rotations without the need to break the radical/surface interactions (e.g., the Rc reactions on W18 of the type CH$_3$ + X with X = CH$_3$, NH$_2$, HCO, CH$_3$O, CH$_2$OH), then we estimate the energy barriers to be low (lower than about 4 kJ mol$^{-1}$).
    \item If the reaction involves the breaking of strong radical/surface interactions (e.g., Rc and dHa for CH$_3$O + CH$_3$O), or the translation of a radical somehow trapped by the ice (e.g., CH$_3$ in CH$_3$ + CH$_3$/CH$_3$O on W33), then we estimate the energy barriers to be high (higher than about 10 kJ mol$^{-1}$).
    \item For dHa channels only, if the reaction involves the cleavage of intrinsically stable chemical bonds (e.g., the CH$_2$O--H bond), then we consider the energy barriers to be high ($\geq 10$ kJ mol$^{-1}$). If the reaction involves the opposite situation (e.g., the H--CO bond), then we consider the energy barriers to be low ($\leq 4$ kJ mol$^{-1}$).
\end{itemize}
Given the fundamental role that the binding energies play in such surface reactions, we have calculated the binding energies of NH (in its triplet electronic ground state) and OH radicals on the water ice clusters. 
They are: 13.0 and 24.2 kJ mol$^{-1}$ on the W18 model, and 32.5 and 44.7 kJ mol$^{-1}$ on the cavity of the W33 model, for NH and OH respectively (see Figure \ref{fig:ANNEX_OH_NH_ads} in the annex). 
Thus, NH has a binding energy that lies between HCO and NH$_2$, while OH is almost the same as NH$_2$.
With this information, we obtain the $E_a$-- and $T$-- dependent efficiencies (of either Rc and dHa reactions) for each system with OH + X, NH + X (X = CH$_3$, HCO, OH, NH$_2$, CH$_3$O, CH$_2$OH), NH$_2$ + Y (Y = NH$_2$, CH$_3$O, CH$_2$OH) and CH$_3$O + CH$_2$OH. Figure \ref{fig:guesses_subset} contains a subset of them: OH + CH$_3$, HCO, CH$_3$O on both W33 and W18. 
The figures relative to the other systems are available in the \S \ref{sec:appendix:extra_efficiencies} in the annex.

It can be rapidly noticed that there is a limit under which the efficiencies take values of unity; this is the point at which the reaction energy barriers coincide with the value of the diffusion barrier of the fastest hopper from each couple (i.e. $E_{diff}=$0.35$\times E_{des}$) and it is surface and radical dependent.
Of course, the higher the binding energy (CH$_3$ $<$ HCO $<$ CH$_3$O $<$ OH $<$ CH$_2$OH) the higher this limit, meaning that for activation energy values below this threshold, radicals have enough time to react before they separate due to thermal hopping. On the contrary, above this energetic limit, the dependence on temperature becomes more and more important, so that at higher temperatures, higher efficiencies are obtained. Eventually, for sufficiently high activation energies, the reactivity between two radicals become inefficient.

While these plots are very informative, they lack three key points in the Langmuir-Hinshelwood-like surface reactions: (i) the temperature limit after which radicals will certainly not be available on the surface anymore (e.g. a theoretical limit can be set at the "desorption temperature", see \S ~\ref{sec:react_efficiencies} for more details, while it could also be the point at which some of the two radicals has been consumed), (ii) the effects of quantum tunneling, that may be important for direct H-abstraction reactions, so that the rate constants related to the activation energy barriers become less dependent on temperature and, finally, (iii) the meeting rates, which will modulate the efficiency according to the meeting probability of radicals on the grain surfaces. 
Such effects are only attainable by more detailed, dedicated modelling.

In summary, in addition to the three considerations from above, one must also consider that the higher the binding energies of the radical couple, the higher the range of activation energy barriers that the reaction can have in order to present high efficiencies, at the expense of a lower meeting rate.

With this information, and bearing in mind the three above considerations, we propose the guessed reactivity properties for each one of these systems in the following paragraphs.

\subsubsection{Radical coupling reactions}

\paragraph{OH + X: (X = CH$_3$, HCO, OH, NH, NH$_2$, CH$_3$O, CH$_2$OH)} OH has a high binding energy on W33 (44.7 kJ mol$^{-1}$) and an intermediate one on W18 (24.2 kJ mol$^{-1}$).
Therefore, its diffusion barrier is rather high. 
For reactions with a radical X that have small diffusion barriers, like CH$_3$ or NH (the latter only on W18) reactions will take place only if they have small activation energy barriers. 
For such low binding energy radicals, this might hold as long as they do not experience trapping. 
On the other hand, if the radical X has also a high diffusion barrier, the reaction efficiencies will be unity even for relatively high activation energy barriers. 
There might be two cases where reactions could have very high reaction energies, and as a consequence low efficiencies: X=OH/CH$_3$O, since they both have their radical atom (oxygen) establishing H-bonding.

\paragraph{NH + X: (X = CH$_3$, HCO, OH, NH, NH$_2$, CH$_3$O, CH$_2$OH)} NH has an intermediate to low binding energy depending on the surface environment (13.0 kJ mol$^{-1}$ on W18 and 32.5 kJ mol$^{-1}$ on W33).
Therefore, one would expect low energy activation energy barriers on flat surfaces where NH can easily reorient, meaning high efficiencies.
On the contrary, the high binding energy on W33 renders NH + X reactions efficient as long as the barrier is not very high, a scenario which is likely not given by any of the possible X radicals based on the experience with the systems we have studied. 

\paragraph{NH$_2$ + X: (X = NH$_2$, CH$_3$O, CH$_2$OH)} The NH$_2$ radical sports high binding energy regardless of the surface environment. Therefore, reactions with other radicals need to have very high energy barriers in order to have low efficiencies.
Hence, we expect high efficiencies for any radical X.

\paragraph{CH$_3$O + CH$_2$OH} Given the special binding structure of CH$_2$OH with its C atom free from surface interaction, we expect these reactions to have high efficiencies, also given their high binding energies. 

\subsubsection{Direct H-abstraction reactions}

From the list of reactions, only HCO, CH$_3$O and CH$_2$OH can donate an H atom. 
From our experience with the systems in Tab. \ref{tab:sum-best-Eact}, we know that HCO is a very good H-donor, while CH$_3$O and CH$_2$OH are not. 
Reactions where HCO is the H-donor will most likely sport competition between Rc and dHa channels, while tunneling effects will be much more important for reactions where CH$_3$O is the H-donor. 
Regarding CH$_2$OH, its high stability makes energy barriers to be high and mechanisms to be more complex, likely with many energetic reorientation steps. 
Therefore, overall it is expected to be a non efficient process.

\section{Conclusions} \label{sec:concl}

In this work, we have carried out DFT computations of the reactions on icy surfaces between nine radical--radical systems, postulated to lead to the formation of iCOMs by several astrochemical models based on the \citet{Garrod2006} scheme.
The set of studied systems are HCO + X and CH$_3$ + X, where X is equal to CH$_3$, HCO, NH$_2$, CH$_2$OH and CH$_3$O, plus the systems CH$_3$O + CH$_3$O and CH$_2$OH + CH$_2$OH.
We considered both the combination between radicals, leading to the iCOM, and the H-abstraction from one of them, leading to simpler molecules.

 In order to simulate the interstellar icy surfaces, we employed two ice cluster models made of 18 and 33 water molecules (W18 and W33), respectively, which we tested in previous works \citep{rimola2014, ER2019, ER2021}. 
 The W33 ice model presents a cavity structure, which likely makes it a better representation of interstellar ices than the W18 ice model, which only possesses a rather flat surface because of its limited size.
 Therefore, in the following, we will only report the conclusions based on the results obtained with the W33 model.

We computed the binding energy of the involved radicals and, for all the possible reactions between the nine radical--radical systems, the reaction energy barriers.
We also computed the diffusion energy of each radical, assuming that it is 0.35 times the binding energy.
Then, using the definition of reaction efficiency that takes into account the reaction activation energy barrier as well as the radicals diffusion and desorption timescales \citep{ER2021}, we provided a rough estimate of the reaction efficiency of each reaction using the Eyring equation approximation.
The computed reaction efficiencies allows to predict which reactions will lead to iCOMs or to a competition with the H-abstraction channels, or to nothing.

The main conclusions of this work are the following.\\

\noindent
(1) Radical--radical reactions on icy surfaces are not straightforward nor barrierless in most of the studied systems. 
    Very often, we find that two channels, radical coupling and H-abstraction, are in competition. 
    In a few cases, we find that no reaction can occur between the two radicals.
    Specifically:
    
    (i) Ethane (C$_2$H$_6$), methylamine (CH$_3$NH$_2$) and ethylene glycol (CH$_2$OHCH$_2$OH) are the only products of their respective radical--radical reactions.
    
    (ii) The formation of glyoxal (HCOCHO), formamide (NH$_2$CHO), methyl formate (CH$_3$OCHO) and glycolaldehyde (CH$_2$OHCHO) is in competition with the H-abstraction products (CO + H$_2$CO, NH$_3$ + CO, CH$_3$OH + CO and CH$_3$OH + CO, respectively).
    Very likely, the branching ratio is 1:1, thanks to the capacity of HCO to become an H-donor in the H-abstraction reactions.
    
    (iii) Acetaldehyde (CH$_3$CHO) and dimethyl peroxide (CH$_3$OOCH$_3$) are unlikely to be formed.\\
    
\noindent
(2) The effect of the surface structure on the reaction output is best represented by the different binding energies on the two ice models. 
    On the cavity structure of the W33 model, the binding energies are $\sim$10--80\% higher than on the W18 model, due to the larger number/efficiency of intermolecular interactions. This effect is higher for weakly bound species like CH$_3$, evidencing its capacity to get trapped. 
    Nevertheless, the same trend on the binding energy of the different radicals is observed on both ice models: CH$_3$ $<$ HCO $<$ NH$_2$ $<$ CH$_3$O $<$ CH$_2$OH.\\

In addition, some radicals present features worth to emphasize.\\

\noindent
(3) CH$_3$ is usually a very reactive species due to its low binding energy (in many cases its reaction mechanisms comprise a low energy torsion), although there are some exceptions where the mobility of CH$_3$ is much restricted by the cavity in W33, so that the activation energy barriers can rise up to $\sim$7 kJ mol$^{-1}$.\\

\noindent
(4) CH$_2$OH presents an interesting binding pattern to the ice surface, which makes its C atom very reactive. 
The strong interaction of its OH group with the water molecules of the surface fixes its adsorption geometry, leaving the C atom unprotected and highly reactive. 
We predict that its reactivity with other radicals (with very low energy barriers) and specially with atomic hydrogen will be a major destruction route for this radical on the icy surfaces.\\

\noindent
(5) CH$_3$O has its radical electron on the O atom, which in turn establishes H-bonds with the surface water molecules. 
This makes this radical to be slightly less reactive than expected and, therefore, high energy barriers appear for the CH$_3$O + CH$_3$O reactions. 
On the other hand,  CH$_3$O can still perform direct H-abstraction reactions as a donor in other situations.
However, the likely high reaction energy barriers, due to its intrinsic H--C bond stability (and for the cavity, the higher number of intermolecular interactions), suggest that the H-abstraction reactions are efficient only when considering quantum tunneling effects.\\

\noindent
(6) Here we have studied in detail only a subset of radical--radical reactions present in \citet{Garrod2008b} scheme.
For the systems involving the same set of radicals investigated in this work (CH$_3$, HCO, NH$_2$, CH$_2$OH, CH$_3$O) and, additionally, those involving OH and NH, we have discussed the possible outcomes, based on what we learned from the in-depth studied systems.\\

As conclusive remarks, we emphasize again that the assumption of the radical--radical combination leading exclusively and always to iCOM is far to be correct and the need to carry out dedicated studies on each radical--radical system in order to assess the outcome of its possible reactions.
Also, the present study uses simplistic models of the ice structure as well as a very limited number of binding and reaction sites.
More realistic computations should include larger icy grains as well as molecular dynamics simulations involving the encountering plus the reaction of the two radicals, probably possible in a near future thanks to the fast increase of the high performance super-computing facilities.
In conclusion, our study probably just scratched the surface of the surface-chemistry on the icy interstellar grains. 


\section*{Acknowledgments}
This project has received funding within the European Union’s Horizon 2020 research and innovation programme from the European Research Council (ERC) for the projects ``The Dawn of Organic Chemistry” (DOC), grant agreement No 741002 and ``Quantum Chemistry on Interstellar Grains” (QUANTUMGRAIN), grant agreement No 865657, and from the Marie Sklodowska-Curie for the project ``Astro-Chemical Origins” (ACO), grant agreement No 811312.  
AR is indebted to ``Ram{\'o}n y Cajal" program. 
MINECO (project CTQ2017-89132-P) and DIUE (project 2017SGR1323) are acknowledged. 
Finally, we thank Prof. Gretobape for fruitful and stimulating discussions.

Most of the quantum chemistry calculations presented in this paper were performed using the GRICAD infrastructure (https://gricad.univ-grenoble-alpes.fr), which is partly supported by the Equip@Meso project (reference ANR-10-EQPX-29-01) of the programme Investissements d'Avenir supervised by the Agence Nationale pour la Recherche. Additionally this work was granted access to the HPC resources of IDRIS under the allocation 2019-A0060810797 attributed by GENCI (Grand Equipement National de Calcul Intensif). 

We thank Ms. Berta Martínez-Bachs for sharing with us the binding energy values of NH on amorphous water ices.

\bibliography{aa_my_biblio}{}

\begin{thebibliography}{}
\expandafter\ifx\csname natexlab\endcsname\relax\def\natexlab#1{#1}\fi
\providecommand{\url}[1]{\href{#1}{#1}}
\providecommand{\dodoi}[1]{doi:~\href{http://doi.org/#1}{\nolinkurl{#1}}}
\providecommand{\doeprint}[1]{\href{http://ascl.net/#1}{\nolinkurl{http://ascl.net/#1}}}
\providecommand{\doarXiv}[1]{\href{https://arxiv.org/abs/#1}{\nolinkurl{https://arxiv.org/abs/#1}}}

\bibitem[{{Aikawa} {et~al.}(2020){Aikawa}, {Furuya}, {Yamamoto}, \&
  {Sakai}}]{Aikawa2020}
{Aikawa}, Y., {Furuya}, K., {Yamamoto}, S., \& {Sakai}, N. 2020, \apj, 897,
  110, \dodoi{10.3847/1538-4357/ab994a}

\bibitem[{{Balucani} {et~al.}(2015){Balucani}, {Ceccarelli}, \&
  {Taquet}}]{balucani2015}
{Balucani}, N., {Ceccarelli}, C., \& {Taquet}, V. 2015, \mnras, 449, L16,
  \dodoi{10.1093/mnrasl/slv009}

\bibitem[{Becke(1993)}]{becke1993}
Becke, A.~D. 1993, J. Chem. Phys., 98, 1372

\bibitem[{{Bovolenta} {et~al.}(2020){Bovolenta}, {Bovino},
  {V{\"o}hringer-Martinez}, {Saez}, {Grassi}, \& {Vogt-Geisse}}]{Bovolenta2020}
{Bovolenta}, G., {Bovino}, S., {V{\"o}hringer-Martinez}, E., {et~al.} 2020,
  Molecular Astrophysics, 21, 100095, \dodoi{10.1016/j.molap.2020.100095}

\bibitem[{{Ceccarelli} {et~al.}(2017){Ceccarelli}, {Caselli}, {Fontani},
  {Neri}, {L{\'o}pez-Sepulcre}, {Codella}, {Feng}, {Jim{\'e}nez-Serra},
  {Lefloch}, {Pineda}, {Vastel}, {Alves}, {Bachiller}, {Balucani}, {Bianchi},
  {Bizzocchi}, {Bottinelli}, {Caux}, {Chac{\'o}n-Tanarro}, {Choudhury},
  {Coutens}, {Dulieu}, {Favre}, {Hily-Blant}, {Holdship}, {Kahane}, {Jaber
  Al-Edhari}, {Laas}, {Ospina}, {Oya}, {Podio}, {Pon}, {Punanova}, {Quenard},
  {Rimola}, {Sakai}, {Sims}, {Spezzano}, {Taquet}, {Testi}, {Theul{\'e}},
  {Ugliengo}, {Vasyunin}, {Viti}, {Wiesenfeld}, \& {Yamamoto}}]{ceccarelli2017}
{Ceccarelli}, C., {Caselli}, P., {Fontani}, F., {et~al.} 2017, \apj, 850, 176,
  \dodoi{10.3847/1538-4357/aa961d}

\bibitem[{{Cernicharo} {et~al.}(2012){Cernicharo}, {Marcelino}, {Roueff},
  {Gerin}, {Jim{\'e}nez-Escobar}, \& {Mu{\~n}oz Caro}}]{cernicharo2012}
{Cernicharo}, J., {Marcelino}, N., {Roueff}, E., {et~al.} 2012, \apjl, 759,
  L43, \dodoi{10.1088/2041-8205/759/2/L43}

\bibitem[{{Charnley} {et~al.}(1997){Charnley}, {Tielens}, \&
  {Rodgers}}]{Charnley1997}
{Charnley}, S.~B., {Tielens}, A.~G.~G.~M., \& {Rodgers}, S.~D. 1997, \apjl,
  482, L203, \dodoi{10.1086/310697}

\bibitem[{{Cheung} {et~al.}(1968){Cheung}, {Rank}, {Townes}, {Thornton}, \&
  {Welch}}]{Cheung1968}
{Cheung}, A.~C., {Rank}, D.~M., {Townes}, C.~H., {Thornton}, D.~D., \& {Welch},
  W.~J. 1968, \prl, 21, 1701, \dodoi{10.1103/PhysRevLett.21.1701}

\bibitem[{{Cheung} {et~al.}(1969){Cheung}, {Rank}, {Townes}, {Thornton}, \&
  {Welch}}]{Cheung1969}
---. 1969, \nat, 221, 626, \dodoi{10.1038/221626a0}

\bibitem[{De~Duve(2005)}]{deDuve2005Book}
De~Duve, C. 2005, Singularities: landmarks on the pathways of life (Cambridge
  University Press)

\bibitem[{{de Duve}(2011)}]{deDuve2011}
{de Duve}, C. 2011, Philosophical Transactions of the Royal Society of London
  Series A, 369, 620, \dodoi{10.1098/rsta.2010.0312}

\bibitem[{{Duflot} {et~al.}(2021){Duflot}, {Toubin}, \&
  {Monnerville}}]{Duflot2021}
{Duflot}, D., {Toubin}, C., \& {Monnerville}, M. 2021, Frontiers in Astronomy
  and Space Sciences, 8, 24, \dodoi{10.3389/fspas.2021.645243}

\bibitem[{{Enrique-Romero} {et~al.}(2021){Enrique-Romero}, {Ceccarelli},
  {Rimola}, {Skouteris}, {Balucani}, \& {Ugliengo}}]{ER2021}
{Enrique-Romero}, J., {Ceccarelli}, C., {Rimola}, A., {et~al.} 2021, A\&A
  submitted

\bibitem[{{Enrique-Romero} {et~al.}(2019){Enrique-Romero}, {Rimola},
  {Ceccarelli}, {Ugliengo}, {Balucani}, \& {Skouteris}}]{ER2019}
{Enrique-Romero}, J., {Rimola}, A., {Ceccarelli}, C., {et~al.} 2019, ACS Earth
  and Space Chemistry, 3, 2158, \dodoi{10.1021/acsearthspacechem.9b00156}

\bibitem[{Enrique-Romero {et~al.}(2021)Enrique-Romero, Rimola, Ceccarelli,
  Ugliengo, Balucani, \& Skouteris}]{zenodo_data}
Enrique-Romero, J., Rimola, A., Ceccarelli, C., {et~al.} 2021, {XYZ files \&
  Spin densities and Molecular Orbitals for "Quantum mechanical simulations of
  the radical-radical chemistry on icy surfaces"},  Zenodo,
  \dodoi{10.5281/zenodo.5723996}

\bibitem[{{Enrique-Romero} {et~al.}(2020){Enrique-Romero},
  {{\'A}lvarez-Barcia}, {Kolb}, {Rimola}, {Ceccarelli}, {Balucani}, {Meisner},
  {Ugliengo}, {Lamberts}, \& {K{\"a}stner}}]{ER2020}
{Enrique-Romero}, J., {{\'A}lvarez-Barcia}, S., {Kolb}, F.~J., {et~al.} 2020,
  \mnras, 493, 2523, \dodoi{10.1093/mnras/staa484}

\bibitem[{{Fermann} \& {Auerbach}(2000)}]{FermannAuerbach2000_Tc}
{Fermann}, J.~T., \& {Auerbach}, S. 2000, \jcp, 112, 6787,
  \dodoi{10.1063/1.481318}

\bibitem[{{Ferrero} {et~al.}(2020){Ferrero}, {Zamirri}, {Ceccarelli}, {Witzel},
  {Rimola}, \& {Ugliengo}}]{Ferrero_2020_BE}
{Ferrero}, S., {Zamirri}, L., {Ceccarelli}, C., {et~al.} 2020, \apj, 904, 11,
  \dodoi{10.3847/1538-4357/abb953}

\bibitem[{Frisch {et~al.}(2016)Frisch, Trucks, Schlegel, Scuseria, Robb,
  Cheeseman, Scalmani, Barone, Petersson, Nakatsuji, Li, Caricato, Marenich,
  Bloino, Janesko, Gomperts, Mennucci, Hratchian, Ortiz, Izmaylov, Sonnenberg,
  Williams-Young, Ding, Lipparini, Egidi, Goings, Peng, Petrone, Henderson,
  Ranasinghe, Zakrzewski, Gao, Rega, Zheng, Liang, Hada, Ehara, Toyota, Fukuda,
  Hasegawa, Ishida, Nakajima, Honda, Kitao, Nakai, Vreven, Throssell,
  Montgomery, Peralta, Ogliaro, Bearpark, Heyd, Brothers, Kudin, Staroverov,
  Keith, Kobayashi, Normand, Raghavachari, Rendell, Burant, Iyengar, Tomasi,
  Cossi, Millam, Klene, Adamo, Cammi, Ochterski, Martin, Morokuma, Farkas,
  Foresman, \& Fox}]{g16}
Frisch, M.~J., Trucks, G.~W., Schlegel, H.~B., {et~al.} 2016, Gaussian˜16
  {R}evision {C}.01

\bibitem[{{Garrod} \& {Herbst}(2006)}]{Garrod2006}
{Garrod}, R.~T., \& {Herbst}, E. 2006, \aap, 457, 927,
  \dodoi{10.1051/0004-6361:20065560}

\bibitem[{{Garrod} \& {Pauly}(2011)}]{GP2011}
{Garrod}, R.~T., \& {Pauly}, T. 2011, \apj, 735, 15,
  \dodoi{10.1088/0004-637X/735/1/15}

\bibitem[{{Garrod} {et~al.}(2008){Garrod}, {Widicus Weaver}, \&
  {Herbst}}]{Garrod2008b}
{Garrod}, R.~T., {Widicus Weaver}, S.~L., \& {Herbst}, E. 2008, \apj, 682, 283,
  \dodoi{10.1086/588035}

\bibitem[{Grimme {et~al.}(2010)Grimme, Antony, Ehrlich, \&
  Krieg}]{D3-grimme2010}
Grimme, S., Antony, J., Ehrlich, S., \& Krieg, H. 2010, J. Chem. Phys., 132,
  154104

\bibitem[{Grimme {et~al.}(2011)Grimme, Ehrlich, \& Goerigk}]{d3bj_grimme}
Grimme, S., Ehrlich, S., \& Goerigk, L. 2011, Journal of Computational
  Chemistry, 32, 1456, \dodoi{https://doi.org/10.1002/jcc.21759}

\bibitem[{{Guti{\'e}rrez-Quintanilla}
  {et~al.}(2021){Guti{\'e}rrez-Quintanilla}, {Layssac}, {Butscher}, {Henkel},
  {Tsegaw}, {Grote}, {Sander}, {Borget}, {Chiavassa}, \&
  {Duvernay}}]{2021_Gutierrez_Quintanilla_iCOM}
{Guti{\'e}rrez-Quintanilla}, A., {Layssac}, Y., {Butscher}, T., {et~al.} 2021,
  \mnras, 506, 3734, \dodoi{10.1093/mnras/stab1850}

\bibitem[{Hariharan \& Pople(1973)}]{hariharan_influence_1973}
Hariharan, P.~C., \& Pople, J.~A. 1973, Theoret. Chim. Acta, 28, 213,
  \dodoi{10.1007/BF00533485}

\bibitem[{{Hasegawa} \& {Herbst}(1993)}]{Hasegawa1993}
{Hasegawa}, T.~I., \& {Herbst}, E. 1993, \mnras, 263, 589,
  \dodoi{10.1093/mnras/263.3.589}

\bibitem[{{Hasegawa} {et~al.}(1992){Hasegawa}, {Herbst}, \& {Leung}}]{HHL1992}
{Hasegawa}, T.~I., {Herbst}, E., \& {Leung}, C.~M. 1992, \apjs, 82, 167,
  \dodoi{10.1086/191713}

\bibitem[{{He} {et~al.}(2018){He}, {Emtiaz}, \& {Vidali}}]{He_Vidali_2018}
{He}, J., {Emtiaz}, S., \& {Vidali}, G. 2018, \apj, 863, 156,
  \dodoi{10.3847/1538-4357/aad227}

\bibitem[{Hehre {et~al.}(1972)Hehre, Ditchfield, \&
  Pople}]{hehre_selfconsistent_1972}
Hehre, W.~J., Ditchfield, R., \& Pople, J.~A. 1972, J. Chem. Phys., 56, 2257,
  \dodoi{10.1063/1.1677527}

\bibitem[{{Herbst} \& {van Dishoeck}(2009)}]{Herbst2009}
{Herbst}, E., \& {van Dishoeck}, E.~F. 2009, \araa, 47, 427,
  \dodoi{10.1146/annurev-astro-082708-101654}

\bibitem[{{Jensen} {et~al.}(2021){Jensen}, {J{\o}rgensen}, {Furuya},
  {Haugb{\o}lle}, \& {Aikawa}}]{Jensen2021}
{Jensen}, S.~S., {J{\o}rgensen}, J.~K., {Furuya}, K., {Haugb{\o}lle}, T., \&
  {Aikawa}, Y. 2021, \aap, 649, A66, \dodoi{10.1051/0004-6361/202040196}

\bibitem[{{Jin} \& {Garrod}(2020)}]{JinGarrod2020}
{Jin}, M., \& {Garrod}, R.~T. 2020, \apjs, 249, 26,
  \dodoi{10.3847/1538-4365/ab9ec8}

\bibitem[{{Kalv{\={a}}ns}(2018)}]{kalvans2018}
{Kalv{\={a}}ns}, J. 2018, \mnras, 478, 2753, \dodoi{10.1093/mnras/sty1172}

\bibitem[{{Karssemeijer} \& {Cuppen}(2014)}]{Karssemeijer_2014}
{Karssemeijer}, L.~J., \& {Cuppen}, H.~M. 2014, \aap, 569, A107,
  \dodoi{10.1051/0004-6361/201424792}

\bibitem[{Krishnan {et~al.}(1980)Krishnan, Binkley, Seeger, \&
  Pople}]{krishnan_selfconsistent_1980}
Krishnan, R., Binkley, J.~S., Seeger, R., \& Pople, J.~A. 1980, J. Chem. Phys.,
  72, 650, \dodoi{10.1063/1.438955}

\bibitem[{{Lamberts} {et~al.}(2019){Lamberts}, {Markmeyer}, {Kolb}, \&
  {K{\"a}stner}}]{Lamberts2019}
{Lamberts}, T., {Markmeyer}, M.~N., {Kolb}, F.~J., \& {K{\"a}stner}, J. 2019,
  ACS Earth and Space Chemistry, 3, 958,
  \dodoi{10.1021/acsearthspacechem.9b00029}

\bibitem[{Lee {et~al.}(1988)Lee, Yang, \& Parr}]{LYP88}
Lee, C., Yang, W., \& Parr, R.~G. 1988, Phys. Rev. B, 37, 785,
  \dodoi{10.1103/PhysRevB.37.785}

\bibitem[{Mart{\'\i}nez-Bachs {et~al.}(2020)Mart{\'\i}nez-Bachs, Ferrero, \&
  Rimola}]{martinez-bachs2020ICCSA}
Mart{\'\i}nez-Bachs, B., Ferrero, S., \& Rimola, A. 2020, in International
  Conference on Computational Science and Its Applications, Springer, 683--692

\bibitem[{McQuarrie(1976)}]{mcquarrie2000}
McQuarrie, D. 1976, Statistical Mechanics (New York: Harper and Row)

\bibitem[{{Minissale} {et~al.}(2016){Minissale}, {Congiu}, \&
  {Dulieu}}]{Minissale2016}
{Minissale}, M., {Congiu}, E., \& {Dulieu}, F. 2016, \aap, 585, A146,
  \dodoi{10.1051/0004-6361/201526702}

\bibitem[{{Neese}(2004)}]{Neese2004}
{Neese}, F. 2004, Journal of Physics and Chemistry of Solids, 65, 781,
  \dodoi{10.1016/j.jpcs.2003.11.015}

\bibitem[{{Penteado} {et~al.}(2017){Penteado}, {Walsh}, \&
  {Cuppen}}]{Penteado2017}
{Penteado}, E.~M., {Walsh}, C., \& {Cuppen}, H.~M. 2017, \apj, 844, 71,
  \dodoi{10.3847/1538-4357/aa78f9}

\bibitem[{Rimola {et~al.}(2021)Rimola, Ceccarelli, Balucani, \&
  Ugliengo}]{Rimola2021_ions}
Rimola, A., Ceccarelli, C., Balucani, N., \& Ugliengo, P. 2021, Frontiers in
  Astronomy and Space Sciences, 8, 38, \dodoi{10.3389/fspas.2021.655405}

\bibitem[{{Rimola} {et~al.}(2018){Rimola}, {Skouteris}, {Balucani},
  {Ceccarelli}, {Enrique-Romero}, {Taquet}, \& {Ugliengo}}]{Rimola2018}
{Rimola}, A., {Skouteris}, D., {Balucani}, N., {et~al.} 2018, ACS Earth and
  Space Chemistry, 2, 720, \dodoi{10.1021/acsearthspacechem.7b00156}

\bibitem[{{Rimola} {et~al.}(2014){Rimola}, {Taquet}, {Ugliengo}, {Balucani}, \&
  {Ceccarelli}}]{rimola2014}
{Rimola}, A., {Taquet}, V., {Ugliengo}, P., {Balucani}, N., \& {Ceccarelli}, C.
  2014, \aap, 572, A70, \dodoi{10.1051/0004-6361/201424046}

\bibitem[{{Ruaud} {et~al.}(2015){Ruaud}, {Loison}, {Hickson}, {Gratier},
  {Hersant}, \& {Wakelam}}]{Ruaud_ER_2015}
{Ruaud}, M., {Loison}, J.~C., {Hickson}, K.~M., {et~al.} 2015, \mnras, 447,
  4004, \dodoi{10.1093/mnras/stu2709}

\bibitem[{Ruaud {et~al.}(2016)Ruaud, Wakelam, \& Hersant}]{ruaud2016_nautilus}
Ruaud, M., Wakelam, V., \& Hersant, F. 2016, Monthly Notices of the Royal
  Astronomical Society, 459, 3756

\bibitem[{{Rubin} {et~al.}(1971){Rubin}, {Swenson}, {Benson}, {Tigelaar}, \&
  {Flygare}}]{Rubin1971}
{Rubin}, R.~H., {Swenson}, G.~W., J., {Benson}, R.~C., {Tigelaar}, H.~L., \&
  {Flygare}, W.~H. 1971, \apjl, 169, L39, \dodoi{10.1086/180810}

\bibitem[{Sameera {et~al.}(2017)Sameera, Senevirathne, Andersson, Maseras, \&
  Nyman}]{Sameera_2017_BE}
Sameera, W. M.~C., Senevirathne, B., Andersson, S., Maseras, F., \& Nyman, G.
  2017, The Journal of Physical Chemistry C, 121, 15223,
  \dodoi{10.1021/acs.jpcc.7b04105}

\bibitem[{{Senevirathne} {et~al.}(2017){Senevirathne}, {Andersson}, {Dulieu},
  \& {Nyman}}]{Senevirathne2017}
{Senevirathne}, B., {Andersson}, S., {Dulieu}, F., \& {Nyman}, G. 2017,
  Molecular Astrophysics, 6, 59, \dodoi{10.1016/j.molap.2017.01.005}

\bibitem[{{Simons} {et~al.}(2020){Simons}, {Lamberts}, \&
  {Cuppen}}]{Simons2020}
{Simons}, M.~A.~J., {Lamberts}, T., \& {Cuppen}, H.~M. 2020, \aap, 634, A52,
  \dodoi{10.1051/0004-6361/201936522}

\bibitem[{{Skouteris} {et~al.}(2019){Skouteris}, {Balucani}, {Ceccarelli},
  {Faginas Lago}, {Codella}, {Falcinelli}, \& {Rosi}}]{Skouteris2019}
{Skouteris}, D., {Balucani}, N., {Ceccarelli}, C., {et~al.} 2019, \mnras, 482,
  3567, \dodoi{10.1093/mnras/sty2903}

\bibitem[{{Skouteris} {et~al.}(2018){Skouteris}, {Balucani}, {Ceccarelli},
  {Vazart}, {Puzzarini}, {Barone}, {Codella}, \& {Lefloch}}]{Skouteris2018}
---. 2018, \apj, 854, 135, \dodoi{10.3847/1538-4357/aaa41e}

\bibitem[{{Snyder}(2006)}]{Snyder2006}
{Snyder}, L.~E. 2006, Proceedings of the National Academy of Science, 103,
  12243, \dodoi{10.1073/pnas.0601750103}

\bibitem[{{Snyder} {et~al.}(1969){Snyder}, {Buhl}, {Zuckerman}, \&
  {Palmer}}]{Snyder1969}
{Snyder}, L.~E., {Buhl}, D., {Zuckerman}, B., \& {Palmer}, P. 1969, \prl, 22,
  679, \dodoi{10.1103/PhysRevLett.22.679}

\bibitem[{{Taquet} {et~al.}(2016){Taquet}, {Wirstr{\"o}m}, \&
  {Charnley}}]{Taquet2016}
{Taquet}, V., {Wirstr{\"o}m}, E.~S., \& {Charnley}, S.~B. 2016, \apj, 821, 46,
  \dodoi{10.3847/0004-637X/821/1/46}

\bibitem[{{Tielens} \& {Hagen}(1982)}]{Tielens_Hagen_1982}
{Tielens}, A.~G.~G.~M., \& {Hagen}, W. 1982, \aap, 114, 245

\bibitem[{{Vasyunin} {et~al.}(2017){Vasyunin}, {Caselli}, {Dulieu}, \&
  {Jim{\'e}nez-Serra}}]{Vasyunin2017}
{Vasyunin}, A.~I., {Caselli}, P., {Dulieu}, F., \& {Jim{\'e}nez-Serra}, I.
  2017, \apj, 842, 33, \dodoi{10.3847/1538-4357/aa72ec}

\bibitem[{{Vasyunin} \& {Herbst}(2013)}]{Vasyunin2013}
{Vasyunin}, A.~I., \& {Herbst}, E. 2013, \apj, 769, 34,
  \dodoi{10.1088/0004-637X/769/1/34}

\bibitem[{{Vazart} {et~al.}(2020){Vazart}, {Ceccarelli}, {Balucani}, {Bianchi},
  \& {Skouteris}}]{Vazart2020MNRAS}
{Vazart}, F., {Ceccarelli}, C., {Balucani}, N., {Bianchi}, E., \& {Skouteris},
  D. 2020, \mnras, 499, 5547, \dodoi{10.1093/mnras/staa3060}

\bibitem[{{Viti} {et~al.}(2004){Viti}, {Collings}, {Dever}, {McCoustra}, \&
  {Williams}}]{viti2004}
{Viti}, S., {Collings}, M.~P., {Dever}, J.~W., {McCoustra}, M. R.~S., \&
  {Williams}, D.~A. 2004, \mnras, 354, 1141,
  \dodoi{10.1111/j.1365-2966.2004.08273.x}

\bibitem[{{Wakelam} {et~al.}(2017){Wakelam}, {Loison}, {Mereau}, \&
  {Ruaud}}]{Wakelam_2017_BE}
{Wakelam}, V., {Loison}, J.~C., {Mereau}, R., \& {Ruaud}, M. 2017, Molecular
  Astrophysics, 6, 22, \dodoi{10.1016/j.molap.2017.01.002}

\bibitem[{{Zamirri} {et~al.}(2019){Zamirri}, {Ugliengo}, {Ceccarelli}, \&
  {Rimola}}]{Zamirri2019Review}
{Zamirri}, L., {Ugliengo}, P., {Ceccarelli}, C., \& {Rimola}, A. 2019, ACS
  Earth and Space Chemistry, 3, 1499, \dodoi{10.1021/acsearthspacechem.9b00082}

\end{thebibliography}
\bibliographystyle{aasjournal}




\appendix
\renewcommand\thefigure{\thesection.\arabic{figure}}    

\section{Radical--water interactions}

\begin{figure}[!htbp]
    \centering
    \includegraphics[width=0.8\columnwidth]{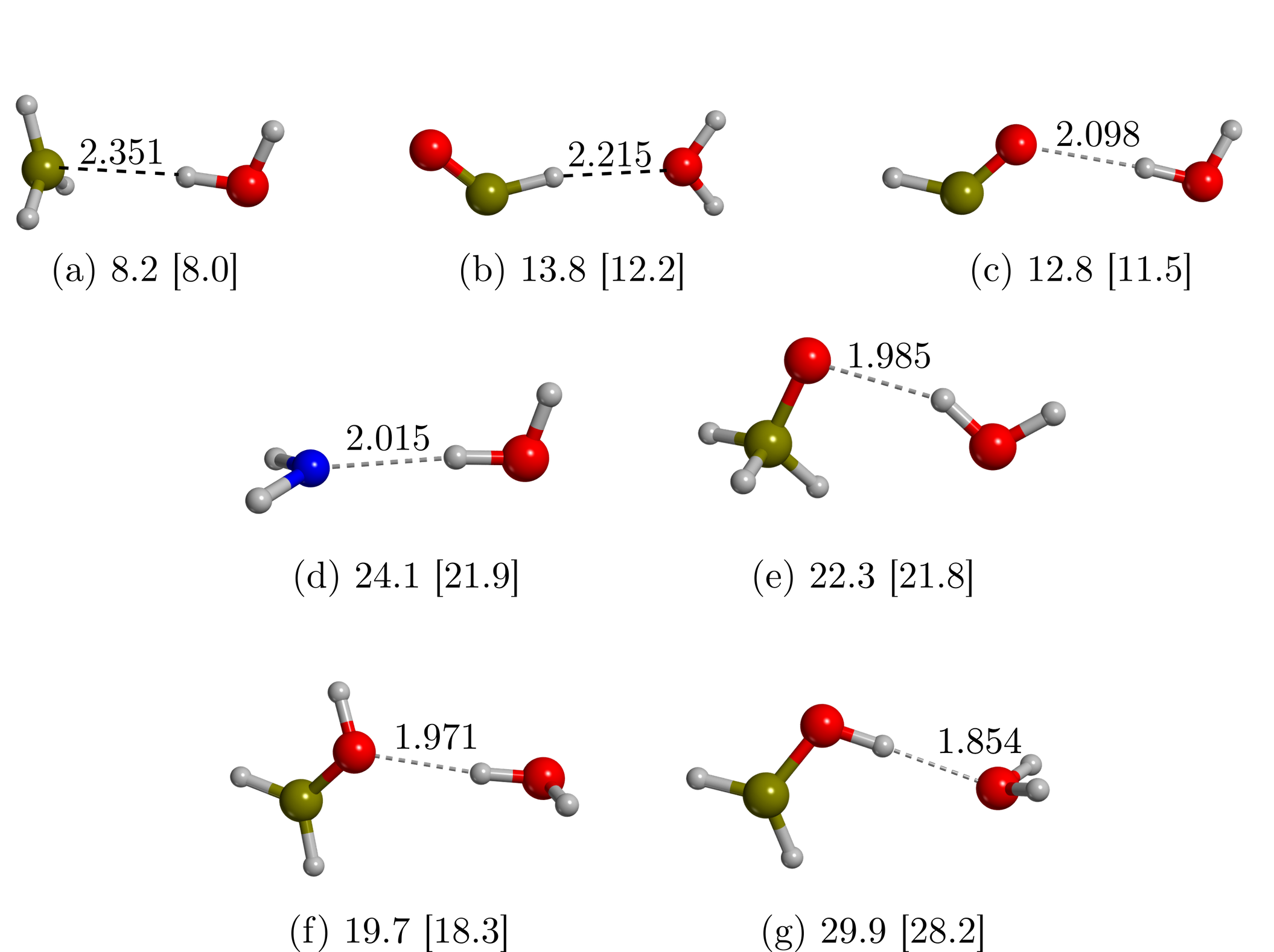}
    \caption{ZPE- and BSSE- non-corrected binding energies of CH$_3$ (a), HCO (b,c), NH$_2$ (d), CH$_3$O (e) and CH$_2$OH (f,g) with a single water molecule at BHLYP-D3(BJ)/6-311++G(2df,2pd) and CCSD(T)/aug-cc-pVTZ//BHLYP-D3(BJ)/6-311++G(2df,2pd) in brackets. Distances in \r{A}.}
    \label{fig:interaction_energies}
\end{figure}

In order to trace the origin of CH$_2$OH and CH$_3$O binding energies we have run optimisations at BHLYP-D3-BJ/6-311++G(2df,2pd) level and single point calculations at CCSD(T)/aug-cc-pVTZ level, similarly to what we did for CH$_3$, HCO and NH$_2$ in \cite{ER2019}. For the three latter, the main differences are the bond distances (due to the change of DFT method) with minor changes in the interaction energies, smaller than 1 kJ mol$^{-1}$ (comparing the new values to those computed at BHLYP-D3/6-311++G(2df,2pd)//B3LYP-D3/6-311++G(2df,2pd) level) and the NH--H$\cdots$OH$_2$ geometry from \cite{ER2019} evolves into the NH$_2\cdots$H$_2$O case shown in Figure \ref{fig:interaction_energies}(d) during optimisation.
Regarding the two new radicals, CH$_3$O mainly interacts with the water molecules via a strong H-bond on its O atom while CH$_2$OH can interact \textit{via} two strong H-bonds on the --OH group, one as an H-donor and another as H-acceptor. 
The resulting geometries and energetics are shown in Figure \ref{fig:interaction_energies}. 
It is worth mentioning that different initial radical-water orientations were tried for CH$_3$O and CH$2$OH, which, after optimisation converged to the ones in Figure \ref{fig:interaction_energies}.
All in all, the differences between DFT and CCSD(T) values ie below 2.2 kJ mol$^{-1}$, corresponding to the NH$_2\cdots$H$_2$O case (Figure \ref{fig:interaction_energies}(d)).
\\

\begin{figure}[!htbp]
    \centering
    \includegraphics[width=0.99\columnwidth]{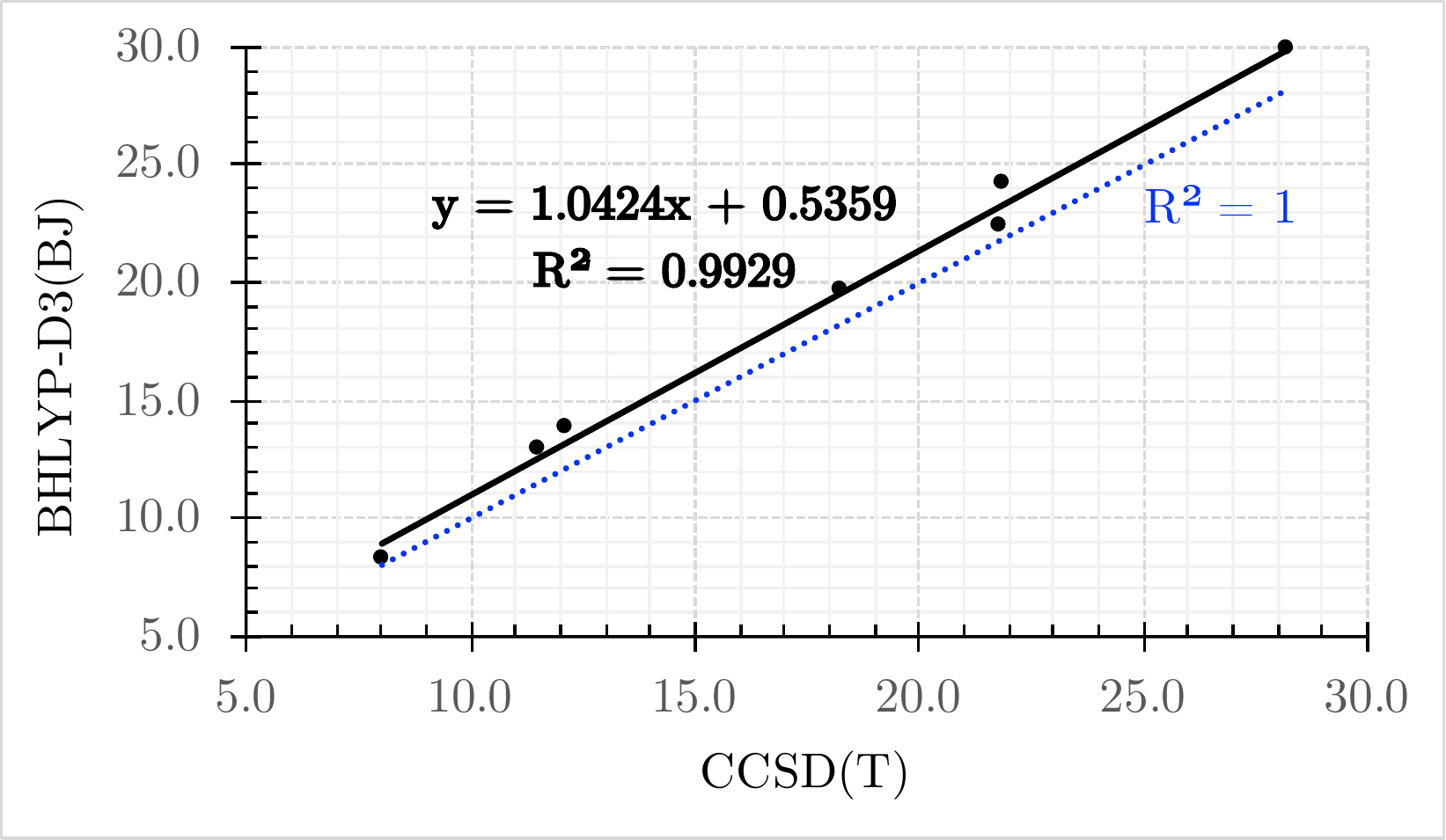}
    \caption{Correlation between ZPE- and BSSE- non-corrected binding energies of the radicals in Figure \ref{fig:interaction_energies} BHLYP-D3(BJ)/6-311++G(2df,2pd) and CCSD(T)/aug-cc-pVTZ//BHLYP-D3(BJ)/6-311++G(2df,2pd) (black-filled points) with their trend line, and for the sake of comparison, the line corresponding to a perfect correlation with CCSD(T) data.}
    \label{fig:interaction_energies_correlation}
\end{figure}

\clearpage
\section{Adsorption geometries on W18}
\label{sec:appendix:adsW18}

\begin{figure}[!htbp]
    \centering
    \includegraphics[width=0.99\columnwidth]{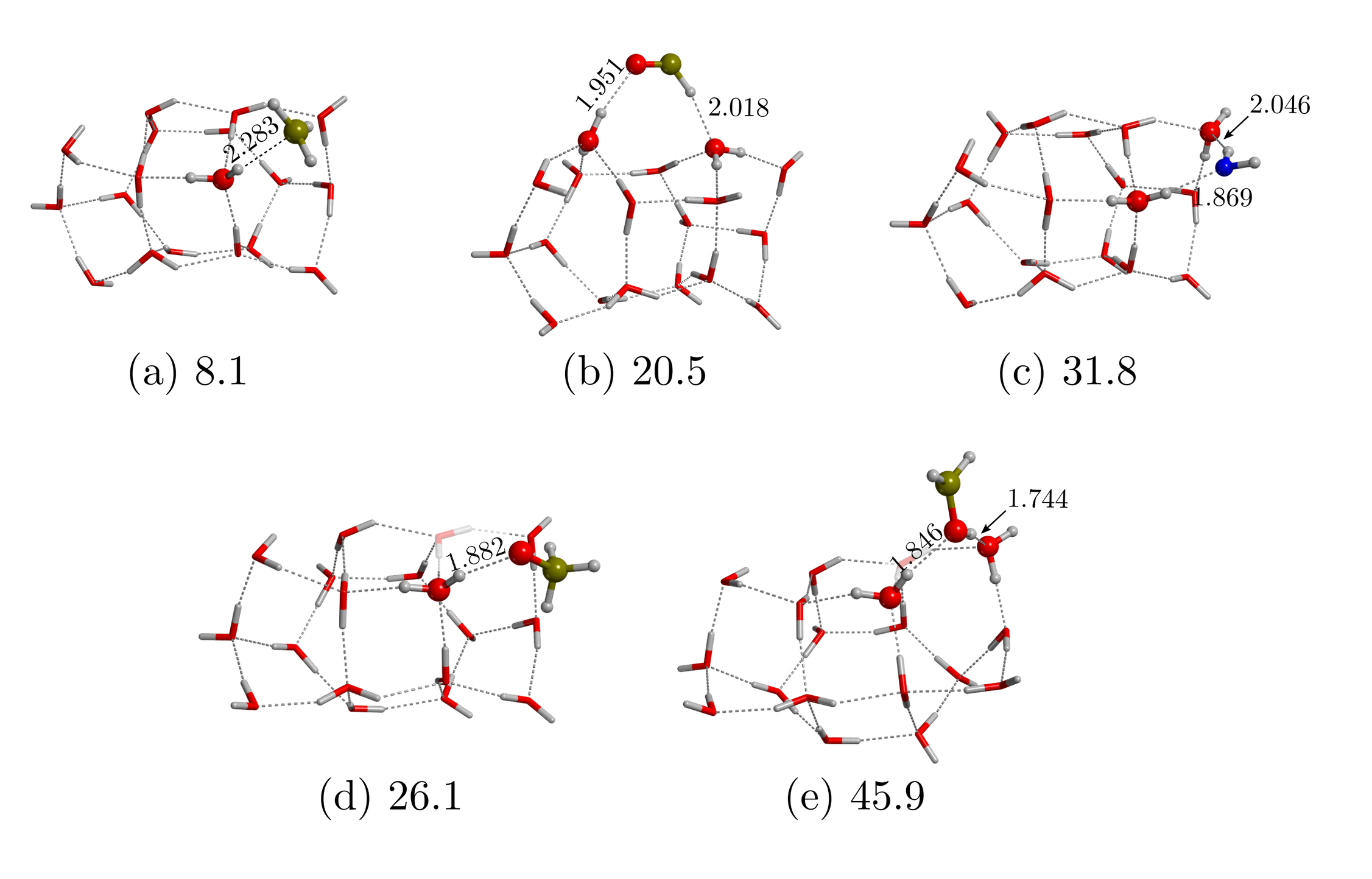}
    \caption{Geometries of the five studied radicals, CH$_3$ (a), HCO (b), NH$_2$ (c), CH$_3$O (d) and CH$_2$H (e); adsorbed on W18 fully optimised at the UBHLYP-D3(BJ)/6-31+G(d,p) theory level. Energy values in kJ mol$^{-1}$ are those refined at UBHLYP-D3(BJ)/6-311++G(2df,2pd) level with the ZPE- (at 6-31+G(d,p) level) and BSSE- corrections. Distances in \r{A}.}
    \label{fig:ANNEX_W18_ads}
\end{figure}

\clearpage
\section{Adsorption geometries of NH and OH on W18 and W33}

\begin{figure}[!htbp]
    \centering
    \includegraphics[width=0.99\columnwidth]{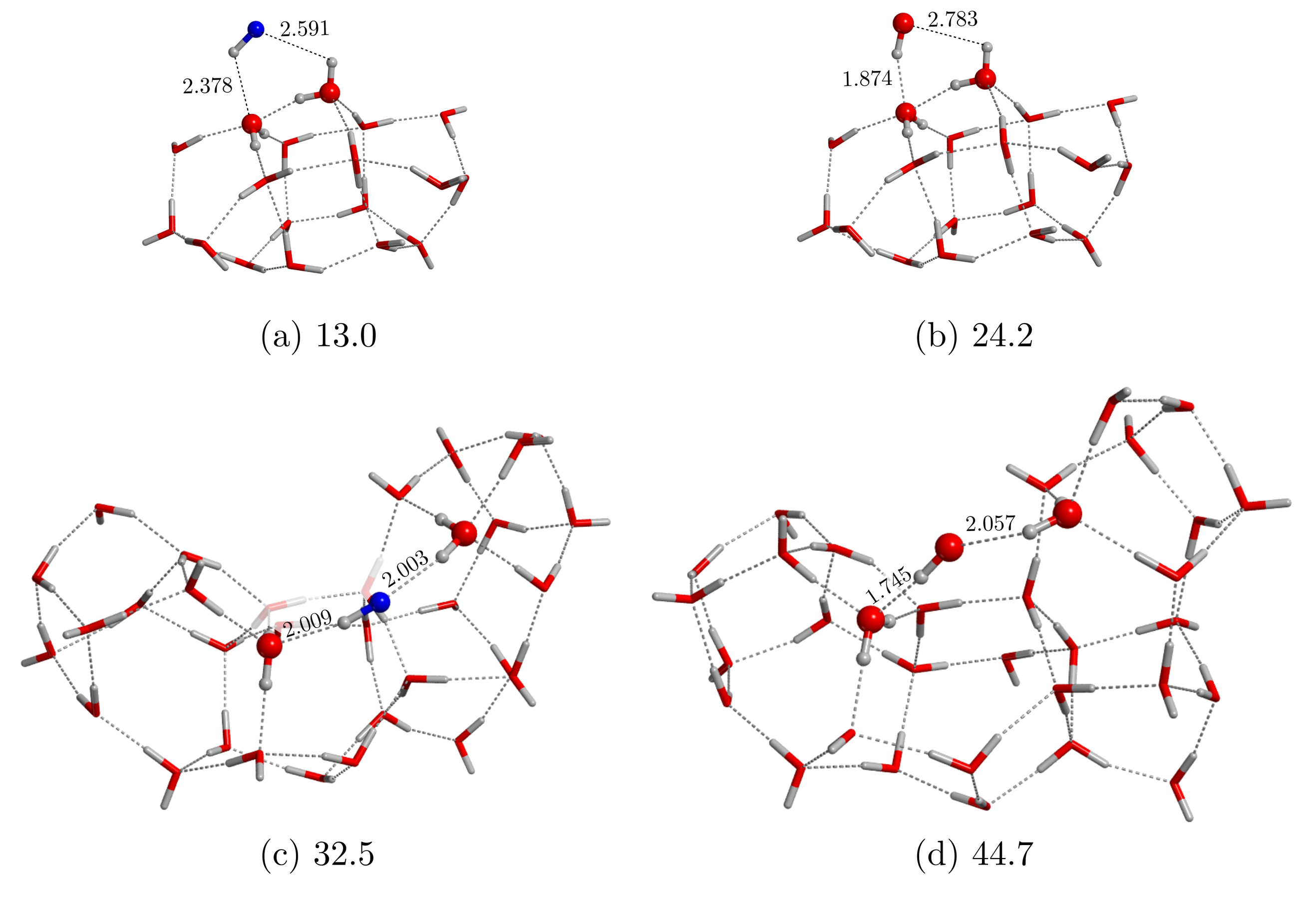}
    \caption{Geometries of NH (in its ground triplet electronic state) and OH (doublet electronic state) on W18 (a, b) and on the cavity of W33 (c, d). These geometries were fully optimised at the UBHLYP-D3(BJ)/6-31+G(d,p) theory level. Energy values in kJ mol$^{-1}$ are those refined at UBHLYP-D3(BJ)/6-311++G(2df,2pd) level with the ZPE- (at 6-31+G(d,p) level) and BSSE- corrections. Distances in \r{A}.}
    \label{fig:ANNEX_OH_NH_ads}
\end{figure}

We have calculated the binding energies (BE) of NH (ground triplet electronic state) and OH (doublet electronic state) radicals following the same methodology as for the other radicals in this works. As it can be seen from Figure \ref{fig:ANNEX_OH_NH_ads}, we find again that the BE on the W18 amorphous solid water (ASW) ice model, 13.0 and 24.2 kJ mol$^{-1}$ for NH and OH, are roughly half those on the cavity of W33, 32.5 and 44.7 kJ mol$^{-1}$ for NH and OH.
This values are in good agreement with those in the literature.
There are several works reporting the BE for OH on water surfaces: \cite{Sameera_2017_BE} report a range of BEs in between 19.3 and 64.6 kJ mol$^{-1}$ on top of a crystalline ice structure, \cite{Wakelam_2017_BE} recommend a value of 38.2 kJ mol$^{-1}$ for astrochemical models and, most recently, \cite{Ferrero_2020_BE} report a range in between 12.9 and 44.2 kJ mol$^{-1}$ on amorphous water ice. On the other hand, for NH \cite{Wakelam_2017_BE} recommends a value of 21.6 kJ mol$^{-1}$, which lies in between the values we report for this nitrene. On the other hand, \cite{martinez-bachs2020ICCSA} reported a binding energy of 35.1 kJ mol$^{-1}$ on top of a crystalline water ice surface. On top of an ASW ice surface, the BE cover a wider range, from  $\sim$ 11 to 45 kJ mol$^{-1}$ (private communication), centered at around 20 kJ mol$^{-1}$, so that our BE values are well within the limits and close to this central value.

\clearpage
\section{Components of the binding energies}

Here we report the components of the BEs. We recall that we define $\Delta E_{bind} = -\Delta E_{ads}$.

\begin{table}[!htbp]
\centering 
\caption{Components of the adsorption energies energies (we remind that $\Delta E_{ads} = -\Delta E_{BE}$). U are pure DFT energies, D are dispersion corrections, ZPE are zero point energies, BSSE are the basis set superposition error energies.
Energy units are kJ mol$^{-1}$.}
\label{tab:components_BE}
\begin{tabular}{|l|cccc|}
\hline
W18 &  $\Delta U_{ads}$ & $\Delta D_{ads}$ & $\Delta ZPE$ & BSSE \\ \hline 
CH$_3$   & -9.5  & -6.7  & 7.3  & 0.8 \\
HCO      & -26.1 & -2.8  & 6.2  & 2.3 \\
NH$_2$   & -40.0 & -7.2  & 13.2 & 2.2 \\
CH$_3$O  & -26.8 & -8.4  & 6.6  & 2.4 \\
CH$_2$OH & -50.1 & -7.0  & 8.0  & 3.2 \\ \hline
NH       & -14.5 & -4.0 & 4.4 & 1.1   \\
OH       & -28.7 & -3.7 & 6.0 & 2.2   \\ \hline
\end{tabular}%
\vspace{1cm}
\begin{tabular}{|l|cccc|}
\hline
W33-cav & $\Delta U_{ads}$ & $\Delta D_{ads}$ & $\Delta ZPE$ & BSSE \\ \hline 
CH$_3$   & -15.3 & -0.5  & 0.5  & 1.1 \\
HCO      & -28.6 & -5.0  & 1.7  & 2.4 \\
NH$_2$   & -51.3 & -2.8  & 7.0  & 2.7 \\
CH$_3$O  & -35.8 & -14.9 & 8.9  & 3.7 \\
CH$_2$OH & -38.6 & -26.3 & 9.3  & 4.4 \\\hline
NH       & -35.5 & -10.3 & 10.8 & 2.5 \\
OH       &  -49.3 & -10.1 & 11.4 & 3.4 \\\hline
\end{tabular}%
\end{table}

\clearpage
\section{Radical-radical reaction energetics}

\begin{table}[!htbp]
\centering
\caption{Energetics of the investigated radical-radical reactions on W18 (left) and W33 (right). DFT and dispersion energies were calculated at BHLYP-D3(BJ)/6-311++G(2df,2pd), while ZPE corrections were calculated at BHLYP-D3(BJ)/6-31+G(d,p) level. Column number 2 indicates the reaction type, and for dHa wether it is case 1 (dHa1) or 2 (dHa2). Some reactions have more than one step, as indicated by column number 3 (Step \#).  Energy units in kJ mol$^{-1}$.}
\label{tab:EnergiesW18}
\resizebox{0.49\textwidth}{!}{%
\begin{tabular}{ccccc}
\hline
X + Y···W18                    & RX   & Step \# & $\Delta H^{\ddagger}$ & $\Delta H_{rx}$ \\ \hline
\multirow{4}{*}{CH$_2$OH + CH$_2$OH} & Rc   &         & 4.4                   & -288.7          \\
                             & dHa1 &         & 39.1                  & -223.7          \\
                             & dHa2 & 1       & 20.6                  & 14.9            \\
                             & dHa2 & 2       & 27.6                  & -226.3          \\ \hline
\multirow{3}{*}{CH$_3$O + CH$_3$O}   & Rc   &         & 10.3                  & -64.1           \\
                             & dHa1 &         & 15.9                  & -312.5          \\
                             & dHa2 &         & 18.2                  & -296.0          \\ \hline
\multirow{3}{*}{CH$_3$ + CH$_2$OH}   & Rc   &         & 1.9                   & -320.4          \\
                             & dHa & 1  & 23.9                  & 24.2            \\
                             & dHa & 2  & 32.2                  & -255.1          \\ \hline
CH$_3$ + CH$_3$                      & Rc   &         & -0.1                  & -333.4          \\ \hline
\multirow{2}{*}{CH$_3$ + CH$_3$O}    & Rc   &         & 0.2                   & -299.5          \\
                             & dHa  &         & 1.0                   & -302.7          \\ \hline
CH$_3$ + NH$_2$              & Rc   &         & 1.6                  & -316.8          \\ \hline
\multirow{3}{*}{HCO + CH$_2$OH}   & Rc   &         & 1.6                   & -286.2          \\
                             & dHa1 &         & -0.6                  & -290.2          \\
                             & dHa2 &         & 30.7  &  -182.9 \\ \hline
\multirow{3}{*}{HCO + CH$_3$O}    & Rc   &         & 5.1                   & -358.5          \\
                             & dHa1 &         & 3.2                   & -323.4          \\
                             & dHa2 &         & 9.6                   & -246.3          \\ \hline
\multirow{3}{*}{HCO + HCO}     & Rc   &         & 4.0                   & -260.3          \\
                             & dHa1 &         & 2.7                   & -272.7          \\
                             & dHa2 &         & 8.3                   & -273.3          \\ \hline
\end{tabular}%
}%
\resizebox{0.49\textwidth}{!}{%
\begin{tabular}{ccccc}
\hline
X + Y···W33                        & RX   & Step \# & $\Delta H^{\ddagger}$ & $\Delta H_{rx}$ \\ \hline
\multirow{8}{*}{CH$_2$OH + CH$_2$OH} & Rc   &         & 2.6                   & -299.5          \\
                                     & dHa1 & 1       & 1.4                   & -10.2           \\
                                     & dHa1 & 2       & 11.6                  & 7.2             \\
                                     & dHa1 & 3       & 9.1                   & 8.0             \\
                                     & dHa1 & 4       & 24.9                  & 25.4            \\
                                     & dHa1 & 5       & 29.1                  & -218.3          \\
                                     & dHa2 & 1       & 9.0                   & -0.7            \\
                                     & dHa2 & 2       & 7.4                   & -245.8          \\ \hline
\multirow{3}{*}{CH$_3$O + CH$_3$O}   & Rc   &         & 20.1                  & -58.3           \\
                                     & dHa1 &         & 11.7                  & -292.3          \\
                                     & dHa2 &         & 21.2                  & -298.6          \\ \hline
\multirow{3}{*}{CH$_3$ + CH$_2$OH}   & Rc   & 1       & 2.5                   & 0.5             \\
                                     & Rc   & 2       & 0.4                   & -310.0          \\
                                     & dHa  &         & 39.0                  & -258.5          \\ \hline
CH$_3$ + CH$_3$                      & Rc   &         & 4.6                   & -323.6          \\ \hline
\multirow{2}{*}{CH$_3$ + CH$_3$O}    & Rc   &         & 3.1                   & -290.1          \\
                                     & dHa  &         & 9.5                   & -301.9          \\ \hline
CH$_3$ + NH$_2$                      & Rc   &         & 0.4                   & -319.2          \\ \hline
\multirow{4}{*}{HCO + CH$_2$OH}      & Rc   &         & 1.7                   & -288.6          \\
                                     & dHa1 &         & 0.8                   & -295.8          \\
                                     & dHa2 & 1       & 10.2                  & 10.4            \\
                                     & dHa2 & 2       & 18.4                  & -228.6          \\ \hline
\multirow{3}{*}{HCO + CH$_3$O}       & Rc   &         & 3.5                   & -351.6          \\
                                     & dHa1 &         & 2.0                   & -322.2          \\
                                     & dHa2 &         & 13.2                  & -242.4          \\ \hline
\multirow{2}{*}{HCO + HCO}           & Rc   &         & 4.1                   & -268.0          \\
                                     & dHa1 &         & 4.0                   & -279.5          \\ \hline
\end{tabular}%
}
\end{table}

\clearpage

\section{W18 and W33 transition state energetics}

\begin{table*}[!htbp]
\centering
\caption{Data of the transition states found on the W18 cluster model at BHLYP-D3(BJ)/6-31+G(d,p) level. Pure DFT energies were refined at BHLYP-D3(BJ)/6-311++G(2df,2pd) level (TZ). U, D and ZPE stand for pure DFT, dispersion and zero point (vibrational) energies, H is the combination of them three (i.e. enthalpies at zero Kelvin). $T_c$ is the tunneling crossover temperature. Column number 2 indicates the reaction type, and for dHa wether it is case 1 (dHa1) or 2 (dHa2). Some reactions have more than one step, as indicated by column number 3 (Step \#). Energies kJ mol$^{-1}$.}
\label{tab:W18_energies}
\resizebox{\textwidth}{!}{%
\hskip-2.5cm \begin{tabular}{ccccccccc|cc|c}
\hline
                                     &          &       & \multicolumn{7}{c}{$\Delta E_{TS}$}                           \\ \cline{2-12} 
X/Y$\cdots$W18 & RX & TS \# &$\Delta U$& $\Delta D$ & $\Delta U+D$ & $\Delta ZPE$ & $\Delta H$    & i$\nu$ (cm$^{-1}$) & $\Delta U$ (TZ) & $\Delta H$ (TZ) & $T_c$ (K) \\ \hline
\multirow{4}{*}{CH$_2$OH + CH$_2$OH} 
            & Rc       &       & 9.1   & -4.8   & 4.3  & 0.3   & \textbf{4.6}  & -82.50   & 8.9  & \textbf{4.4}   & --  \\
            & dHa1     &       & 61.9  & -7.7   & 54.2 & -14.4 & \textbf{39.8} & -1837.06 & 61.1 & \textbf{39.1}  & 448.5 \\
            & dHa2     & 1     & 25.5  & -0.8   & 24.7 & -2.2  & \textbf{22.5} & -60.69   & 23.6 & \textbf{20.6}  & --  \\
            & dHa2     & 2     & 26.8  & -5.7   & 21.1 & -11.4 & \textbf{9.7}  & -1461.84 & 44.4 & \textbf{27.6}  & 394.3  \\ \hline
\multirow{3}{*}{CH$_3$O + CH$_3$O} 
            & Rc         &       & 16.2  & -4.0   & 12.2 & -0.5  & \textbf{11.7} & -31.29 & 14.8 & \textbf{10.3}  & --  \\
            & dHa1     &       & 29.3  & -3.6   & 25.7 & -7.4  & \textbf{18.3} & -863.13  & 26.9 & \textbf{15.9}  & 212.9  \\
            & dHa2     &       & 33.6  & -4.4   & 29.2 & -8.2  & \textbf{21.0} & -822.52  & 30.8 & \textbf{18.2}  & 352.2  \\ \hline
\multirow{3}{*}{CH$_3$ + CH$_2$OH}   
            & Rc       &       & 2.8   & -0.7   & 2.1  & -0.1  & \textbf{2.0}  & -69.97   & 2.7  & \textbf{1.9}   & --   \\
            & dHa      & 1     & 28.2  & -2.0   & 26.2 & -0.6  & \textbf{25.6} & -21.93   & 26.5 & \textbf{23.9}  & --   \\
            & dHa      & 2     & 47.0  & -2.7   & 44.3 & -11.7 & \textbf{32.6} & -1226.28 & 46.6 & \textbf{32.2}  & 295.7  \\ \hline
CH$_3$ + CH$_3$                           
            & Rc       &       & 7.4   & -5.4   & 2.1  & -1.4  & \textbf{0.7}  & -36.520  & 6.6  & \textbf{-0.1}  & --  \\ \hline
\multirow{2}{*}{CH$_3$ + CH$_3$O}         
            & Rc       &       & 3.0   & -1.5   & 1.5  & -0.8  & \textbf{0.7}  & -88.820  & 2.5  & \textbf{0.2}   & --  \\
            & dHa      &       & 3.9   & -1.9   & 2.0  & -0.6  & \textbf{1.3}  & -129.30  & 3.5  & \textbf{1.0}   & 36.1  \\ \hline
CH$_3$ + NH$_2$                           
            & Rc       &       & 1.4   & -1.0   & 0.4  & -0.3  & \textbf{0.0}  & -19.83   & 1.2  & \textbf{-0.1}  & --  \\ \hline
\multirow{3}{*}{HCO + CH$_2$OH}           
            & Rc       &       & 6.8   & -4.7   & 2.1  & 0.0   & \textbf{2.2}  & -55.33   & 6.3  & \textbf{1.6}   & --  \\
            & dHa1     &       & 5.0   & -3.3   & 1.7  & -1.6  & \textbf{0.1}  & -154.61  & 4.3 & \textbf{-0.6}   & 26.2  \\ 
            & dHa2     &       & 46.9  & -0.6   & 46.2 & -13.7 & \textbf{32.6} & -652.3997& 45.0& \textbf{30.7}   & 153.7 \\ \hline
\multirow{3}{*}{HCO + CH$_3$O}
           & Rc       &       & 8.9  & -0.5   & 8.4 & -2.1  & \textbf{6.3}  & -106.15   & 7.7 & \textbf{5.1}      & --  \\
           & dHa1     &       & 12.1  & -5.3   & 6.8  & -2.3  & \textbf{4.5}  & -228.50  & 10.8 & \textbf{3.2}    & 57.8 \\
           & dHa2     &       & 23.0  & -3.0   & 19.9 & -6.6  & \textbf{13.3} & -387.25  & 19.2 & \textbf{9.6}    & 93.7 \\ \hline
\multirow{3}{*}{HCO + HCO}           
           & Rc       &       & 8.3   & -2.7   & 5.6  & -0.9  & \textbf{4.7}  & -75.94   & 7.6  & \textbf{4.0}    & -- \\
           & dHa1     &       & 6.3   & -1.6   & 4.7  & -1.4  & \textbf{3.3}  & -46.31   & 5.7  & \textbf{2.7}    & 10.8 \\
           & dHa2     &       & 15.8  & -0.9   & 14.9 & -4.1  & \textbf{10.8} & -102.29  & 13.3 & \textbf{8.3}    & 8.3 \\ \hline
\end{tabular}%
}
\end{table*}

\begin{table*}[!htbp]
\centering
\caption{Summary of all the transition states found in this work at BHLYP-D3(BJ)/6-31+G(d.p) level, see the results section for the full PESs. Pure DFT energies were refined at BHLYP-D3(BJ)/6-311++G(2df,2pd) level (TZ). U, D and ZPE stand for pure DFT, dispersion and zero point (vibrational) energies, H is the combination of them three (i.e. enthalpies at zero Kelvin). $T_c$ is the tunneling crossover temperature. Column number 2 indicates the reaction type, and for dHa wether it is case 1 (dHa1) or 2 (dHa2). Some reactions have more than one step, as indicated by column number 3 (Step \#). Energies in kJ mol$^{-1}$.}
\label{tab:All_TS}
\resizebox{\textwidth}{!}{%
\hskip-2.5cm \begin{tabular}{cccccccccc|cc|c}
\hline
                            &                            &      &   & \multicolumn{6}{c}{$\Delta$ E TS}                     \\ \cline{3-12} 
\multicolumn{2}{c}{X/Y···W33-cav} &
  RX &
  \# TS &
  $\Delta$U &
  $\Delta$D &
  $\Delta$(U+D) &
  $\Delta$ZPE &
  $\Delta$H &
  i$\nu$ (cm$^{-1}$) & 
  $\Delta$U (TZ) & 
  $\Delta$H (TZ) & 
  $T_c$\\ \hline
\multicolumn{2}{c}{\multirow{8}{*}{CH$_2$OH + CH$_2$OH}} & Rc   & - & 0.5  & 4.4  & 4.9  & -1.8  & \textbf{3.1}  & 28.78   &0.0  &\textbf{2.6}    & -- \\
\multicolumn{2}{c}{}                                     & dHa1 & 1 & 0.8  & 2.0  & 2.8  & -1.5  & \textbf{1.3}  & 71.45   &1.0  &\textbf{1.4}    & -- \\
\multicolumn{2}{c}{}                                     & dHa1 & 2 & 1.3  & 13.3 & 14.5 & -3.9  & \textbf{10.7} & -46.04  &2.1  &\textbf{11.6}   & -- \\
\multicolumn{2}{c}{}                                     & dHa1 & 3 & -1.4 & 14.4 & 13.0 & -3.7  & \textbf{9.2}  & 79.17   &-1.6 &\textbf{9.1}    & -- \\
\multicolumn{2}{c}{}                                     & dHa1 & 4 & 10.7 & 18.5 & 29.2 & -3.4  & \textbf{25.7} & 68.37   &9.8  &\textbf{24.9}   & -- \\
\multicolumn{2}{c}{}                                     & dHa1 & 5 & 23.8 & 15.2 & 39.0 & -10.9 & \textbf{28.1} & 542.54  &24.7 &\textbf{29.1}   & 154.7 \\
\multicolumn{2}{c}{}                                     & dHa2 & 1 & 10.6 & -1.3 & 9.3  & -0.4  & \textbf{8.9}  & 95.87   & 10.6 &\textbf{9.0}   & -- \\
\multicolumn{2}{c}{}                                     & dHa2 & 2 & 26.6 & -8.9 & 17.7 & -11.0 & \textbf{6.7}  & 1068.76 & 27.3 &\textbf{7.4}   & 296.2 \\ \hline
\multicolumn{2}{c}{\multirow{3}{*}{CH$_3$O + CH$_3$O}}   & Rc   & - & 20.4 & -2.3 & 18.1 & 2.8   & \textbf{21.0} & 140.15  &19.5  &\textbf{20.1}  & -- \\
\multicolumn{2}{c}{}                                     & dHa1 & - & 22.1 & -3.0 & 19.1 & -5.9  & \textbf{13.2} & 886.62  &20.6  &\textbf{11.7}  & 225.7 \\
\multicolumn{2}{c}{}                                     & dHa2 & - & 32.2 & -1.6 & 30.6 & -4.9  & \textbf{25.7} & 788.36  &27.8  &\textbf{21.2}  & 189.8 \\ \hline
\multicolumn{2}{c}{\multirow{3}{*}{CH$_3$ + CH$_2$OH}}   & Rc   & 1 & -1.6 & 4.5  & 3.0  & -0.6  & \textbf{2.4}  & 27.48   &-1.4  &\textbf{2.5}   & --  \\
\multicolumn{2}{c}{}                                     & Rc   & 2 & -2.8 & 1.6  & -1.2 & 1.6   & \textbf{0.4}  & 59.97   &-2.7  &\textbf{0.4}   & -- \\
\multicolumn{2}{c}{}                                     & dHa  & - & 52.1 & -0.4 & 51.7 & -11.8 & \textbf{39.9} & 1020.12 &51.2  &\textbf{39.0}  & 242.0 \\ \hline
\multicolumn{2}{c}{CH$_3$ + CH$_3$}                      & Rc   & - & 9.8  & -2.5 & 7.2  & -1.3  & \textbf{5.9}  & 110.49  &8.4   &\textbf{4.6}   & --  \\ \hline
\multicolumn{2}{c}{\multirow{2}{*}{CH$_3$ + CH$_3$O}}    & Rc   & - & 8.1  & -3.8 & 4.3  & 0.1   & \textbf{4.4}  & 123.81  &6.8   &\textbf{3.1}   & --  \\
\multicolumn{2}{c}{}                                     & dHa  & - & 17.1 & -2.9 & 14.2 & -2.3  & \textbf{11.9} & 200.13  &14.7  &\textbf{9.5}   & 47.1  \\ \hline
\multicolumn{2}{c}{CH$_3$ + NH$_2$}                      & Rc   & - & 1.4  & -0.3 & 1.1  & -0.4  & \textbf{0.7}  & 92.25   &1.1   &\textbf{0.4}   & --  \\ \hline
\multicolumn{2}{c}{\multirow{4}{*}{HCO + CH$_2$OH}}      & Rc   & - & 0.2  & 2.8  & 3.0  & -1.3  & \textbf{1.7}  & 55.18   &0.2   &\textbf{1.7}   & --  \\
\multicolumn{2}{c}{}                                     & dHa1 & - & 1.9  & 1.2  & 3.2  & -1.6  & \textbf{1.6}  & 90.00   &1.2   &\textbf{0.8}   & 24.2  \\
\multicolumn{2}{c}{}                                     & dHa2 & 1 & 13.6 & 3.8  & 17.4 & -2.2  & \textbf{15.2} & 35.32   &10.6  &\textbf{10.2}  & --  \\
\multicolumn{2}{c}{}                                     & dHa2 & 2 & 37.4 & 0.5  & 37.9 & -16.1 & \textbf{21.7} & 1321.28 &36.0  &\textbf{18.4}  & 387.0  \\ \hline
\multicolumn{2}{c}{\multirow{3}{*}{HCO + CH$_3$O}}       & Rc   & - & 2.4  & 2.3  & 4.7  & -0.2  & \textbf{4.5}  & 27.20   &1.4   &\textbf{3.5}   & --  \\
\multicolumn{2}{c}{}                                     & dHa1 & - & 13.3 & -4.8 & 8.6  & -4.6  & \textbf{4.0}  & 137.45  &11.4  &\textbf{2.0}   & 34.6  \\
\multicolumn{2}{c}{}                                     & dHa2 & - & 15.4 & 3.4  & 18.8 & -3.6  & \textbf{15.1} & 188.68  &13.4  &\textbf{13.2}  & 35.4  \\ \hline
\multicolumn{2}{c}{\multirow{3}{*}{HCO + HCO}}           & Rc   & - & 6.7  & -1.0 & 5.7  & -0.7  & \textbf{4.9}  & 64.27   &5.8   &\textbf{4.1}   & --  \\
\multicolumn{2}{c}{}                                     & dHa1  & 1 & 9.0  & -1.2 & 7.8  & -1.5  & \textbf{6.3}  & 119.03  &6.7   &\textbf{4.0}  & 28.4  \\
\multicolumn{2}{c}{}                                     & dHa2*  & 2 & 1.5  & 0.0  & 1.5  & -1.7  & \textbf{-0.3} & 164.89  &0.9 &\textbf{-0.8 } & 29.6  \\ \hline
\end{tabular}%
}
\hspace{2cm}
\resizebox{\textwidth}{!}{%
\begin{tabular}{l}
* The energy reference is point 24 from the backwards ICR calculation, see the Figure Sets available in the online version of the journal. \\
\end{tabular}%
}
\end{table*}

\clearpage
\section{Spin densities of each radical in reactant structures}

\begin{table}[!htbp]
\centering
\caption{Spin densities of the reactant radical radical structures. Computed employing natural bond population analysis. Figures of the spin densities and molecular orbitals are available online in Zenodo \cite{zenodo_data}.}
\label{tab:initial_spin_dens}
\resizebox{0.99\textwidth}{!}{%
\begin{tabular}{lllllllllll}
\cline{1-5} \cline{7-11}
\cline{1-5} \cline{7-11}
\multicolumn{5}{c}{W18}                       & $\qquad\qquad$ & \multicolumn{5}{c}{W33}                       \\ \cline{1-5} \cline{7-11} 
Reaction & R1    & R2    & SD R1    & SD R2   &  & Reaction & R1    & R2    & SD R1    & SD R2   \\ \cline{1-5} \cline{7-11} 
Rc       & CH3O  & CH3O  & -0.99655 & 1.00061 &  & Rc       & CH3O  & CH3O  & -1.00019 & 1.00021 \\
Rc       & CH2OH & CH3   & -0.99561 & 0.98991 &  & Rc       & CH3O  & HCO   & -1.00020 & 0.99241 \\
Rc       & CH3O  & HCO   & -1.00005 & 0.98816 &  & Rc       & CH2OH & HCO   & -0.98984 & 0.98981 \\
Rc       & CH2OH & CH2OH & -0.99358 & 0.99107 &  & Rc       & CH2OH & CH2OH & -0.98789 & 0.99624 \\
Rc       & HCO   & HCO   & -0.99323 & 0.99112 &  & Rc       & CH2OH & CH3   & -0.99694 & 0.98891 \\
Rc       & CH3   & CH3O  & -0.99476 & 0.99810 &  & Rc       & CH2OH & CH3   & -0.98978 & 0.98025 \\
Rc       & HCO   & CH2OH & -0.98802 & 0.99912 &  & Rc       & HCO   & HCO   & -0.97371 & 0.99443 \\
Rc       & CH3   & NH2   & -0.99307 & 1.00575 &  & Rc       & CH3   & CH3   & -0.97903 & 0.98453 \\
dHa2     & HCO   & CH3O  & -0.98813 & 1.00073 &  & Rc       & CH3   & CH3O  & -0.98658 & 0.99996 \\
dHa      & CH3   & CH3O  & -0.99352 & 0.99509 &  & Rc       & CH3   & NH2   & -0.98649 & 1.00420 \\
dHa2     & HCO   & CH2OH & -0.99078 & 0.99727 &  & dHa1     & HCO   & HCO   & -0.97365 & 0.99442 \\
dHa2     & CH3O  & CH3O  & -0.99985 & 0.99948 &  & dHa1     & CH2OH & HCO   & -0.98988 & 0.98980 \\
dHa2     & CH3O  & HCO   & -0.99904 & 0.98953 &  & dHa      & CH3   & CH3O  & -0.98652 & 0.99999 \\
dHa1     & HCO   & CH2OH & -0.98788 & 0.99898 &  & dHa2     & HCO   & CH2OH & -0.98932 & 0.99573 \\
dHa2     & HCO   & HCO   & -0.98843 & 0.99094 &  & dHa2     & CH3O  & CH3O  & -0.99722 & 0.99942 \\
dHa1     & CH2OH & CH2OH & -0.99982 & 0.99592 &  & dHa2     & CH2OH & CH2OH & -0.99802 & 0.99996 \\
dHa1     & CH3O  & CH3O  & -1.00116 & 0.99809 &  & dHa1     & CH3O  & HCO   & -0.99924 & 0.97255 \\
dHa1     & CH3O  & HCO   & -1.00104 & 0.98860 &  & dHa1     & CH3O  & HCO   & -0.99884 & 0.98571 \\ \cline{1-5}
         &       &       &          &         &  & dHa      & CH2OH & CH3   & -0.99693 & 0.98892 \\
         &       &       &          &         &  & dHa1     & CH2OH & CH2OH & -0.98764 & 0.99148 \\
         &       &       &          &         &  & dHa1     & CH3O  & CH3O  & -0.99668 & 0.99872 \\
         &       &       &          &         &  & dHa2     & HCO   & CH3O  & -0.98176 & 1.00135 \\
         \cline{7-11} 
         \cline{7-11} 
\end{tabular}%
}
\end{table}

\clearpage
\section{S2 values of reactant and TS structures}

\begin{table}[!htbp]
\centering
\caption{$\langle S^{2} \rangle$ values before ($\langle S^{2} \rangle_{bef}$) and after ($\langle S^{2} \rangle_{aft}$) the spin annihilation step of Gaussian for the reactant structures of each studied reactions on the W18 ASW ice model (left) and on the cavity of the W33 ASW ice model (right).}
\label{tab:S2_w18_w33}
\resizebox{0.49\textwidth}{!}{%
\begin{tabular}{lllll}
\hline \hline
\multicolumn{5}{c}{W18}\\
\hline
Rad1     & Rad2     & Reaction & $\langle S^{2} \rangle_{bef}$ & $\langle S^{2} \rangle_{aft}$  \\
\hline
CH$_3$   & CH$_3$   & Rc       & 1.009  & 0.0720  \\
CH$_3$   & HCO      & Rc/dHa   & 1.0101 & 0.0810  \\
CH$_3$   & NH$_2$   & Rc       & 1.0089 & 0.0710  \\
CH$_3$   & CH$_3$O  & Rc       & 1.0105 & 0.0841 \\
CH$_3$   & CH$_3$O  & dHa      & 1.0093 & 0.0849 \\
CH$_3$   & CH$_2$OH & Rc       & 1.0086 & 0.0797 \\
CH$_3$   & CH$_2$OH & dHa      & 1.0086 & 0.0797 \\
         &          &          &        &        \\
HCO      & HCO      & Rc/dHa1  & 1.0114 & 0.0913 \\
HCO      & HCO      & dHa2     & 1.0113 & 0.0904 \\
HCO      & NH$_2$   & Rc/dHa   & 1.0099 & 0.0797 \\
HCO      & CH$_3$O  & Rc       & 1.0115 & 0.0920  \\
HCO      & CH$_3$O  & dHa1     & 1.0113 & 0.0910  \\
HCO      & CH$_3$O  & dHa2     & 1.0115 & 0.0919 \\
HCO      & CH$_2$OH & dHa2     & 1.0113 & 0.0910  \\
HCO      & CH$_2$OH & Rc       & 1.0113 & 0.0903 \\
HCO      & CH$_2$OH & dHa1     & 1.0113 & 0.0903 \\
         &          &          &        &        \\
CH$_3$O  & CH$_3$O  & dHa1     & 1.0111 & 0.0937 \\
CH$_3$O  & CH$_3$O  & dHa2     & 1.0118 & 0.0950  \\
CH$_3$O  & CH$_3$O  & Rc       & 1.0112 & 0.0948 \\
CH$_2$OH & CH$_2$OH & dHa1     & 1.0113 & 0.0906 \\
CH$_2$OH & CH$_2$OH & Rc       & 1.0108 & 0.0894 \\
CH$_2$OH & CH$_2$OH & dHa2     & 1.0114 & 0.0912 \\
\hline \hline
\end{tabular}%
}
\centering
\resizebox{0.49\textwidth}{!}{%
\begin{tabular}{lllll}
\hline \hline
\multicolumn{5}{c}{W33}\\
\hline
Rad1     & Rad2     & Reaction & $\langle S^{2} \rangle_{bef}$ & $\langle S^{2} \rangle_{aft}$ \\
\hline
CH$_3$   & CH$_3$   & Rc       & 1.0089 & 0.0723 \\
CH$_3$   & HCO      & Rc/dHa   & 1.0102 & 0.0823 \\
CH$_3$   & NH$_2$   & Rc       & 1.0082 & 0.0705 \\
CH$_3$   & CH$_3$O  & Rc       & 1.0106 & 0.0847 \\
CH$_3$   & CH$_3$O  & dHa      & 1.0106 & 0.0847 \\
CH$_3$   & CH$_2$OH & Rc       & 1.0100 & 0.0808 \\
CH$_3$   & CH$_2$OH & dHa      & 1.0100 & 0.0808 \\
         &          &          &        &        \\
HCO      & HCO      & Rc       & 1.0114 & 0.0923 \\
HCO      & HCO      & dHa1     & 1.0114 & 0.0923 \\
HCO      & NH$_2$   & Rc/dHa   & 1.0095 & 0.0807 \\
HCO      & CH$_3$O  & Rc       & 1.0116 & 0.0931 \\
HCO      & CH$_3$O  & dHa1     & 1.0117 & 0.0937 \\
HCO      & CH$_3$O  & dHa2     & 1.0116 & 0.0936 \\
HCO      & CH$_2$OH & Rc       & 1.0107 & 0.0880 \\
HCO      & CH$_2$OH & dHa1     & 1.0107 & 0.0880 \\
HCO      & CH$_2$OH & dHa2     & 1.0112 & 0.0923 \\
         &          &          &        &        \\
CH$_3$O  & CH$_3$O  & Rc       & 1.0118 & 0.0951 \\
CH$_3$O  & CH$_3$O  & dHa1     & 1.0096 & 0.0941 \\
CH$_3$O  & CH$_3$O  & dHa2     & 1.0119 & 0.0955 \\
CH$_2$OH & CH$_2$OH & Rc       & 1.0109 & 0.0871 \\
CH$_2$OH & CH$_2$OH & dHa1     & 1.0021 & 0.0854 \\
CH$_2$OH & CH$_2$OH & dHa2     & 1.0110 & 0.0887 \\
\hline \hline
\end{tabular}%
}
\end{table}

\begin{table}[!htbp]
\centering
\caption{$\langle S^{2} \rangle$ values before ($\langle S^{2} \rangle_{bef}$) and after ($\langle S^{2} \rangle_{aft}$) the spin annihilation step of Gaussian for the transition structures of each studied reaction on the W18 ASW ice model (left) and on the cavity of the W33 ASW ice model (right). Values in parenthesis for columns 3 in each table correspond to the reaction step number if the channel is multistep.}
\label{tab:S2_TSs}
\resizebox{0.49\textwidth}{!}{%
\begin{tabular}{lllll}
\hline \hline
\multicolumn{5}{c}{W33}\\
\hline
Rad1     & Rad2     & Reaction & $\langle S^{2} \rangle_{bef}$ & $\langle S^{2} \rangle_{aft}$  \\
\hline
CH$_3$   & CH$_3$   & Rc       & 1.0072 & 0.0713 \\
CH$_3$   & HCO      & Rc       & 1.0095 & 0.0813 \\
CH$_3$   & HCO      & dHa      & 1.0104 & 0.0830  \\
CH$_3$   & NH$_2$   & Rc       & 1.0088 & 0.0709 \\
CH$_3$   & CH$_3$O  & Rc       & 1.0019 & 0.0807 \\
CH$_3$   & CH$_3$O  & dHa      & 0.938  & 0.0820  \\
CH$_3$   & CH$_2$OH & Rc       & 0.996  & 0.0776 \\
CH$_3$   & CH$_2$OH & dHa (1)  & 1.0099 & 0.0793 \\
CH$_3$   & CH$_2$OH & dHa (2)  & 0.8359 & 0.0604 \\
         &          &          &        &        \\
HCO      & HCO      & Rc       & 0.9927 & 0.0880  \\
HCO      & HCO      & dHa1     & 1.0058 & 0.0921 \\
HCO      & HCO      & dHa2     & 0.9993 & 0.0893 \\
HCO      & NH$_2$   & Rc       & 0.9822 & 0.0733 \\
HCO      & NH$_2$   & dHa      & 1.0054 & 0.0801 \\
HCO      & CH$_3$O  & Rc       & 0.9949 & 0.0857 \\
HCO      & CH$_3$O  & dHa1     & 0.9474 & 0.0766 \\
HCO      & CH$_3$O  & dHa2     & 0.8249 & 0.0623 \\
HCO      & CH$_2$OH & Rc       & 0.9911 & 0.0848 \\
HCO      & CH$_2$OH & dHa1     & 1.0052 & 0.0885 \\
HCO      & CH$_2$OH & dHa2     & 0.9831 & 0.0831 \\
         &          &          &        &        \\
CH$_3$O  & CH$_3$O  & Rc       & 0.9475 & 0.0717 \\
CH$_3$O  & CH$_3$O  & dHa1     & 0.8639 & 0.0935 \\
CH$_3$O  & CH$_3$O  & dHa2     & 0.8652 & 0.0942 \\
CH$_2$OH & CH$_2$OH & Rc       & 0.9717 & 0.0774 \\
CH$_2$OH & CH$_2$OH & dHa1     & 0.7698 & 0.0528 \\
CH$_2$OH & CH$_2$OH & dHa2 (1) & 1.0109 & 0.0882 \\
CH$_2$OH & CH$_2$OH & dHa2 (2) & 0.8024 & 0.0563 \\
\hline \hline
\end{tabular}%
}
\hfill
\resizebox{0.49\textwidth}{!}{%
\begin{tabular}{lllll}
\hline \hline
\multicolumn{5}{c}{W33}\\
\hline
Rad1     & Rad2     & Reaction & $\langle S^{2} \rangle_{bef}$ & $\langle S^{2} \rangle_{aft}$  \\
\hline
CH$_3$   & CH$_3$   & Rc       & 0.9875 & 0.0675 \\
CH$_3$   & NH$_2$   & Rc       & 0.9941 & 0.0677 \\
CH$_3$   & HCO      & Rc       & 0.9866 & 0.0760  \\
CH$_3$   & HCO      & dHa      & 0.9532 & 0.0726 \\
CH$_3$   & CH$_3$O  & Rc       & 0.9872 & 0.0779 \\
CH$_3$   & CH$_3$O  & dHa      & 0.9296 & 0.0822 \\
CH$_3$   & CH$_2$OH & Rc (1)   & 1.0101 & 0.0811 \\
CH$_3$   & CH$_2$OH & Rc (2)   & 0.9852 & 0.0739 \\
CH$_3$   & CH$_2$OH & dHa      & 0.8373 & 0.0582 \\
         &          &          &        &        \\
HCO      & HCO      & Rc       & 0.9835 & 0.0864 \\
HCO      & HCO      & dHa1     & 0.9753 & 0.0858 \\
HCO      & HCO      & dHa2     & 1.0048 & 0.0912 \\
HCO      & CH$_3$O  & Rc       & 1.0057 & 0.0898 \\
HCO      & CH$_3$O  & dHa1 (1) & 1.0117 & 0.0939 \\
HCO      & CH$_3$O  & dHa1 (2) & 0.9344 & 0.0772 \\
HCO      & CH$_3$O  & dHa2     & 0.8876 & 0.0766 \\
HCO      & CH$_2$OH & Rc       & 1.011  & 0.0886 \\
HCO      & CH$_2$OH & dHa1     & 0.9695 & 0.0790  \\
HCO      & CH$_2$OH & dHa2 (1) & 1.0111 & 0.0893 \\
HCO      & CH$_2$OH & dHa2 (2) & 0.8068 & 0.0578 \\
         &          &          &        &        \\
CH$_3$O  & CH$_3$O  & Rc       & 0.8303 & 0.0476 \\
CH$_3$O  & CH$_3$O  & dHa1     & 0.8644 & 0.0940  \\
CH$_3$O  & CH$_3$O  & dHa2     & 0.8697 & 0.0949 \\
CH$_2$OH & CH$_2$OH & Rc       & 1.0109 & 0.0879 \\
CH$_2$OH & CH$_2$OH & dHa1 (1) & 1.0038 & 0.0856 \\
CH$_2$OH & CH$_2$OH & dHa1 (2) & 1.0113 & 0.0908 \\
CH$_2$OH & CH$_2$OH & dHa1 (3) & 1.0114 & 0.0919 \\
CH$_2$OH & CH$_2$OH & dHa1 (4) & 1.0084 & 0.0886 \\
CH$_2$OH & CH$_2$OH & dHa1 (5) & 0.7961 & 0.0473 \\
CH$_2$OH & CH$_2$OH & dHa2 (1) & 1.0093 & 0.0859 \\
CH$_2$OH & CH$_2$OH & dHa2 (2) & 0.7997 & 0.0530  \\
\hline \hline
\end{tabular}%
}
\end{table}

\clearpage

\section{Activation energy and temperature dependent efficiencies for those systems not explicitly studied in this work}
\label{sec:appendix:extra_efficiencies}

In this section we present all the $E_a$ and $T$ dependent efficiencies for the systems: OH + CH$_3$/HCO/OH/NH$_2$/CH$_3$O/CH$_2$OH, NH + CH$_3$/HCO/OH/NH$_2$/CH$_3$O/CH$_2$OH, NH$_2$ + NH$_2$/CH$_3$O/CH$_2$OH and CH$_3$O + CH$_2$OH. More details can be found in the main body of the article.

\begin{figure}[!htbp]
    \centering
    \includegraphics[width=\textwidth]{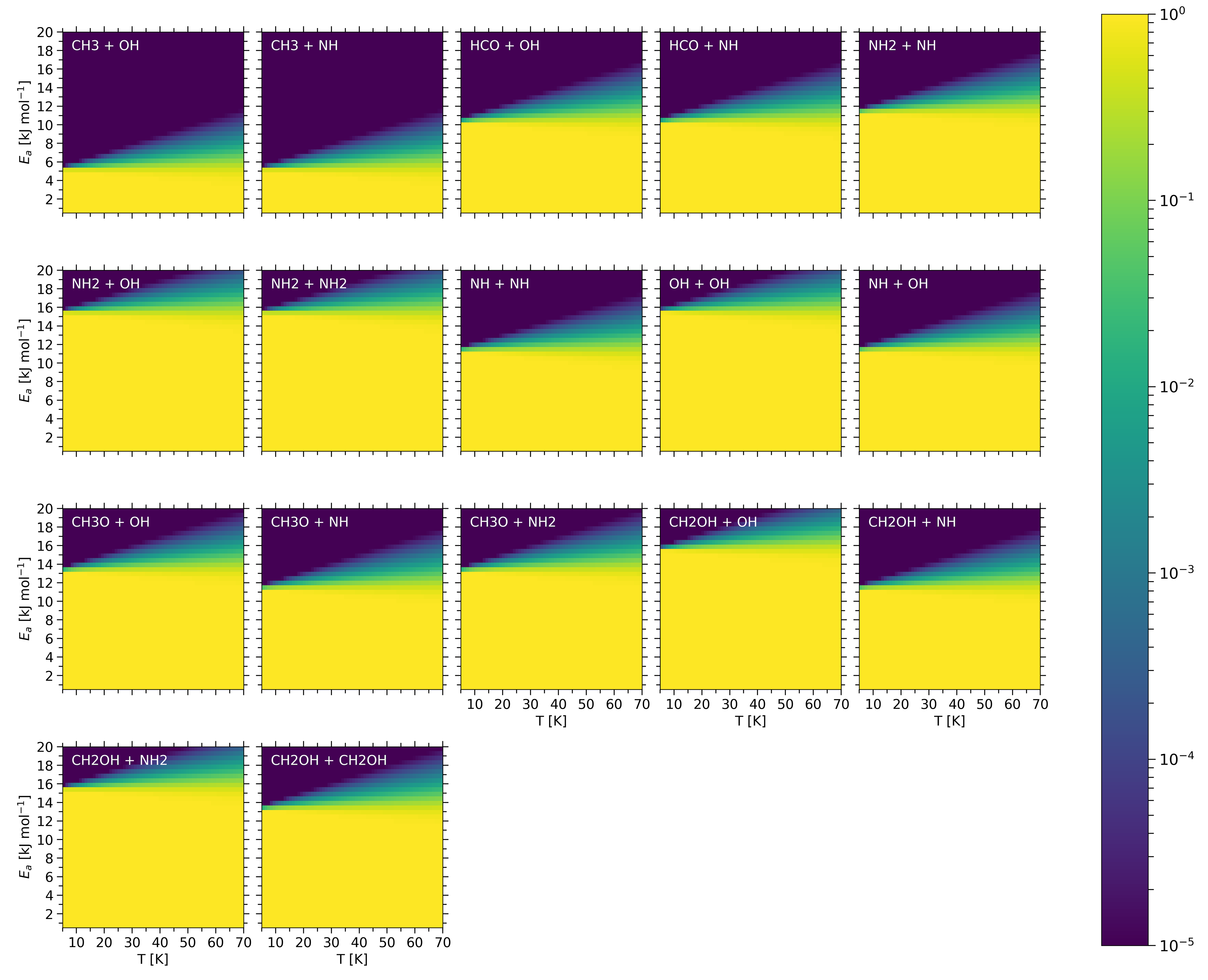}
    \caption{Reaction efficiencies on W33 as a function of activation energy and temperature of those radical--radical reactions in \cite{Garrod2008b} not explicitly studied in this work.}
    \label{fig:supp:guesses}
\end{figure}

\begin{figure}[!htbp]
    \centering
    \includegraphics[width=\textwidth]{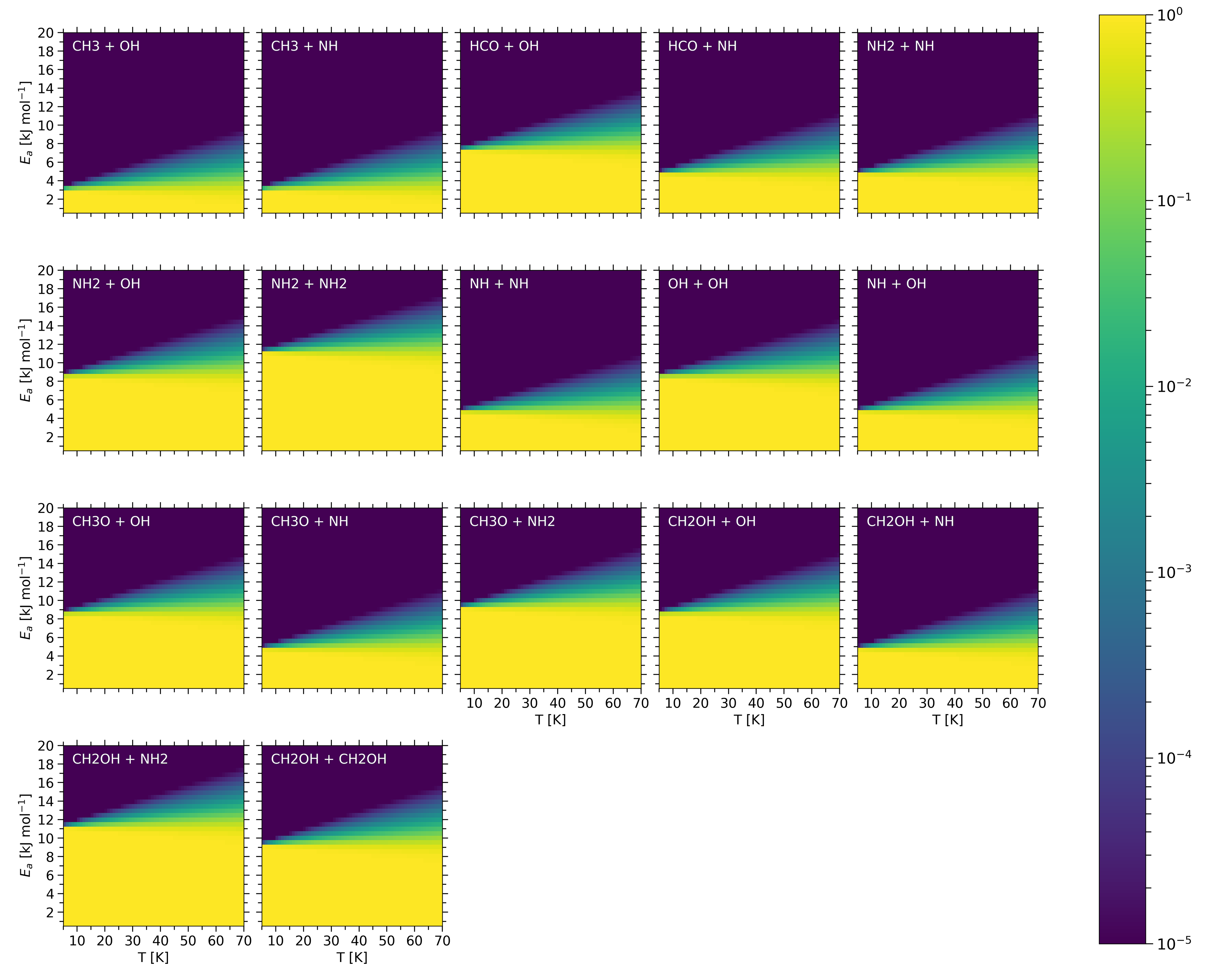}
    \caption{Reaction efficiencies on W18 as a function of activation energy and temperature of those radical--radical reactions in \cite{Garrod2008b} not explicitly studied in this work.}
    \label{fig:supp:guesses_w18}
\end{figure}

\clearpage
\section{Crossover temperature formula}
\label{sec:appendix:crossover_temp}

In order to calculate the crossover temperatures ($T_c$), we have used equation \ref{eqn:CriticalTemperature} \citep{FermannAuerbach2000_Tc}. At temperatures below $T_c$ tunnelling effects become dominant, and above it tunnelling is negligible.

\begin{equation}
    T_c = \frac{\hbar\omega^{\ddagger}\Delta H^{\ddagger}/k_B}{2\pi \Delta H^{\ddagger}-\hbar\omega^{\ddagger}\ln(2)}
    \label{eqn:CriticalTemperature}
\end{equation}

\noindent where $\hbar$ is the reduced Plank constant, $\omega^{\ddagger}=2\pi \nu^{\ddagger}$ with $\nu^{\ddagger}$ the frequency (in absolute value) associated to the transition state, $\Delta H^{\dagger}$ is the ZPE-corrected energy barrier at 0 K and $k_B$ is the Boltzmann constant.

\section{XYZ structures}
The XYZ data is available in a separated file uploaded to Zenodo: \cite{zenodo_data}.

\end{document}